\documentclass[11pt]{article}
\usepackage{amssymb,latexsym, amsmath}
\newcommand{\bfi}{\bfseries\itshape}

\makeatletter
\@addtoreset{figure}{section}
\def\thefigure{\thesection.\@arabic\c@figure}
\def\fps@figure{h, t}
\@addtoreset{table}{bsection}
\def\thetable{\thesection.\@arabic\c@table}
\def\fps@table{h, t}
\@addtoreset{equation}{section}

\makeatother

\newtheorem{thm}{Theorem}[section]

\newtheorem{lem}[thm]{Lemma}
\newtheorem{cor}[thm]{Corollary}

\newtheorem{cnj}[thm]{Conjecture}

\begin{document}

\title{The Euler--Poincar\'{e} Equations and Semidirect Products
with Applications to Continuum Theories}
\author{Darryl D. Holm
\\Theoretical Division and Center for Nonlinear Studies
\\Los Alamos National
Laboratory, MS B284
\\ Los Alamos, NM 87545
\\ {\footnotesize dholm@lanl.gov}
\and
Jerrold E. Marsden\thanks{Research partially supported by
NSF grant DMS 96--33161.}\\Control and Dynamical Systems\\
California Institute of Technology 107-81\\ Pasadena, CA 91125
\\ {\footnotesize  marsden@cds.caltech.edu}
\and
Tudor S. Ratiu
\thanks{Research partially supported by NSF Grant DMS-9503273 and DOE
contract DE-FG03-95ER25245-A000.}
\\Department of Mathematics
\\University of California, Santa Cruz, CA 95064
\\ {\footnotesize  ratiu@math.ucsc.edu} }
\date{}

\maketitle

\begin{abstract}
We study Euler--Poincar\'e systems (i.e., the Lagrangian
analogue of Lie-Poisson Hamiltonian systems) defined on
semidirect product Lie algebras. We first give a derivation of
the Euler--Poincar\'e equations for a parameter dependent
Lagrangian by using a variational principle of Lagrange
d'Alembert type. Then we derive an abstract Kelvin-Noether
theorem for these equations. We also explore their relation with
the theory of Lie-Poisson Hamiltonian systems defined on the
dual of a semidirect product Lie algebra. The Legendre
transformation in such cases is often not invertible; so it does
not produce a corresponding Euler--Poincar\'e system on that Lie
algebra. We avoid this potential difficulty by developing the
theory of Euler--Poincar\'e systems entirely within the
Lagrangian framework. We apply the general theory to a number of
known examples, including the heavy top, ideal compressible
fluids and MHD. We also use this framework to derive higher
dimensional Camassa-Holm equations, which have many potentially
interesting analytical properties. These equations are
Euler-Poincar\'e equations for geodesics on diffeomorphism
groups (in the sense of the Arnold program) but where the metric
is $H^1$ rather than $L^2$.

\vspace{0.25in}
\footnoterule
\noindent
{\it Submitted to Advances in Mathematics}

\end{abstract}
\tableofcontents

\section{History and Background} \label{sect-history}

To put our paper in context, we shall
pick up the thread of the history of mechanics in the later
part of the 1800's. By that time, through the work of many
people, including Euler, Lagrange, Hamilton, Jacobi, and
Routh, it was well understood that the equations
of mechanics are expressible in either Hamiltonian or
Lagrangian form.

The Lagrangian formulation of mechanics can be based on the
variational principles behind Newton's fundamental laws of force
balance $\mathbf{F}= m\mathbf{a}$. One chooses a configuration space
$Q$ (a manifold, assumed to be of finite dimension $n$ to start the
discussion) with coordinates denoted $ q^i,i = 1,\ldots, n$, that
describe the configuration of the system under study. One then forms
the velocity phase space $TQ $ (the tangent bundle of $Q$).
Coordinates on $TQ$ are denoted $(q^1, \dots, q^n, \dot q^1,\ldots,
\dot q ^n),$ and the Lagrangian is regarded as a function $L: TQ \to
\mathbb{R}$. In coordinates, one writes $L(q^i,\dot {q}^i, t)$, which
is shorthand notation for $L(q ^1 , \ldots , q ^n , \dot{q} ^1 ,
\ldots , \dot{q} ^n,t )$.  Usually, $L$ is the kinetic  minus the
potential energy of the system and one takes
$\dot {q}^i = dq^i/dt$ to be  the system velocity. The variational
principle of Hamilton states that the variation of the action is
stationary at a solution:
\begin{equation}\label{vp1}
 \delta \mathfrak{S} = \delta \int _a ^b  L(q ^i,\dot{ q} ^i ,t) \,
dt = 0.
\end{equation}
In this principle, one chooses curves $q^i(t)$ joining two fixed
points in $Q$ over a fixed time interval $[a, b]$, and
calculates the action $\mathfrak{S}$, which is the time integral of
the Lagrangian, regarded as a function of this curve. Hamilton's
principle states that the action $\mathfrak{S}$ has a critical point
at a solution in the space of curves.  As is well known, Hamilton's
principle is equivalent to the Euler--Lagrange equations:
\begin{equation}\label{el}
\frac{d}{dt} \frac{\partial L}{\partial \dot{q} ^i}  -\frac{
\partial L }{ \partial q^i} = 0, \quad i = 1, \ldots , n.
\end{equation}

If the system is subjected to external forces, these are to be
added to the right hand side of the Euler-Lagrange equations.
For the case in which $L$ comprises kinetic minus potential
energy, the Euler-Lagrange equations reduce to a geometric form
of Newton's second law.  For Lagrangians that are purely
kinetic energy, it was already known in Poincar\'{e}'s time
that the corresponding solutions of the Euler-Lagrange equations are
geodesics. (This fact was certainly known to Jacobi by 1840, for
example.)

To pass to the Hamiltonian formalism, one introduces
the conjugate momenta
\begin{equation}\label{cm} p_i = \frac{\partial L}{\partial
\dot{q}^i} ,	 \quad i=1, \ldots , n,
\end{equation}
and makes the change of variables $ (q^i,\dot{q}^i) \mapsto
(q^i, p_i)$, by a Legendre transformation. The Lagrangian is called
{\it regular\/} when this change of variables is invertible. The
Legendre transformation introduces the Hamiltonian
\begin{equation}\label{ham1}
H(q^i, p_i, t)  =  \sum _{ j=1} ^n p_j
\dot{q}^j - L(q^i, \dot{q} ^i , t)	.
\end{equation}
One shows that the Euler--Lagrange equations are equivalent
to Hamilton's equations:
\begin{equation}\label{hameq1}
\frac{dq^i}{dt}  =  \frac{\partial H}{\partial p_i},
\quad
\frac{dp_i}{dt}  =  -\frac{ \partial H }{ \partial q ^i},
\end{equation}
where $i=1, \ldots , n$.
There are analogous Hamiltonian partial differential equations
for field theories such as Maxwell's equations and the
equations of fluid and solid mechanics.

Hamilton's equations can be recast
in Poisson bracket form as
\begin{equation}\label{pbf}
\dot{ F} = \{F,H \},
\end{equation}
where the canonical Poisson brackets are given by
\begin{equation}\label{pb1}
\{F, G\} = \sum ^n _{ i = 1}
\left( \frac{ \partial F }{ \partial q ^i}
\frac{ \partial G }{ \partial p _i}
- \frac{ \partial F }{ \partial p_i}
\frac{ \partial G }{ \partial q ^i} \right).
\end{equation}

Associated to any configuration space $Q$ is a phase space
$T^{*}Q$ called the cotangent bundle of $Q$, which has
coordinates $(q ^1, \dots , q ^n, p _1, \dots, p _n$).  On this
space, the canonical Poisson bracket is intrinsically
defined in the sense that the value of $\{ F, G \}$ is
independent of the choice of coordinates. Because the Poisson
bracket satisfies $\{ F,G \} = - \{G, F\}$ and in particular
$\{H,H\}=0$, we see that $\dot H = 0$; that is, energy is
conserved along solutions of Hamilton's equations.  This is the
most elementary of many deep and beautiful conservation
properties of mechanical systems.

\paragraph{Poincar\'{e} and the Euler equations.} Poincar\'{e}
played an enormous role in the topics treated in the present
paper. We mention a few of Poincar\'{e}'s contributions that are relevant
here. First is his work on the gravitating fluid problem, continuing the
line of investigation begun by MacLaurin, Jacobi and Riemann. Some
solutions of this problem still bear his name today. This work is
summarized in Chandrasekhar [1967, 1977] (see Poincar\'{e} [1885,
1890, 1892, 1901a] for the original treatments).  This background led
to his famous paper, Poincar\'{e} [1901b], in which he laid out the
basic equations of Euler type, including the rigid body, heavy top and
fluids as special cases. Abstractly, these equations are determined
once one is given a Lagrangian on a Lie algebra. We shall make
some additional historical comments on this situation below,
after we present a few more mechanical preliminaries. It is because
of the paper Poincar\'{e} [1901b] that the name {\bfi Euler--Poincar\'e
equations} is now used for these equations.

To state the Euler--Poincar\'e equations, let $\mathfrak{g}$ be
a given Lie algebra and let $ l : \mathfrak{g} \rightarrow
\mathbb{R} $ be a given function (a La\-gran\-gian), let $\xi$ be a
point in $\mathfrak{g}$ and let $f \in \mathfrak{g}^\ast$ be
given forces (whose nature we shall explicate later).
Then the evolution of the variable $\xi$ is determined by the
Euler--Poincar\'e equations. Namely,
$$
\frac{d}{dt} \frac{\delta l}{\delta \xi} =
\operatorname{ad}_{\xi}^{*}\frac{\delta l}{\delta \xi} + f.
$$
The notation is as follows: $\partial l / \partial \xi \in
\mathfrak{g}^\ast$ (the dual vector space) is the derivative of
$l$ with respect to $\xi$; we use partial derivative notation
because $l$ is a function of the vector
$\xi$ and because shortly $l$ will be a function of other
variables as well. The map $\operatorname{ad}_{\xi} : \mathfrak{g} \to
\mathfrak{g}$ is the linear map $\eta \mapsto [\xi, \eta]$,
where $[\xi, \eta ]$ denotes the Lie bracket of $\xi$ and
$\eta$, and where $\operatorname{ad}_{\xi}^{*}: \mathfrak{g}^{*} \to
\mathfrak{g}^{*}$ is its dual (transpose) as a linear map. In the
case that $ f = 0 $, we will call these equations the
{\bfi basic Euler--Poincar\'e equations}.

These equations are valid for either finite or infinite
dimensional Lie algebras. For fluids, Poincar\'{e} was aware that
one needs to use infinite dimensional Lie algebras, as is clear
in his paper Poincar\'{e} [1910]. He was aware that one has to
be careful with the signs in the equations; for example, for rigid
body dynamics one uses the equations as they stand, but for
fluids, one needs to be careful about the conventions for the Lie
algebra operation $\operatorname{ad}_{\xi}$; cf. Chetayev [1941].

To state the equations in the finite dimensional case in
coordinates, one must choose a basis $e_1 , \ldots, e_r$   of
$\mathfrak{g}$ (so dim$\,{\mathfrak{g}} = r)$.
Define, as usual, the structure constants
$C^d_{ab}$ of the Lie algebra by
\begin{equation}\label{sc}
[e_a, e_b]  = \sum _{ d=1} ^r C^d_{ab}e_d,
\end{equation}
where $a,b$ run from $1$ to $r$. If $\xi \in \mathfrak{g} $, its
components relative to this basis are denoted $\xi ^a$.   If
$e^1, \ldots, e^n$ is the corresponding dual basis, then the
components of the differential of the Lagrangian
$l$ are  the partial derivatives $\partial l / \partial \xi ^a$.
The Euler--Poincar\'e equations in this basis are
\begin{equation}\label{aepe}
\frac{d }{d t} \frac{\partial l}{\partial \xi ^b } = \sum _{a,d
= 1} ^r C ^d _{ a b}
\frac{\partial l}{\partial \xi ^d }  \xi ^a + f _b .
\end{equation}

For example, consider the Lie algebra $\mathbb{R}^3$ with the
usual vector cross product. (Of course, this is the Lie algebra of the
proper rotation group in $ \mathbb{R}^3$.) For $l: \mathbb{R}^3
\rightarrow \mathbb{R}$, the Euler--Poincar\'{e} equations become
\[
  \frac{ d}{dt} \frac{\partial l}{\partial \boldsymbol{\Omega}}
     = \frac{\partial l}{\partial \boldsymbol{\Omega}} \times
\boldsymbol{\Omega} + \mathbf{f},
\]
which generalize the Euler equations for rigid body motion.

These equations were written down for a certain class of
Lagrangians $l$ by Lagrange [1788, Volume 2, Equation A on p.
212], while it  was Poincar\'{e} [1901b] who generalized them
(without reference to the ungeometric Lagrange!) to an arbitrary Lie
algebra. However, it was La\-gran\-ge who was grappeling with
the derivation and deeper understanding of the nature of these
equations. While Poincar\'{e} may have understood how to derive
them from other principles, he did not reveal this.

Of course, there was a lot of mechanics going on in the decades
leading up to Poincar\'{e}'s work and we shall comment on some
of it below. However, it is a curious historical fact that the
Euler--Poincar\'e equations were not pursued extensively until
quite recently. While many authors mentioned these equations and
even tried to understand them more deeply (see, e.g., Hamel [1904,
1949] and Chetayev [1941]), it was not until the Arnold school that
this understanding was at least partly achieved (see Arnold [1966a,c]
and Arnold [1988]) and was used for diagnosing hydrodynamical
stability (e.g., Arnold [1966b]).

It was already clear in the last century that certain
mechanical systems resist the usual canonical formalism,
either Hamiltonian or Lagrangian, outlined in the first
paragraph. The rigid body provides an elementary
example of this. In another example, to obtain a Hamiltonian
description for ideal fluids, Clebsch [1857, 1859] found it
necessary to introduce certain nonphysical potentials\footnote{For
modern accounts of Clebsch potentials and further references, see
Holm and Kupershmidt [1983], Marsden and Weinstein [1983], Marsden,
Ratiu, and Weinstein [1984a,b], Cendra and Marsden [1987],
Cendra, Ibort, and Marsden [1987] and Goncharov and Pavlov [1997].}.

\paragraph{More about the rigid body.} In the absence of external
forces, the rigid body equations are usually written as follows:
\begin{equation}\label{rbe1}
\begin{aligned}
I_1\dot{\Omega}{_1}  =  (I_2 -I_3)\Omega_2\Omega_3,
\\
I_2\dot{\Omega}{_2}  =  (I_3 -I_1)\Omega_3\Omega_1,
\\
I_3\dot{\Omega}{_3}  =  (I_1 -I_2)\Omega_1\Omega_2,
\end{aligned}
\end{equation}
where $\boldsymbol{\Omega} = (\Omega_1, \Omega_2, \Omega_3)$ is the
body angular velocity vector and  $I_1,  I_2, I_3$  are the moments
of inertia of the rigid body.  Are these equations as written
Lagrangian or Hamiltonian in any sense?  Since there are an odd
number of equations, they cannot be put in canonical
Hamiltonian form.

One answer is to reformulate the equations on $T{\rm SO(3)}$
or $T ^\ast{\rm SO(3)}$, as is classically done in  terms of
Euler angles and their velocities or conjugate momenta,
relative to which the equations {\em are\/} in Euler--Lagrange
or canonical Hamiltonian form.  However, this reformulation answers a
different question for a {\it six} dimensional system. We are
interested in these structures for the equations as given
above.

The Lagrangian answer is easy: these equations
have Euler--Poincar\'e form on the Lie algebra $ \mathbb{R}^3$
using the Lagrangian
\begin{equation}\label{rbl-1}
l (\boldsymbol{\Omega}) = \frac{1}{2} (I _1 \Omega ^2 _1 + I _2
\Omega ^2 _2 + I _3 \Omega ^2 _3 ).
\end{equation}
which is the (rotational) kinetic energy of the
rigid body.

One of our main messages is that the Euler--Poincar\'e
equations possess a natural variational principle. In fact, the
Euler rigid body equations are equivalent to the {\bfi rigid body
action principle}
\begin{equation}\label{rbvp1}
\delta \mathfrak{S}_{\rm red}=\delta \int^b_a l \, d t = 0,
\end{equation}
where variations of $\boldsymbol{\Omega} $ are restricted to
be of the form
\begin{equation}\label{rpvp2}
\delta \boldsymbol{\Omega} = \dot{ \boldsymbol{\Sigma}} +
\boldsymbol{\Omega} \times \boldsymbol{\Sigma},
\end{equation}
in which $\boldsymbol{\Sigma}$ is a curve in $\mathbb{R}^3$ that
vanishes at the endpoints. As before, we regard the {\bfi reduced
action} $\mathfrak{S}_{\rm red}$ as a function on the space of
curves, but only consider variations of the form described. The
equivalence of the rigid body equations and the rigid body action
principle may be proved in the same way as one proves that Hamilton's
principle is equivalent to the Euler--Lagrange equations: Since
$ l (\boldsymbol{\Omega}) = \frac{1}{2} \langle  {\mathbb I}
\boldsymbol{\Omega}, \boldsymbol{\Omega} \rangle $, and ${\mathbb I}$
is symmetric, we obtain
\begin{align*}
\delta \int^b_a l \, d t
& =  \int^b_a \langle  {\mathbb I} \boldsymbol{\Omega},\delta
\boldsymbol{\Omega}\rangle \, dt \\
& =  \int^b_a \langle  {\mathbb I} \boldsymbol{\Omega},
\dot{\boldsymbol{\Sigma}}
      + \boldsymbol{\Omega} \times \boldsymbol{\Sigma}\rangle \, dt\\
& =  \int^b_a \left[ \left\langle
      - \frac{d }{d t}{\mathbb I} \boldsymbol{\Omega} ,
      \boldsymbol{\Sigma}\right\rangle + \left\langle {\mathbb I}
\boldsymbol{\Omega},
      \boldsymbol{\Omega} \times
\boldsymbol{\Sigma}\right\rangle\right] \\
& = \int^b_a \left\langle
- \frac{d }{d t} {\mathbb I} \boldsymbol{\Omega}
      + {\mathbb I} \boldsymbol{\Omega}\times \boldsymbol{\Omega},
\boldsymbol{\Sigma} \right\rangle d t,
\end{align*}
upon integrating by parts and using the
endpoint conditions, $ \boldsymbol{\Sigma} (b) =
\boldsymbol{\Sigma} (a) = 0 $. Since $\boldsymbol{\Sigma}$ is
otherwise arbitrary,~(\ref{rbvp1}) is equivalent to
\[ - \frac{d }{d t} ({\mathbb I} \boldsymbol{\Omega})
+ {\mathbb I} \boldsymbol{\Omega} \times \boldsymbol{\Omega}
= 0 , \]
which are Euler's equations.

Let us explain in concrete terms (that will be abstracted
later) how to {\it derive\/} this variational principle from the
{\it standard} variational principle of Hamilton.

We regard an element $\mathbf{ R} \in {\rm SO}(3)$
giving the configuration of the body as a map of a reference
configuration ${\mathcal B}\subset \mathbb{R}^3$ to the current
configuration $\mathbf{ R}({\mathcal B})$; the map
$\mathbf{ R}$ takes a reference or label point $ X \in
{\mathcal B} $ to a current point $ x = \mathbf{ R}(X) \in
\mathbf{ R}({\mathcal B} )$. When the rigid body is in motion,
the matrix $\mathbf{ R}$ is time-dependent and the velocity of a
point of the body is $ \dot{ x} = \dot{ \mathbf{ R}} X = \dot{
\mathbf{ R}} \mathbf{ R} ^{-1} x. $ Since $\mathbf{ R}$ is an
orthogonal matrix, $\mathbf{ R} ^{-1} \dot{ \mathbf{ R}}$ and
$\dot{\mathbf{ R}}\mathbf{ R}^{-1}$ are skew matrices, and so
we can write
\begin{equation}\label{sav}
\dot{x} = \dot{ \mathbf{ R}}\mathbf{ R} ^{-1} x = \boldsymbol{\omega}
\times x,
\end{equation}
which defines the {\bfi spatial angular velocity
vector\/} $\boldsymbol{\omega}$.  Thus, $\boldsymbol{\omega}$ is
essentially given by {\em right\/} translation of $\dot{\mathbf{ R}
}$ to the identity.

The corresponding body angular velocity is defined by
\begin{equation}\label{bav}
\boldsymbol{\Omega} = \mathbf{ R} ^{ -1} \boldsymbol{\omega} ,
\end{equation}
so that $\boldsymbol{\Omega}$ is the angular velocity
relative to a body fixed frame. Notice that
\begin{align}\label{lefttranslation}
\mathbf{ R} ^{ -1} \dot{\mathbf{ R} } X
&= \mathbf{ R} ^{ -1} \dot \mathbf{ R}
        \mathbf{ R} ^{ -1} x = \mathbf{ R} ^{ -1}
        (\boldsymbol{\omega} \times x) \nonumber \\
&= \mathbf{ R} ^{ -1} \boldsymbol{\omega} \times \mathbf{ R}
^{ -1} x = \boldsymbol{\Omega} \times X,
\end{align}
so that $\boldsymbol{\Omega}$ is given by {\em left\/} translation
of $\dot \mathbf{R} $ to the identity.
The kinetic energy is obtained by summing up $m | \dot{x}
| ^2 / 2$ (where $|{\bullet}|$ denotes the Euclidean norm) over the body:
\begin{equation}\label{ke}
 K = \frac{1}{2} \int _{\mathcal B} \rho(X) |
\dot{ \mathbf{ R} } X | ^2 \, d^3 X,
\end{equation}
in which $\rho$ is a given mass density in the reference
configuration. Since
\[
| \dot{ \mathbf{ R}} X |  = | \boldsymbol{\omega} \times x | = |
\mathbf{ R} ^{-1}(\boldsymbol{\omega} \times x ) | = |
\boldsymbol{\Omega} \times X |,
\]
$K$ is a quadratic function of $\boldsymbol{\Omega}$. Writing
\begin{equation}\label{mit}
K = \frac{1}{2} \boldsymbol{\Omega} ^T {\mathbb I}\boldsymbol{\Omega}
\end{equation}
defines the {\bfi moment of inertia
tensor\/} ${\mathbb I},$ which, provided the
body does not degenerate to a line, is a positive-definite
$(3\times 3)$ matrix, or better, a quadratic form.  This quadratic
form can be diagonalized by a change of basis; thereby defining the
principal axes and moments of inertia. In this basis, we write
$ {\mathbb I} = {\rm diag}(I _1, I _2, I_3). $ The function $K$ is
taken to be the Lagrangian of the system on
$ T{\rm SO}(3) $ (and by means of the Legendre transformation we obtain
the corresponding Hamiltonian description on
$ T ^{\ast} {\rm SO}(3)$). Notice that $K$ in equation (\ref{ke}) is
{\em left\/} (not right) invariant on $T{\rm SO}(3)$.  It follows
that the corresponding Hamiltonian is also {\em left\/} invariant.

In the Lagrangian framework, the relation between
motion in $\mathbf{ R}$ space and motion in body angular velocity
(or $\boldsymbol{\Omega}$) space is as follows: The curve
$ \mathbf{R} (t) \in {\rm SO}(3) $ satisfies the Euler-Lagrange
equations for
\begin{equation}\label{rbl-2}
 L (\mathbf{ R}, \dot{ \mathbf{ R}} )
= \frac{1}{2} \int_{{\mathcal B}}
\rho (X)
|  \dot{ \mathbf{ R}} X | ^2 \, d ^3 X,
\end{equation}
if and only if $ \boldsymbol{\Omega} (t) $ defined by
$\mathbf{ R} ^{-1} \dot{ \mathbf{ R}} \mathbf{v} = \boldsymbol{\Omega}
\times \mathbf{v}$ for all $\mathbf{v} \in \mathbb{R}^3$ satisfies
Euler's equations
\begin{equation}\label{ee}
{\mathbb I} \dot{ \boldsymbol{\Omega}} = {\mathbb
I}\boldsymbol{\Omega}\times \boldsymbol{\Omega}.
\end{equation}

An instructive proof of this relation involves understanding how
to reduce variational principles using their symmetry groups.
By Hamilton's principle, $\mathbf{R}(t)$
satisfies the Euler-Lagrange equations, if and only if
\[
\delta\int L \, dt = 0 .
\]
 Let $ l (\boldsymbol{\Omega}) = \frac{1}{2} ( {\mathbb I}
\boldsymbol{\Omega})
\cdot \boldsymbol{\Omega} $, so that
$ l (\boldsymbol{\Omega}) = L (\mathbf{ R}, \dot{ \mathbf{ R}} ) $
if $\mathbf{ R}$ and
$\boldsymbol{\Omega}$ are related as above. To see how we should
transform Hamilton's principle, define the skew matrix
$\hat {\boldsymbol{\Omega}}$  by
$\hat{\boldsymbol{\Omega}}\mathbf{v} = \boldsymbol{\Omega} \times
\mathbf{v}$ for any $\mathbf{v} \in \mathbb{R}^3$, and
differentiate the relation
$\mathbf{ R} ^{-1} \dot{ \mathbf{ R}} = \hat{\boldsymbol{\Omega} }$
with respect to $\mathbf{ R}$ to get
\begin{equation}\label{rda}
- \mathbf{ R} ^{-1} (\delta \mathbf{ R})\mathbf{ R}^{-1}
\dot{\mathbf{R}}  + \mathbf{ R}^{-1} (\delta
\dot{ \mathbf{ R}}) = \widehat{\delta \boldsymbol{\Omega}} .
\end{equation}
Let the skew matrix $\hat{\boldsymbol{\Sigma}}$ be defined by
\begin{equation}\label{dsig}
 \hat{\boldsymbol{\Sigma}} = \mathbf{R}^{-1} \delta \mathbf{R},
\end{equation}
and define the vector $\boldsymbol{\Sigma}$ by
\begin{equation}\label{dhat}
\hat{\boldsymbol{\Sigma}} \mathbf{v} = \boldsymbol{\Sigma} \times
\mathbf{v} .
\end{equation}
Note that
\[
\dot{ \hat{\boldsymbol{\Sigma}}}
= - \mathbf{ R}^{-1} \dot{ \mathbf{ R}} \mathbf{ R} ^{-1}
\delta \mathbf{ R} + \mathbf{ R} ^{-1}
\delta\dot{\mathbf{ R}},
\]
so
\begin{equation}\label{dela}
\mathbf{ R} ^{-1} \delta \dot{ \mathbf{ R}}
= \dot{ \hat{\boldsymbol{\Sigma}}} + \mathbf{ R} ^{-1}
\dot{ \mathbf{ R}} \hat{\boldsymbol{\Sigma}}\,.
\end{equation}
Substituting~(\ref{dela}) and~(\ref{dsig})
into~(\ref{rda}) gives
\[ - \hat{\boldsymbol{\Sigma}} \hat{\boldsymbol{\Omega}} + \dot{
\hat{\boldsymbol{\Sigma}}} + \hat{
\boldsymbol{\Omega}} \hat{\boldsymbol{\Sigma}}
= \widehat{ \delta \boldsymbol{\Omega}}, \]
that is,
\begin{equation}\label{domh}
\widehat{ \delta \boldsymbol{\Omega}} = \dot{ \hat{
\boldsymbol{\Sigma} }} + [
\hat{\boldsymbol{\Omega}},
\hat{\boldsymbol{\Sigma}} ] .
\end{equation}
The identity
$ [ \hat{\boldsymbol{\Omega}}, \hat{\boldsymbol{\Sigma}} ] =
(\boldsymbol{\Omega} \times
\boldsymbol{\Sigma}) \hat{\ } $ holds by Jacobi's identity for the
cross product and so
\begin{equation}\label{dom}
\delta \boldsymbol{\Omega} = \dot{\boldsymbol{\Sigma}} +
\boldsymbol{\Omega}
\times \boldsymbol{\Sigma} .
\end{equation}

\noindent
These calculations prove the following:
\begin{thm}  \label{red/variational/prpl}
Hamilton's variational principle
\begin{equation}\label{hprin}
\delta \mathfrak{S} = \delta \int^b_a L \, d t = 0
\end{equation}   on $T{\rm SO}(3)$ is equivalent to the {\bfi
reduced variational principle\/}
\begin{equation}\label{rvprin}
\delta \mathfrak{S}_{\rm red}= \delta \int^b_a l \, d t = 0
\end{equation}  on $\mathbb{R}^3$ where the variations $ \delta
\boldsymbol{\Omega} $ are of the form~(\ref{dom}) with
$\boldsymbol{\Sigma}(a) =\boldsymbol{\Sigma}(b) = 0 $.
\end{thm}

\paragraph{Hamiltonian Form.}
If, instead of variational principles, we concentrate on
Poisson brackets and drop the requirement that they be in the
canonical form, then there is also a simple and
beautiful Hamiltonian structure for the rigid body equations
that is now well known\footnote{See Marsden and Ratiu [1994]
for details, references, and the history of this structure.}.  To
recall this, introduce the angular momenta
\begin{equation}\label{rbm}
\Pi_i  =  I_i\Omega_i = \frac{\partial L}{\partial \Omega _i } ,
\quad i = 1, 2, 3,
\end{equation}
so that the Euler equations become
\begin{equation}\label{rbe2}
\begin{aligned}
\dot {\Pi}_1  &=  \frac{I_2 - I_3}{I_2I_3} \Pi_2\Pi_3,\\
		\dot {\Pi}_2  &=  \frac{I_3 - I_1}{I_3I_1} \Pi_3\Pi_1,\\
		\dot {\Pi}_3  &=  \frac{I_1 - I_2}{I_1I_2} \Pi_1\Pi_2,
\end{aligned}
\end{equation}
that is,
\begin{equation}\label{rbe3}
\dot{{\boldsymbol\Pi }} = {\boldsymbol\Pi}
\times {\boldsymbol \Omega}.
\end{equation}
Introduce the following rigid body Poisson bracket on
functions of the ${\boldsymbol \Pi}$'s:
\begin{equation}\label{rbb}
 \{F,G\}({\boldsymbol  \Pi})
= -{\boldsymbol  \Pi} \cdot (\nabla_{\Pi} F \times \nabla_{\Pi} G)
 \end{equation} and the Hamiltonian
\begin{equation}\label{rbh} H
=\frac {1}{2} \left(\frac{\Pi^2_1}{I_1}
+ \frac{\Pi^2_2}{I_2}
+ \frac{\Pi^2_3}{I_3} \right).
\end{equation}
One checks that Euler's equations are equivalent to
$ \dot{F} = \{F, H\}.$

The rigid body variational principle and the rigid body Poisson
bracket are special cases of general constructions associated
to any Lie algebra $\mathfrak{g}$.  Since we have already
described the general Euler--Poincar\'e construction on
$\mathfrak{g}$, we turn next to the Hamiltonian counterpart on
the dual space.

\paragraph{The Lie-Poisson Equations.}
Let
$F, G$ be real valued functions on the dual space  $\mathfrak{g}
^{\ast}$.  Denoting elements of $\mathfrak{g} ^{\ast}$ by
$\mu$, let the functional derivative of
$F$   at $\mu$ be the unique element $\delta  F/ \delta \mu$ of
$\mathfrak{g}$ defined by
\begin{equation}\label{fd}
\lim_{\varepsilon \rightarrow 0}
\frac{1}{\varepsilon }[F(\mu
+ \varepsilon \delta \mu) - F(\mu )]
=  \left\langle \delta  \mu , \frac{
\delta  F}{\delta  \mu } \right\rangle ,
\end{equation}
for all $\delta\mu \in \mathfrak{g}^*$, where $\left\langle \,,
\right\rangle$ denotes the pairing between
$\mathfrak{g} ^{\ast} $ and  $\mathfrak{g} $. Define the
$(\pm)$  Lie-Poisson brackets by
\begin{equation}\label{lpb}
\{F, G\}_\pm (\mu )  =  \pm \left\langle \mu , \left[ \frac{ \delta
F}{\delta  \mu},
\frac{\delta  G}{\delta \mu } \right] \right\rangle .
\end{equation}
Using the coordinate notation introduced above, the
$ (\pm)$ Lie-Poisson brackets become
\begin{equation}\label{lpbc}
\{F, G\} _\pm (\mu) =  \pm \sum _{ a,b,d=1} ^r  C^d_{ab} \mu_d
\frac{\partial F}{\partial
\mu_a}\frac{ \partial G}{\partial \mu_b},
\end{equation}
where $\mu = \sum _{d=1} ^r\mu_d e ^d$.

The Lie-Poisson equations, determined by $\dot F = \{ F,H\} $
read
$$
\dot \mu _a = \pm \sum _{b,d=1} ^r  C^d_{ab} \mu_d
\frac{\partial H}{\partial \mu _b},
$$
or intrinsically,
\begin{equation}\label{alpe}
\dot \mu  = \mp \operatorname{ad} ^{* }_{\partial H/\partial \mu} \mu
.
\end{equation}

This setting of mechanics is a special case of the general
theory of systems on Poisson manifolds, for which there is now
an extensive theoretical development. (See Guillemin and Sternberg
[1984] and Marsden and Ratiu [1994] for a start on this literature.)
There is an especially important  feature of the rigid body bracket
that carries  over to general Lie algebras,  namely, {\em Lie-Poisson
brackets arise from canonical brackets on the cotangent bundle\/}
(phase space) $T^\ast G$ associated with a Lie group $G$ which has
$\mathfrak{g}$ as its associated Lie algebra.

For a rigid body which is free to rotate about its center of mass, $G$
is the (proper) rotation group ${\rm SO}(3)$. The choice of
$T^\ast G  $ as the primitive phase space is made according to
the classical procedures of mechanics described earlier.  For
the description using Lagrangian mechanics, one forms the
velocity-phase  space $T{\rm SO}(3)$.  The Hamiltonian
description on $T^\ast G$ is then obtained by standard
procedures.

The passage from $T^\ast G$ to the space of ${\boldsymbol
\Pi}$'s (body angular momentum space) is determined by {\em
left\/} translation on the group.  This mapping is  an example
of a {\em momentum map\/}; that is, a
mapping whose components are the ``Noether  quantities''
associated with a symmetry group.  The map from $T^\ast G$ to
$\mathfrak{g}^\ast$ being a Poisson (canonical) map {\em is a
general fact about momentum maps\/}.  The Hamiltonian point of
view of all this is again a well developed subject.

\paragraph{Geodesic motion.} As emphasized by Arnold [1966a], in
many interesting cases, the Euler--Poincar\'e equations on
a Lie algebra $ \mathfrak{g} $ correspond to {\it geodesic motion}
on the corresponding group $G$. We shall explain the
relationship between the equations on $\mathfrak{g}$ and on $G$
shortly, in theorem \ref{bep.theorem}. Similarly, on the
Hamiltonian side, the preceding paragraphs explained the relation
between the Hamiltonian equations on $ T ^\ast G $ and the
Lie--Poisson equations on $ \mathfrak{g}^\ast $.  However, the issue
of geodesic motion is simple: if the Lagrangian or Hamiltonian on $
\mathfrak{g} $ or $\mathfrak{g}^\ast$ is purely quadratic, then
the corresponding motion on the group is geodesic motion.

\paragraph{More History.}
The Lie-Poisson bracket  was discovered by Sophus Lie (Lie [1890],
Vol. II, p. 237). However, Lie's bracket and his related work was
not given much attention until the work of Kirillov, Kostant, and
Souriau (and others) revived it in the mid-1960s. Meanwhile, it was
noticed by Pauli and Martin around 1950 that the rigid body
equations are in Hamiltonian form using the rigid body bracket, but
they were apparently unaware of the underlying Lie theory. It would
seem that while Poincar\'{e} was aware of Lie theory, in his work on
the Euler equations he was unaware of Lie's work on Lie-Poisson
structures. He also seems not to have been be aware of the
variational structure of the Euler equations.

\paragraph{The heavy top.}

Another system important to Poincar\'{e} and also for us in
this paper is the heavy top; that is, a rigid
body with a fixed point in a gravitational field.  For the
Lie-Poisson description, the underlying Lie algebra,
surprisingly, consists of the algebra of infinitesimal Euclidean
motions in $\mathbb{R}^3$. These do {\em not\/} arise as actual
Euclidean motions of the body since the body has a fixed
point!  As we shall see, there is a close parallel with the
Poisson structure for compressible fluids.

The basic phase space we start with is again $  T^\ast {\rm
SO}(3)$.  In this space, the equations are in canonical
Hamiltonian form. Gravity breaks the symmetry and the system is
no longer ${\rm SO}(3)$ invariant, so it cannot be written
entirely in terms of the body angular momentum $\boldsymbol{\Pi}$.
One also needs to keep track of $\boldsymbol{\Gamma}$, the
``direction of gravity'' as seen from the body $
(\boldsymbol{\Gamma}  = \mathbf{R}^{-1}
\mathbf{k} $ where the unit vector $\mathbf{k}$ points upward and
$\mathbf{R}$ is the element of ${\rm SO}(3) $ describing the
current configuration of the  body).  The equations of motion
are
\begin{align}\label{hte1}
\dot {\Pi}_1  & =    \frac{ I _2 -I _3 }{ I _2 I _3 } \Pi_2 \Pi_3 +
{Mg}\ell\, (\Gamma ^2 \chi ^3 - \Gamma ^3 \chi ^2)  ,\nonumber
\\
\dot {\Pi}_2  & =   \frac{ I _3 -I _1 }{ I _3 I _1} \Pi_3 \Pi_1 +
{Mg}\ell\, (\Gamma ^3 \chi ^1 - \Gamma ^1 \chi ^3 ) ,
\\
\dot {\Pi}_3  & =   \frac{ I _1 - I _2}{ I_1 I _2 } \Pi_1 \Pi_2 +
{Mg}\ell\, (\Gamma ^1
\chi ^2 - \Gamma ^2 \chi ^1 ) ,
\nonumber
\end{align}
or, in vector notation,
\begin{equation} \label{Pi-dot}
\dot{ \boldsymbol{\Pi}}
= \boldsymbol{\Pi}\times\boldsymbol{\Omega}
+ {Mg}\ell\, \boldsymbol{\Gamma}\times\boldsymbol{\chi}\,,
\end{equation}
and
\begin{equation}\label{hte2}
\dot{ \boldsymbol{\Gamma}} = \boldsymbol{\Gamma}\times
\boldsymbol{\Omega},
\end{equation}
where $M$ is the body's mass, $g$ is the
acceleration of gravity, $\boldsymbol{\chi}$ is the unit vector
on the line connecting the fixed point with the body's center
of mass, and $\ell$ is the length of this segment.

The Lie algebra of the Euclidean group is $\mathfrak{se}(3) =
\mathbb{R}^3  \times \mathbb{R}^3$ with the  Lie bracket
\begin{equation}\label{htla} [(\boldsymbol{\xi}, \mathbf{u} ),
(\boldsymbol{\eta}, \mathbf{ v} )]  =  (\boldsymbol{\xi}
\times \boldsymbol{\eta},
\boldsymbol{\xi} \times \mathbf{ v} - \boldsymbol{\eta}
\times
\mathbf{u})  .
\end{equation}
%
We identify the dual space with pairs $({\boldsymbol
\Pi},\boldsymbol{\Gamma} )$; the corresponding
$  (-)  $ Lie-Poisson bracket called the {\bfi heavy top
bracket\/} is
\begin{multline}\label{htpb}
\qquad\{F,G\}({\boldsymbol  \Pi}, \boldsymbol{\Gamma} )
 =  - {\boldsymbol  \Pi}
\cdot (\nabla _\Pi F
\times \nabla _\Pi G)  \\
 - \boldsymbol{\Gamma} \cdot (\nabla
_\Pi F \times \nabla _\Gamma G -
\nabla _\Pi G \times \nabla _\Gamma F).
\qquad
\end{multline}
%
The above equations for ${\boldsymbol  \Pi},
\boldsymbol{\Gamma} $ can be
checked to be equivalent to
\begin{equation}\label{htbbe}
\dot{ F}  =  \{F, H\},
\end{equation}  where the {\bfi heavy top Hamiltonian\/}
\begin{equation}\label{hth} H(\boldsymbol{\Pi},
\boldsymbol{\Gamma} )  =
\frac {1}{2}\left( \frac{ \Pi ^2 _1 }{ I _1}
+ \frac{ \Pi ^2 _2 }{ I _2} +
\frac{ \Pi ^2 _3}{ I _3}\right)
+ {Mg}\ell\, \boldsymbol{\Gamma} \cdot \boldsymbol{\chi}
\end{equation}
is the total energy of the body (see, for example, Sudarshan and
Mukunda [1974]).

The Lie algebra of the Euclidean group has a structure which is
a special case of what is called a {\em semidirect product\/}.
Here it is the product of the group of rotations with the
translation  group. It turns out that semidirect products
occur under rather general circumstances when the symmetry in
$T^\ast G $ is broken.  In particular, there are
similarities in structure between the Poisson bracket
for compressible flow and that for the heavy top.
The general theory for semidirect products will be reviewed
shortly.

\paragraph{A Kaluza-Klein form for the heavy top.} We make a
remark about the heavy top equations that is relevant for later
purposes. Namely, since the equations have a Hamiltonian that is
of the form kinetic plus potential, it is clear that the
equations are {\it not of Lie-Poisson form on
$\mathfrak{so}(3)^\ast$, the dual of the Lie algebra of ${\rm
SO(3)}$} and correspondingly, are not geodesic equations on
${\rm SO(3)}$. While the equations {\it are Lie--Poisson} on
$\mathfrak{se}(3)^\ast$, the Hamiltonian is not quadratic, so again
the equations are {\it not geodesic equations on ${\rm SE}(3)$}.

However, they can be viewed a different way so that they become
Lie-Poisson equations for a different group and with a {\it
quadratic Hamiltonian}. In particular, they are the reduction of
geodesic motion. To effect this, one changes the Lie algebra from
$\mathfrak{se}(3)$ to the product $\mathfrak{se}(3) \times
\mathfrak{so}(3)$. The dual variables are now denoted
$\boldsymbol{\Pi}, \boldsymbol{\Gamma},
\boldsymbol{\chi}$. We regard the variable $\boldsymbol{\chi}$
as a momentum conjugate to a new variable, namely a {\it ghost}
element of the rotation group in such a way that
$\boldsymbol{\chi}$ is a constant of the motion; in
Kaluza-Klein theory for charged particles on thinks of the charge
this way, as being the momentum conjugate to a (ghost) cyclic
variable.

We modify the Hamiltonian by replacing $\boldsymbol{\Gamma} \cdot
\boldsymbol{\chi} $ by, for example,
$\boldsymbol{\Gamma} \cdot \boldsymbol{\chi} +
\|\boldsymbol{\Gamma} \| ^2 + \| \boldsymbol{\chi} \|^2 $, or
any other terms of this sort that convert the potential energy
into a positive definite quadratic form in $\boldsymbol{\Gamma}$
and $\boldsymbol{\chi}$. The added terms, being Casimir
functions, do not affect the equations of motion. However, now the
Hamiltonian is purely quadratic and hence comes from geodesic motion
on the group $SE(3) \times {\rm SO(3)}$. Notice that this
construction is quite different from that of the well known Jacobi
metric method.

Later on in our study of
continuum mechanics, we shall repeat this construction to achieve
geodesic form for some other interesting continuum models. Of course
one can also treat a heavy top that is charged or has a magnetic
moment using these ideas.

\paragraph{Incompressible Fluids.}
Arnold [1966a] showed that the Euler equations for an
incompressible fluid could be given a Lagrangian and Hamiltonian
description similar to that for the rigid body.  His
approach\footnote{Arnold's approach is consistent with what
appears in the thesis of Ehrenfest from around 1904; see Klein
[1970]. However, Ehrenfest bases his principles on the more
sophisticated curvature principles of Gauss and Hertz.} has
the appealing feature that one sets things up just the way
Lagrange and Hamilton would have done: one begins with a
configuration space $Q$, forms a Lagrangian $L$ on the
velocity phase space $TQ$ and then Legendre transforms to a
Hamiltonian $H$ on the momentum phase space $T^\ast Q$.  Thus, one
automatically has variational principles, etc.  For ideal fluids,
$Q=G$ is the group
$\operatorname{Diff}_{\operatorname{vol}}(\mathcal{D})$ of
volume preserving transformations of the fluid container (a
region $\mathcal{D}$ in $\mathbb{R}^2$ or
$\mathbb{R}^3$, or a Riemannian manifold in general,
possibly with  boundary).  Group multiplication in $G$ is
composition.

The reason we select
$G = \operatorname{Diff}_{\operatorname{vol}}(\mathcal{D})$ as the
configuration space is similar to that for the rigid body; namely,
each $\varphi$ in $G$ is a mapping of
$\mathcal{D}$ to $\mathcal{D}$ which takes a reference point $X
\in \mathcal{D}$ to a current point  $x = \varphi (X)\in
\mathcal{D} $; thus, knowing $\varphi$ tells us where each
particle of fluid goes and hence gives us the current {\bfi
fluid configuration\/}. We ask that $\varphi$  be a
diffeomorphism to exclude discontinuities, cavitation, and fluid
interpenetration, and we ask that $\varphi$ be volume preserving
to correspond to the assumption of incompressibility.

A {\bfi motion\/} of a fluid is a family of time-dependent
elements of $G$, which we write as $x=\varphi(X,t)$. The {\bfi
material velocity\/} field is defined by $ \mathbf{V}(X, t) =
\partial \varphi (X,t)/\partial t$, and the {\bfi spatial
velocity\/} field is defined by $\mathbf{v}(x,t)
=\mathbf{V}(X,t)$ where $x$ and $X$ are related by
$ x = \varphi(X,t)$.  If we suppress ``$ t $'' and write
$\dot{\varphi}$ for $\mathbf{V}$, note that
\begin{equation}\label{svf}  \mathbf{ v}
= \dot{ \varphi } \circ
\varphi^{-1}  \quad   \text{i.e.,\/}
\quad  \mathbf{ v}_t  = \mathbf{V}_t
\circ \varphi^{-1}_t ,
\end{equation}
where $ \varphi_t  (x) = \varphi(X,t) $.  We can regard
(\ref{svf}) as a map from the space of
$(\varphi , \dot{ \varphi })$ (material or Lagrangian
description) to the space of $ \mathbf{ v}$'s (spatial or
Eulerian description). Like the rigid body, the material to
spatial map (\ref{svf}) takes the canonical bracket to a
Lie-Poisson bracket; one of our goals is to understand this
reduction.  Notice that if we replace $\varphi$ by $
\varphi \circ\eta $ for a fixed (time-independent)
$\eta \in \operatorname{Diff}_{\operatorname{vol}}(\mathcal{D}
)$, then $\dot{\varphi} \circ\varphi^{-1}$
is independent of $\eta$; this reflects the {\em right\/}
invariance of the Eulerian description ($\mathbf{ v} $ is
invariant under composition of $  \varphi  $ by $ \eta  $ on the
right).  This is also called the {\bfi particle relabeling
symmetry\/} of fluid dynamics.  The spaces $  TG  $ and $  T^\ast G  $
represent the  Lagrangian (material) description and we pass to
the Eulerian (spatial) description by right translations and use
the $(+)$ Lie-Poisson bracket.  One of the things we shall
explain later is the reason for the switch between right and
left in going from the rigid body to fluids.

The {\bfi Euler equations} for an ideal, incompressible,
homogeneous fluid moving in the region  $\mathcal{D}$ are
\begin{equation}\label{eef}
\frac{\partial \mathbf{ v}}{\partial t} + (\mathbf{ v} \cdot
\nabla )\mathbf{ v}  =
-\nabla p
\end{equation}
with the constraint $\operatorname{div}\,  \mathbf{ v} =  0$ and
boundary conditions: $\mathbf{ v}$ is tangent to
$\partial\mathcal{D}.$

The pressure $p$ is determined implicitly by the
divergence-free (volume preserving) constraint
$\operatorname{div}\, \mathbf{ v}= 0$. The associated Lie
algebra $  \mathfrak{g} $ is  the space of all divergence-free
vector fields tangent to the boundary. This Lie algebra is
endowed with the {\em negative\/} {\bfi  Jacobi-Lie bracket\/} of
vector fields given by
\begin{equation}\label{jlb} [\mathbf{v}, \mathbf{w}]^i_L
= \sum_{j=1} ^n  \left(w^j
\frac{\partial v^i}{\partial x^j} - v^j
\frac{\partial w^i}{\partial x^j} \right) .
\end{equation}
(The subscript $L$ on $  [\cdot \, ,\cdot ]$ refers to
the fact that it is the {\em left\/} Lie algebra bracket on
$\mathfrak{g}$. The most  common convention for the Jacobi-Lie
bracket of vector fields, also the one we adopt, has the
opposite sign.)  We identify $\mathfrak{g}$ and
$\mathfrak{g}^\ast$ by using the pairing
\begin{equation}\label{l2p}
\left\langle \mathbf{ v},\mathbf{w}\right\rangle
=\int_\mathcal{D} \mathbf{ v}
\cdot
\mathbf{w}
\, d^3 x.
\end{equation}

\paragraph{Hamiltonian structure for fluids.}
Introduce the $  (+)  $ Lie-Poisson
bracket, called the {\bfi ideal fluid bracket\/}, on functions of $
\mathbf{ v} $ by
\begin{equation}\label{lpbf}
\{F, G\}(\mathbf{ v} )
= \int_\mathcal{D} \mathbf{ v}\cdot \left[\frac{\delta
F}{\delta \mathbf{ v}}, \frac{\delta G}{\delta \mathbf{ v}}
\right]_L d^3 x ,
\end{equation}
where $ \delta F/\delta \mathbf{ v} $ is defined by
\begin{equation}\label{fdf}
\lim_{\varepsilon \rightarrow 0}\frac{ 1}{ \varepsilon} [F(\mathbf{ v}
+
\varepsilon \delta \mathbf{ v})
 - F(\mathbf{ v})]
= \int_\mathcal{D} \left(\delta \mathbf{ v} \cdot \frac{ \delta
F}{
\delta \mathbf{ v}} \right)  d^3 x  .
\end{equation}
%
With the energy function chosen to be the kinetic energy,
\begin{equation}\label{hf}
H(\mathbf{ v})
=\frac{1}{2}\int_\mathcal{D} | \mathbf{ v}  | ^2 \, d^3 x,
\end{equation}
one can verify that the Euler equations (\ref{eef}) are
equivalent to the Poisson bracket equations
\begin{equation}\label{pbef}
\dot{ F} =  \{F, H\}
\end{equation}
 for all functions $  F  $ on $\mathfrak{g}$.  For this, one
uses the orthogonal decomposition $ \mathbf{w} = {\mathbb
P}\mathbf{w}+ \nabla p  $ of a vector field $ \mathbf{w} $  into
a divergence-free part $  {\mathbb P}\mathbf{w}  $ in $
\mathfrak{g}$ and a gradient.  The Euler equations can be written as
\begin{equation}\label{eepf}
\frac{\partial \mathbf{ v} }{\partial t}  + {\mathbb P}(\mathbf{ v} \cdot
\nabla
\mathbf{ v})  =  0 .
\end{equation}

One can also express the Hamiltonian structure in terms of the
vorticity as a basic dynamic variable and show that the
preservation of coadjoint orbits amounts to Kelvin's circulation
theorem. We shall see a Lagrangian version of this property later
in the paper. Marsden and Weinstein [1983] show that the Hamiltonian
structure in terms of Clebsch potentials fits naturally into this
Lie-Poisson scheme, and that Kirchhoff's Hamiltonian description
of point vortex dynamics, vortex filaments, and vortex patches
can be derived in a natural way from the Hamiltonian structure
described above.

\paragraph{Lagrangian structure for fluids.}
The general  framework of the Euler--Poincar\'e and
the Lie-Poisson equations gives other insights as well. For
example, this general theory shows that the Euler equations
are derivable from the ``variational principle''
\[ \delta \int _a ^b \int_\mathcal{D} \frac{1}{2} |
     \mathbf{ v} | ^2 \, d ^3 x
= 0 \]
which should hold for all variations
$\delta \mathbf{v}$ of the form
\[
   \delta \mathbf{v} = \dot{\mathbf{u}}
   + [\mathbf{u}, \mathbf{v}]_L
\]
where $\mathbf{u}$ is a vector field (representing the
infinitesimal particle displacement) vanishing at the
temporal endpoints. The constraints on the allowed variations
of the fluid velocity field are commonly known as ``Lin
constraints'' and their nature was clarified by Newcomb [1962]
and Bretherton [1970]. This itself has an interesting history,
going back to Ehrenfest, Boltzmann, and Clebsch, but again,
there was little if any contact with the heritage of Lie and
Poincar\'{e} on the subject.

\paragraph{The Basic Euler--Poincar\'e Equations.} We now recall the
abstract derivation of the ``basic'' Euler--Poincar\'e equations
(i.e., the Euler--Poincar\'e equations with no forcing or advected
parameters) for left--invariant Lagrangians on Lie groups (see Marsden
and Scheurle [1993a,b], Marsden and Ratiu [1994] and Bloch et al.
[1996]).

\begin{thm}\label{bep.theorem} Let $G$ be a Lie group and $L : TG
\rightarrow
\mathbb{R}$ a left (respectively, right) invariant Lagrangian.  Let
$l: {\mathfrak{g}}
\rightarrow \mathbb{R}  $ be its restriction to the tangent space at
the identity. For a curve $g(t) \in G,$ let $\xi (t) = g(t) ^{ -1}
\dot{g}(t);$ {\it i.e.,\/} $\xi (t) = T _{ g(t)} L _{ g(t) ^{ -1}}
\dot{g}(t)$ (respectively, $\xi (t) = \dot{g}(t)g(t) ^{ -1}$).  Then
the following are equivalent:
\begin{enumerate}
\item [{\bf i}]
  Hamilton's principle
\begin{equation} \label{eulerpoincare22}
\delta \int _a ^b L(g(t), \dot{g} (t)) dt = 0
\end{equation}  holds, as usual, for variations $\delta g(t)$
of $ g (t) $ vanishing at the endpoints.
\item [{\bf ii}]
      The curve $g(t)$ satisfies the Euler-Lagrange
equations for $L$ on $G$.
\item [{\bf iii}]
      The ``variational'' principle
\begin{equation} \label{eulerpoincare24}
\delta \int _a ^b  l(\xi(t)) dt = 0
\end{equation}  holds on $\mathfrak{g}$, using variations of the
form
\begin{equation} \label{eulerpoincare25}
\delta \xi = \dot{\eta } \pm [\xi , \eta ],
\end{equation}  where $\eta $ vanishes at the endpoints ($+$
corresponds to left invariance and $-$ to right
invariance).\footnote{Because there are constraints on the variations,
this principle is more like a La\-gran\-ge d'Al\-em\-bert principle,
which is why we put ``variational'' in quotes. As we shall explain, such
problems are not literally variational.}
\item [{\bf iv}]
        The {\bfi basic Euler--Poincar\'{e} equations\/} hold
\begin{equation} \label{eulerpoincare23}
\frac{d}{dt} \frac{\delta l}{\delta \xi} = \pm
 \operatorname{ad}_{\xi}^{\ast}
\frac{\delta l}{\delta \xi}\,.
\end{equation}
\end{enumerate}
\end{thm}

\paragraph{Basic Ideas of the Proof.}
First of all, the equivalence of {\bf i} and
{\bf ii} holds on the tangent bundle of any configuration manifold
$Q$, by the general Hamilton principle. To see that {\bf ii} and
{\bf iv} are equivalent, one needs to compute the variations $
\delta \xi $ induced on $\xi = g ^{-1} \dot{ g} = TL _{ g ^{-1}}
\dot{ g} $ by a variation of
$g$.  We will do this for matrix groups; see Bloch, Krishnaprasad,
Marsden, and Ratiu [1994] for the general case. To calculate this, we
need to differentiate
$ g ^{-1}
\dot{ g}
$ in the direction of a variation
$ \delta g $. If $
\delta g = d g / d \epsilon $ at $ \epsilon = 0 $, where $g$ is
extended to a curve $ g _\epsilon $, then,
\[ \delta \xi = \frac{d }{d \epsilon } g ^{-1} \frac{d }{d t} g, \]
while if $ \eta = g ^{-1} \delta g $, then
\[ \dot{ \eta} = \frac{d }{d t} g ^{-1} \frac{d }{d \epsilon } g . \]
The difference $
\delta \xi - \dot{ \eta} $ is thus the commutator $ [\xi, \eta] $.

To complete the proof, we show the equivalence of {\bf iii} and {\bf
iv} in the left-invariant case. Indeed, using the definitions and
integrating by parts produces,
\begin{align*}
\delta \int l (\xi) d t
    & =  \int \frac{ \delta l }{ \delta \xi}\delta \xi \, d t  =
\int \frac{ \delta l }{ \delta \xi} (\dot{ \eta}
             + {\rm a d} _\xi \eta)\, d t \\
    & =  \int\left[ - \frac{d }{d t} \left( \frac{ \delta l }
        { \delta \xi} \right) +\operatorname{ad} ^{\ast} _\xi
        \frac{ \delta l }{ \delta \xi} \right] \eta \, dt\,,
\end{align*}
so the result follows.

There is of course a right invariant version of this theorem in
which $\xi = \dot{g} g^{-1}$ and the Euler--Poincar\'e
equations acquire appropriate minus signs as in equation
(\ref{eulerpoincare23}). We shall go into this in detail later.

Since the Euler-Lagrange and Hamilton equations on $ TQ$ and $
T^\ast Q$ are equivalent in the regular case, it follows that the
Lie-Poisson and Euler--Poincar\'{e} equations are then also equivalent.
To see this {\em directly\/}, we make the following Legendre
transformation from $\mathfrak{g}$ to $\mathfrak{g} ^{\ast}$:
\[
\mu = \frac{ \delta l }{ \delta \xi}, \quad h (\mu) = \langle  \mu, \xi
\rangle - l (\xi) .
\]   Note that
\[
\frac{ \delta h }{ \delta \mu} = \xi + \left\langle \mu, \frac{ \delta
\xi }{ \delta \mu}
\right\rangle - \left\langle \frac{ \delta l }{ \delta \xi}, \frac{
\delta \xi }{ \delta \mu}
\right\rangle = \xi
\]
and so it is now clear that the Lie-Poisson equations (\ref{alpe}) and
the Euler--Poincar\'{e} equations (\ref{eulerpoincare23}) are equivalent.

We close this paragraph by mentioning the geodesic property of
the basic Euler--Poincar\'e form. When $l$ is a metric on $TG$,
the basic Euler--Poincar\'e equations are the {\em geodesic spray
equations} for geodesic motion on the group $G$ with respect to
that metric. For discussions of this property in applications,
see, e.g., Arnold [1966a] for the Euler equations of an
incompressible ideal fluid, and Ovsienko and Khesin [1987] for
the KdV shallow water equation. (An account of the latter case
from the Euler--Poincar\'e viewpoint may also be found in
Marsden and Ratiu [1994].) Zeitlin and Pasmanter [1994] discuss
the geodesic property for certain ideal geophysical fluid flows;
Zeitlin and Kambe [1993] and Ono [1995a, 1995b] discuss it for
ideal MHD; and Kouranbaeva [1997] for the integrable
Camassa-Holm equation. From one viewpoint, casting these systems
into basic Euler--Poincar\'e form explains why they share the
geodesic property.

\paragraph{Lie-Poisson Systems on Semidirect Products.} As we
described above, the heavy top is a basic example of a
Lie-Poisson Hamiltonian system defined on the dual of a
semidirect product Lie algebra. The {\it general} study of
Lie-Poisson equations for systems on the dual of a semidirect
product Lie algebra grew out of the work of many authors
including Sudarshan and Mukunda [1974], Vinogradov and
Kupershmidt [1977], Ratiu [1980], Guillemin and Sternberg
[1980], Ratiu [1981, 1982], Marsden [1982], Marsden, Weinstein,
Ratiu, Schmidt and Spencer [1983], Holm and Kupershmidt [1983],
Kupershmidt and Ratiu [1983], Holmes and Marsden [1983],
Marsden, Ratiu and Weinstein [1984a,b], Guillemin and Sternberg
[1984], Holm, Marsden, Ratiu and Weinstein [1985], Abarbanel,
Holm, Marsden, and Ratiu [1986] and Marsden, Misiolek,
Perlmutter and Ratiu [1997]. As these and related references
show, the Lie-Poisson equations apply to a wide variety of
systems such as the heavy top, compressible flow, stratified
incompressible flow, and MHD (magnetohydrodynamics). We review
this theory in \S2 below.

In each of the above examples as well as in the general theory, one
can view the given Hamiltonian in the material representation as one
that depends on a parameter; this parameter becomes dynamic when
reduction is performed; this reduction amounts in many examples to
expressing the system in the spatial representation.

\paragraph{Goals of this Paper.} The first goal of this paper is to
study a Lagrangian analogue of the Hamiltonian semidirect product
theory. The idea is to carry out a reduction for a Lagrangian that
depends on a parameter and to use the ideas of reduction of
variational principles from Marsden and Scheurle [1993a,b] and
Bloch, Krishnaprasad, Marsden and Ratiu [1996] to directly reduce
the problem to one that parallels Lie-Poisson systems on the duals
of semidirect products. We call the resulting equations the
Euler--Poincar\'e equations since, as we have explained,
Poincar\'{e} [1901b] came rather close to this general picture.
These equations generalize the {\it basic} Euler--Poincar\'e
equations on a Lie algebra in that they depend on a parameter and
this parameter in examples has the interpretation of being advected,
or Lie dragged, as is the density in compressible flow.

One of the reasons this process is interesting and cannot be
derived directly from its Hamiltonian counterpart by means of
the Legendre transformation is that in many examples,  such
as the heavy top, the Hamiltonian describing the Lie-Poisson
dynamics is degenerate; that is, the Legendre transformation
is not invertible.

A second major goal is to prove a version of the Noether theorem in
an action principle formulation that leads immediately to a Kelvin
circulation type theorem for continuum mechanics. We call this
general formulation the Kelvin-Noether theorem.

Finally, we provide a number of applications of the
Euler--Poincar\'e equations in ideal continuum dynamics which
illustrate the power of this approach in unifying various known
models, as well as in formulating new models. We also discuss
some circumstances when the equations can be cast into the form
of geodesics on certain infinite dimensional groups.

\paragraph{Outline of the remainder of this paper.}  In the next
section we review the semidirect product theory for Hamiltonian
systems. Then in section \ref{sec-lag-sdp} we consider the
Lagrangian counterpart to this theory. Section \ref{sec-KN-Theorem}
discusses the Kelvin-Noether theorem for the Euler--Poincar\'e
equations. Section \ref{sec-hvytop} illustrates the general theory in
the example of the heavy top. We introduce the Euler--Poincar\'e
equations for continua in section \ref{sec-EPPC} and consider their
applications to compressible flow (including MHD and adiabatic
Maxwell-fluid plasmas) in section \ref{sec-AppEPCont}. Various
approximate forms of the shallow water equations, such as the
Boussinesq equations, the Camassa-Holm equation and its new
higher-dimensional variants are developed in section
\ref{sec-mod-eqns}. In other publications, the Maxwell-Vlasov
equations will be considered as well as a general framework for the
theory of reduction by stages.

In the remainder of this paper we assume that the reader is familiar
with Lie-Poisson Hamiltonian systems defined on duals of Lie
algebras and the Lie-Poisson reduction theorem, reviewed above. We
refer to Marsden and Ratiu [1994] for a detailed exposition of these
matters.

\paragraph{Acknowledgements.} We thank Hernan Cendra, Shiyi
Chen, Ciprian Foias, Mark Hoyle, David Levermore, Len Margolin,
Gerard Misiolek, Balu Nadiga, Matthew Perlmutter, Steve Shkoller
and Edriss Titi for valuable discussions and remarks.

\section{Hamiltonian Semidirect Product Theory} \label{sec-ham-sdp}
We first recall how the Hamiltonian theory proceeds for systems defined
on semidirect products. We present the abstract theory, but of course
historically this grew out of the examples, especially the heavy top and
compressible flow.

\paragraph{Generalities on Semidirect Products.} We begin by
recalling some definitions and properties of semidirect
products. Let $V$ be a vector space and assume that the Lie
group $G$ acts {\it on the left\/} by linear maps on $V$ (and
hence $G$ also acts on on the left on its dual space
$V^\ast$). As sets, the semidirect product $ S = G
\,\circledS\, V $ is the Cartesian product
$S  = G \times V$ whose group multiplication is given by
\begin{equation}\label{semidirectleft}
(g_1, v_1) (g_2, v_2) = (g_1 g_2, v_1 + g_1 v_2),
\end{equation}
where the action of $g \in G $ on $v  \in V $ is denoted
simply as $gv$. The identity element is
$ (e,0) $ where $e$ is the identity in $G$.  We record for convenience
the inverse of an element:
\begin{equation} \label{sdinverse}
(g,v) ^{-1} = ( g ^{-1} , - g ^{-1} v).
\end{equation}

The Lie algebra of $S$ is the semidirect product
Lie algebra,
$\mathfrak{s}    = \mathfrak{g}  \,\circledS\, V $, whose
bracket has the expression
\begin{equation}\label{semidirectalgebraleft}
[(\xi_1,v _1), (\xi_2, v_2)]
= ([\xi_1,\xi_2],\, \xi_1v_2 - \xi_2 v_1)\,,
\end{equation}
where we denote the induced action of $\mathfrak{g}$ on $V$ by
concatenation, as in $\xi_1 v_2$.

Below we will need the formulae for the adjoint and the
coadjoint actions for semidirect products. We denote these and
other actions by simple concatenation; so they are expressed as (see,
e.g., Marsden, Ratiu and Weinstein [1984a,b])
\begin{equation}\label{adjointleft}
(g,v) (\xi, u) = (g \xi, gu - (g \xi) v ),
\end{equation}
and
\begin{equation}\label{coadjointleft}
(g,v) (\mu, a)  = (g \mu  + \rho_v ^\ast(ga), ga),
\end{equation}
where $(g,v)  \in S  = G  \times V $, $( \xi, u)
\in\mathfrak{s} = \mathfrak{g}  \times V$, $(\mu, a)  \in
\mathfrak{s}^\ast = \mathfrak{g}^{\ast}  \times V ^\ast$,
$g\xi = {\rm Ad}_g\xi$, $g\mu =
{\rm Ad}^\ast_{g^{-1}}\mu$, $ga$ denotes the induced
{\it left\/} action of $g$ on $a$ (the {\it left} action of
$G$ on $V$ induces a {\it left} action of $G$ on $ V ^\ast $
--- the inverse of the transpose of the action on $V$),
$\rho _v: \mathfrak{g}  \rightarrow V$ is the linear map
given by $\rho_v (\xi) = \xi v$, and
$\rho _v^\ast: V ^\ast
\rightarrow \mathfrak{g}^{\ast}$ is its dual.

\paragraph{Important Notation.} For $a \in V ^\ast $, we shall
write, for notational convenience,
\[
   \rho _v^\ast a = v \diamond a  \in \mathfrak{g}^\ast \, ,
\]
which is a bilinear operation in $v$ and $a$. Using this
notation, the above formula for the coadjoint action reads
\[
  (g,v) (\mu, a)  = (g \mu  + v \diamond (ga), ga).
\]
We shall also denote actions of groups and Lie algebras by simple
concatenation. For example, the $\mathfrak{g} $--action on
$\mathfrak{g} ^\ast$ and $V^\ast$, which is
defined as minus the dual map of the $\mathfrak{g} $--action
on $\mathfrak{g} $ and $V$ respectively, is denoted by
$\xi\mu$ and $\xi a$ for $\xi \in \mathfrak{g} $, $\mu \in
\mathfrak{g} ^\ast$, and $a\in V^\ast$.

Using this concatenation notation for Lie algebra actions provides the
following alternative expression of the definition of
$ v \diamond a \in \mathfrak{g}^\ast$: For all $ v \in V$, $ a \in V
^\ast $ and $ \eta \in \mathfrak{g} $, we define
\[
\left \langle \eta a, v
\right \rangle = - \left\langle v \diamond a\,,\eta \right\rangle .
\]

\paragraph{Left Versus Right.} When working with various models of
continuum mechanics and plasmas it is convenient to work with {\it
right\/} representations of $G$ on the vector space $V$ (as in, for
example, Holm, Marsden and Ratiu [1986]). We shall denote the
semidirect product by the same symbol $S= G\,\circledS\, V$, the
action of $G$ on $V$ being denoted by $vg$.  The formulae change
under these conventions as follows. Group multiplication (the analog of
(\ref{semidirectleft})) is given by
\begin{equation}\label{semidirectright}
(g_1, v_1) (g_2, v_2) = (g_1 g_2, v_2 + v_1 g_2),
\end{equation}
and the Lie algebra bracket on $\mathfrak s = \mathfrak{g}
\,\circledS\, V$
(the analog of (\ref{semidirectalgebraleft})) has the expression
\begin{equation}\label{semidirectalgebraright}
[(\xi_1,v _1), (\xi_2, v_2)]
= ([\xi_1,\xi_2],\, v_1 \xi_2 - v_2 \xi_1),
\end{equation}
where we denote the induced action of $\mathfrak{g}$ on $V$ by
concatenation, as in $v_1 \xi_2$. The adjoint and coadjoint actions
have the formulae (analogs of (\ref{adjointleft}) and
(\ref{coadjointleft}))
\begin{equation}\label{adjointright}
(g, v)(\xi, u) = (g\xi, (u + v\xi)g^{-1}),
\end{equation}
\begin{equation}\label{coadjointright}
(g,v)(\mu, a) = (g\mu + (vg^{-1}) \diamond (ag^{-1}), ag^{-1}),
\end{equation}
where, as usual, $g\xi = {\rm Ad}_g\xi$, $g\mu =
{\rm Ad}^\ast_{g^{-1}}\mu$, $ag$ denotes the inverse of the dual
isomorphism defined by $g\in G$ (so that $g \mapsto ag$ is a
{\it right\/} action). Note that the
adjoint and coadjoint actions are {\it left\/} actions. In this
case, the $\mathfrak{g}$--actions on $\mathfrak{g}^\ast$ and $V^\ast$
are defined as before to be minus the dual map given by the
$\mathfrak{g}$--actions on $\mathfrak{g}$ and $V$ and are denoted by
$\xi\mu$ (because it is a left action) and $a\xi$ (because it is a
right action) respectively.

\paragraph{Lie-Poisson Brackets and Hamiltonian Vector Fields.}
For a {\it left\/} representation of $G$ on $V$ the
$\pm$ Lie-Poisson bracket of two functions $f, k : \mathfrak s^\ast
\rightarrow \mathbb{R} $ is given by
\begin{eqnarray}\label{leftLP}
\{f, k\} _\pm (\mu, a) & = & \pm \left\langle \mu,
\left[
\frac{\delta  f}{\delta \mu } , \frac{\delta k}{\delta \mu }
\right] \right\rangle
\pm
\left\langle a, \frac{\delta f}{\delta \mu } \frac{\delta
k }{ \delta a} - \frac{\delta
k}{\delta \mu } \frac{\delta f}{\delta a} \right\rangle
\end{eqnarray}
where $\delta f / \delta \mu  \in \mathfrak{g}$, and
$\delta f / \delta a \in V$
are the functional derivatives of $f$.
The Hamiltonian vector field of $h :
\mathfrak s^\ast \rightarrow \mathbb{R}$ has the expression
\begin{equation}\label{leftham}
X_h (\mu, a) = \mp \left( \operatorname{ad}^\ast_{\delta h /\delta
\mu}\mu -\frac{\delta h }{ \delta a} \diamond  a,\,
-\,\frac{ \delta h}{\delta \mu}\, a \right)\,.
\end{equation}
Thus, Hamilton's equations on the dual of a semidirect product are
given by
\begin{eqnarray} \label{leftsemi1.eqn}
\dot{ \mu } & = &  \mp\,  \operatorname{ad}^\ast_{\delta h /\delta
\mu}\mu
\pm  \frac{\delta h }{ \delta a} \diamond a\,, \\
\dot{ a } & =& \pm\, \frac{\delta h}{\delta \mu}\, a \,,
\label{leftsemi2.eqn}
\end{eqnarray}
where overdot denotes time derivative.
For {\it right\/} representations of $G$ on $V$ the above formulae
change to:
\begin{eqnarray}\label{rightLP}
\{f, k\} _\pm (\mu, a) & = & \pm \left\langle \mu,
\left[
\frac{\delta  f}{\delta \mu } , \frac{\delta k}{\delta \mu }
\right] \right\rangle
\mp \left\langle a, \frac{\delta k }{ \delta a}\frac{\delta f}{\delta
\mu } - \frac{\delta f}{\delta a}\frac{\delta k}{\delta \mu }\,
\right\rangle \, ,
\end{eqnarray}
\begin{equation}\label{righttham}
X_h (\mu, a) = \mp \left( \operatorname{ad}^\ast_{\delta h
/\delta\mu}\mu  +\frac{\delta h }{ \delta a} \diamond a,\,
a\,\frac{ \delta h}{\delta \mu}\right)\,,
\end{equation}
\begin{eqnarray} \label{rightsemi1.eqn}
\dot{ \mu } & = &  \mp\,  \operatorname{ad}^\ast_{\delta h /\delta
\mu}\mu
\mp  \frac{\delta h }{ \delta a} \diamond a\,,
\\
\dot{ a } & = & \mp\, a\, \frac{ \delta h}{\delta \mu}\,.
\label{rightsemi2.eqn}
\end{eqnarray}

\paragraph{Symplectic Actions by Semidirect Products.}
To avoid a proliferation
of signs, in {\it this section} we consider all semidirect products to
come from a left representation. Of course if the
representation is from the right, there are similar formulae.

We consider a symplectic action of
$S$ on a symplectic manifold $P$ and assume that this action has
an equivariant momentum map ${\bf J}_S : P \rightarrow\mathfrak{s}
^{\ast}$.  Since $V$ is a (normal) subgroup of $S$, it also acts
on $P$ and has a momentum map
${\bf J}_V: P  \rightarrow V ^\ast$ given by
\[
{\bf J}_V  = i _V ^\ast \circ {\bf J}_S\,,
\]
where $i _V : V  \rightarrow \mathfrak{s}  $ is the inclusion
$v \mapsto (0,v)$ and $i _V ^\ast: \mathfrak{s}  ^\ast  \rightarrow
V ^\ast$ is its dual. We think of this merely as saying that
${\bf J}_V$ is the second component of ${\bf J}_S$.

We can regard $G$ as a subgroup of $S$ by $g \mapsto (g,0)$.
Thus, $G$ also has a momentum map that is the first component
of ${\bf J}_S$ but this will play a secondary role in what
follows. On the other hand, equivariance of ${\bf J}_S$ under
$G$ implies the following relation for ${\bf J}_V$:
\begin{equation} \label{relation.eq}
{\bf J}_V(gz)  = g {\bf J}_V(z)
\end{equation}
where we denote the appropriate action of  $g \in G$ on an
element by concatenation, as before. To prove (\ref{relation.eq}),
one uses the fact that for the coadjoint action of $S$ on
$\mathfrak{s} ^\ast$ the second component is just the dual of
the given action of $G$ on $V$.

\paragraph{The Classical Semidirect Product Reduction Theorem.}
In a number of interesting applications such as compressible fluids,
the heavy top, MHD, etc., one has two symmetry groups that do not
commute and thus the commuting reduction by stages theorem of Marsden
and Weinstein [1974] does not apply. In this more general
situation, it matters in what order one performs the reduction,
which occurs, in particular for semidirect products. The main result
covering the case of semidirect products has a complicated history,
with important early contributions by many authors, as we have
listed in the introduction. The final version of the theorem as we
shall use it, is due to Marsden, Ratiu and Weinstein [1984a,b].

The semidirect product reduction theorem states, roughly speaking,
that for the semidirect product $S = G\,\circledS\, V$ where $G$ is a
group acting on a vector space $V$ and $S$ is the semidirect product,
one can first reduce $T^{\ast}S$  by $V$ and then by $G$ and thereby
obtain the same result as when reducing by $S$. As above, we let
$\mathfrak{s} = \mathfrak{g}\,\circledS\, V$ denote the Lie algebra of
$S$. The precise statement is as follows.

\begin{thm}[Semidirect Product Reduction Theorem.]
\label{semidirect.thm} \quad Let $S = G \,\circledS\, V$,
choose $\sigma = (\mu, a) \in \mathfrak{g}^{\ast} \times  V
^{\ast}$, and reduce $T ^{\ast}S $ by the action of $S$ at $\sigma$
giving the coadjoint orbit $ \mathcal{O}_\sigma $ through
$\sigma\in\mathfrak{s}^\ast$.  There is a symplectic diffeomorphism
between $\mathcal{O}_\sigma $ and the reduced space obtained by
reducing
$T^{\ast}G$ by the subgroup $G_a$ (the isotropy of $G$ for its
action on $V ^\ast$ at the point $a\in V ^{\ast}$) at the point
$\mu |\mathfrak{g}_a$ where $\mathfrak{g}_a$ is the Lie
algebra of $G_a$.
\end{thm}

\paragraph{Reduction by Stages.} This result is a special case of a
theorem on reduction by stages for semidirect products acting on a
symplectic manifold (see Marsden, Misiolek, Perlmutter and Ratiu
[1997] for this and more general results and see Leonard and Marsden
[1997] for an application to underwater vehicle dynamics).

As above, consider a symplectic action of $S$ on a symplectic
manifold $P$ and assume that this action has an equivariant momentum
map ${\bf J}_S : P \rightarrow\mathfrak{s}^{\ast}$. As we have
explained, the momentum map for the action of
$V$ is the map ${\bf J}_V: P  \rightarrow V ^\ast$ given by
${\bf J}_V  = i _V ^\ast \circ {\bf J}_S$

We carry out the reduction of $P$ by $S$ at
a regular value $\sigma=(\mu, a)$ of the momentum map ${\bf J}_S$
for $S$ in two stages using the following procedure. First, reduce
$P$ by $V$ at the value $a$ (assume it to be a regular value) to
get the reduced space $P_a = {\bf J}_V^{-1} (a)/V$. Second, form
the group $G_a$ consisting of elements of $G$ that leave the point
$a$ fixed using the action of $G$ on $V^\ast$. One shows (and this
step is not trivial) that the group $G_a$ acts on $P_a$ and has an
induced equivariant momentum map ${\bf J}_a : P _a \rightarrow
\mathfrak{g}^{\ast} _a$, where $\mathfrak{g}_a$ is the Lie algebra
of $G_a$, so one can reduce
$P_a$ at the point $\mu_a : = \mu | \mathfrak{g} _a$ to get the
reduced space $(P_a)_{\mu_a}  = {\bf J}_a^{-1}(\mu_a) /
(G_a)_{\mu_a}$.

\begin{thm}[Reduction by Stages for Semidirect Products.] The
reduced space $(P_a)_{\mu_a}$ is symplectically
diffeomorphic to the reduced space $P_\sigma$ obtained by
reducing $P$ by $S$ at the point $\sigma = (\mu,a)$.
\end{thm}

Combined with the cotangent bundle reduction theorem (see Abraham
and Marsden [1978] and Marsden [1992] for an exposition and
references), the semidirect product reduction theorem is a
useful tool. For example, using these tools, one sees readily that
the generic coadjoint orbits for the Euclidean group are cotangent
bundles of spheres with the associated coadjoint orbit symplectic
structure given by the canonical structure plus a magnetic term.

\paragraph{Semidirect Product Reduction of Dynamics.} There is a
technique for reducing dynamics that is associated with the geometry
of the semidirect product reduction theorem. One proceeds as follows:
\begin{itemize}
\item We start with a Hamiltonian $H_{a_0}$  on
$T^{\ast}G$ that depends parametrically on a variable $a _0 \in
V^{\ast}$.
\item The Hamiltonian, regarded as a map
$H : T^{\ast}G \times V ^{\ast} \rightarrow \mathbb{R}$
is assumed to be invariant on $T^{\ast}G$ under the action of $G$
on $T^{\ast}G\times V^{\ast}$.
\item One shows that this condition is equivalent to the
invariance of the function $H$ defined on
$ T^{\ast} S = T ^{\ast} G \times V \times V^{\ast}$
extended to be constant in the variable $V$ under the action of the
semidirect product.
\item By the semidirect product reduction theorem, the dynamics of
$ H _{a_0} $ reduced by $ G_{a_0} $, the isotropy group of $a_0$,
is symplectically equivalent to Lie-Poisson dynamics on
$\mathfrak{s}^\ast = \mathfrak{g}^{\ast}\times V^\ast$.
\item This Lie-Poisson dynamics is given by the equations
(\ref{leftsemi1.eqn}) and (\ref{leftsemi2.eqn}) for the function
$ h( \mu, a ) = H (\alpha _g , g^{-1} a) $ where
$ \mu = g ^{-1} \alpha _g $.
\end{itemize}

\section{Lagrangian Semidirect Product Theory} \label{sec-lag-sdp}
Despite all the activity in the Hamiltonian theory of semidirect
products, little attention has been paid to the corresponding
Lagrangian side. Now that Lagrangian reduction is maturing (see
Marsden and Scheurle [1993a,b]), it is appropriate to consider the
corresponding Lagrangian question. We shall formulate four
versions, depending on the nature of the actions and invariance
properties of the Lagrangian. (Two of them are relegated to the
appendix.)

It should be noted that {\it none of the
theorems below  require that the Lagrangian be nondegenerate}.
The subsequent theory is entirely based on variational principles
with symmetry and is not dependent on any previous Hamiltonian
formulation. We shall, however, show that this purely Lagrangian
formulation is equivalent to the Hamiltonian formulation on duals
of semidirect products, provided an appropriately defined Legendre
transformation happens to be a diffeomorphism.

The theorems that follow are modelled after the reduction theorem
for the basic Euler--Poincar\'e equations given earlier. However,
as we shall explain, they are {\it not\/} literally special cases
of it. To distinguish the two types of results, we shall use
phrases like {\it basic} Euler--Poincar\'e equations for the
equations (\ref{eulerpoincare23}) and simply the Euler--Poincar\'e
equations or the Euler--Poincar\'e equations {\it with advection}
or the Euler--Poincar\'e equations {\it with advected parameters},
for the equations that follow.

The main difference between the left (right) invariant
Lagrangians considered in the theorem above and the ones we shall
work with below is that
$L$ and $l$ depend in addition on another parameter
$a\in V^\ast$, where $V$ is a representation space for the Lie
group $G$ and $L$ has an invariance property relative to both
arguments. As we shall see below, the resulting {\bfi
Euler--Poincar\'e\/} equations are {\it not\/} the
Euler--Poincar\'e equations for the semidirect product Lie algebra
$\mathfrak{g}\,\circledS\, V^\ast$ or on $\mathfrak{g}\,\circledS\,
V$, for that matter.

\paragraph{Upcoming Examples.} As we shall see in the examples, the
parameter $a\in V^\ast$ acquires dynamical meaning under Lagrangian
reduction. For the heavy top, the parameter is the unit vector in the
direction of gravity, which becomes a dynamical variable in the body
representation. For compressible fluids, the parameter is the density of
the fluid in the reference configuration, which becomes a dynamical
variable (satisfying the continuity equation) in the spatial
representation.

\paragraph{Left Representation and Left Invariant Lagrangian.}
We begin with the following ingredients:
\begin{itemize}
\item There is a {\it left\/} representation of Lie group $G$ on
the vector space $V$ and $G$ acts in the natural way on the {\it
left\/} on $TG \times V^\ast$: $h(v_g, a) = (hv_g, ha)$.
\item Assume that the function $ L : T G \times V ^\ast
\rightarrow \mathbb{R}$ is left $G$--invariant.
\item In particular, if $a_0 \in V^\ast$, define the
Lagrangian $L_{a_0} : TG \rightarrow \mathbb{R}$ by
$L_{a_0}(v_g) = L(v_g, a_0)$. Then $L_{a_0}$ is left
invariant under the lift to $TG$ of the left action of
$G_{a_0}$ on $G$, where $G_{a_0}$ is the
isotropy group of $a_0$.

\item  Left $G$--invariance of $L$ permits us to define
$l: {\mathfrak{g}} \times V^\ast \rightarrow \mathbb{R}$ by
\[
l(g^{-1} v_g, g^{-1} a_0) = L(v_g, a_0).
\]
Conversely,  this relation defines for any
$l: {\mathfrak{g}} \times V^\ast \rightarrow
\mathbb{R} $ a left $G$--invariant function
$ L : T G \times V ^\ast
\rightarrow \mathbb{R} $.
\item For a curve $g(t) \in G, $ let
\[ \xi (t) := g(t) ^{ -1} \dot{g}(t)\] and define the curve
$a(t)$ as the unique solution of the following linear
differential equation with time dependent coefficients
\[ \dot a(t) = -\xi(t) a(t), \]
with initial condition $a(0) = a_0$. The solution can be
written as $a(t) = g(t)^{-1}a_0$.
\end{itemize}

\begin{thm} \label{lall}
With the preceding notation, the following are equivalent:
\begin{enumerate}
\item [{\bf i} ] With $a_0$ held fixed, Hamilton's variational
principle
\begin{equation} \label{hamiltonprinciple}
\delta \int _{t_1} ^{t_2} L_{a_0}(g(t), \dot{g} (t)) dt = 0
\end{equation}
holds, for variations $\delta g(t)$
of $ g (t) $ vanishing at the endpoints.
\item [{\bf ii}  ] $g(t)$ satisfies the Euler--Lagrange
equations for $L_{a_0}$ on $G$.
\item [{\bf iii} ]  The constrained variational principle%
\footnote{As with the basic Euler--Poincar\'e equations, this is not
strictly a variational principle in the
same sense as the standard Hamilton's principle.
It is more of a La\-gran\-ge d'Al\-em\-bert
principle, because we impose the stated constraints
on the variations allowed.}
\begin{equation} \label{variationalprinciple}
\delta \int _{t_1} ^{t_2}  l(\xi(t), a(t)) dt = 0
\end{equation}
holds on $\mathfrak{g} \times V ^\ast $, using variations of $ \xi $
and
$a$ of the form
\begin{equation} \label{epvariations}
\delta \xi = \dot{\eta } + [\xi , \eta ], \quad
\delta a =  -\eta a ,
\end{equation}
where $\eta(t) \in \mathfrak{g}$ vanishes at the endpoints.
\item [{\bf iv}] The {\bfi Euler--Poincar\'{e}}
equations\footnote{Note that these equations are not the basic
Euler--Poincar\'e equations
because we are not regarding $\mathfrak{g} \times V ^\ast$ as a Lie
algebra. Rather these
equations are thought of as a generalization of the classical
Euler-Poisson equations
for a heavy top, written in body
angular velocity variables, as we shall see in the examples. Some
authors may prefer the
term Euler-Poisson-Poincar\'{e} equations for these
equations.}hold on $\mathfrak{g} \times V^\ast$
\begin{equation} \label{eulerpoincare}
\frac{d}{dt} \frac{\delta l}{\delta \xi} =
 \operatorname{ad}_{\xi}^{\ast} \frac{ \delta l }{ \delta \xi}
+ \frac{\delta l}{\delta a} \diamond a.
\end{equation}
\end{enumerate}
\end{thm}
\vspace{0.2in}

\noindent {\bf Proof.\,} \quad The equivalence of {\bf i} and {\bf ii}
holds for any configuration manifold and so, in particular, it
holds in this case.

Next we show the equivalence of {\bf iii} and {\bf iv}.
Indeed, using the definitions,
integrating by parts, and taking into
account that $\eta(t_1) = \eta (t_2) = 0$, we compute the
variation of the integral to be
\begin{eqnarray*}
\delta \int_{t_1}^{t_2} l(\xi(t), a(t)) dt &=&
\int_{t_1}^{t_2} \left (\left \langle
\frac{ \delta l}{\delta \xi}\,,
\delta \xi \right \rangle + \left \langle \delta a,
\frac{\delta l}{\delta a}\right \rangle \right ) \, dt\\  &=&
\int_{t_1}^{t_2} \left ( \left \langle
\frac{\delta l}{\delta \xi}\,, \dot{\eta} +
 \operatorname{ad}_{\xi} \eta \right \rangle -
\left \langle \eta a, \frac{\delta l}{\delta a}
\right \rangle \right )dt\\ &=&
\int_{t_1}^{t_2} \left (\left \langle - \frac{ d}{dt}
\left(\frac{\delta l }{ \delta \xi} \right) +
 \operatorname{ad}_{\xi}^{\ast}\frac{\delta l}{\delta \xi}\,, \eta
\right \rangle + \left \langle \frac{\delta l }{\delta a}
\diamond a\,,
\eta \right \rangle \right )\,dt \\ &=&
\int_{t_1}^{t_2} \left \langle - \frac{ d}{dt}
\left(\frac{ \delta l }{ \delta \xi} \right) +
 \operatorname{ad}_{\xi}^{\ast}\frac{\delta l}{\delta \xi} +
\frac{\delta l }{\delta a} \diamond a\,, \eta
\right \rangle \,dt
 \end{eqnarray*}
and so the result follows.

Finally we show that {\bf i} and {\bf iii} are equivalent.
First note that the $G$--invariance of $L:TG
\times V^\ast \rightarrow \mathbb{R}$ and the definition of $a(t) =
g(t)^{-1}a_0$ imply that the
integrands in (\ref{hamiltonprinciple}) and
(\ref{variationalprinciple}) are equal. However, all variations
$\delta g(t) \in TG$ of $g(t)$ with fixed endpoints induce and are
induced by variations $\delta \xi(t) \in
\mathfrak{g}$ of $\xi(t)$ of the form $\delta \xi = \dot{\eta } +
[\xi , \eta ]$ with $\eta(t) \in \mathfrak{g}$ vanishing at the
endpoints; the relation between $\delta g(t)$ and $\eta(t)$
is given by $\eta(t) = g(t)^{-1} \delta g(t)$. This is the content
of the following lemma proved in Bloch et al. [1996].
\footnote{This lemma is simple for matrix groups, as in Marsden and
Ratiu [1994], but it is less elementary for general Lie groups.}

\begin{lem} \label{propvariations} Let $g: U \subset\mathbb{R} ^2
\rightarrow G$ be a smooth map
and denote its partial derivatives by
\[ \xi (t,\varepsilon ) = T L _{ g(t, \varepsilon ) ^{ -1}}
(\partial g(t,\varepsilon) / \partial t)\] and
\[ \eta (t, \varepsilon) =
TL_{g(t,\varepsilon)^{-1}}(\partial g(t,\varepsilon) /
\partial \varepsilon).\]
Then
\begin{equation} \label{eulerpoincare20}
\frac{ \partial \xi }{ \partial \varepsilon} -
\frac{ \partial \eta }{
\partial t} = [\xi , \eta ]\,.
\end{equation}
Conversely, if $U$ is simply connected and $\xi , \eta : U
\rightarrow { \mathfrak{g} }$ are smooth functions
satisfying (\ref{eulerpoincare20})
then there exists a smooth function
$g: U \rightarrow G$ such that $\xi (t,
\varepsilon ) = T L _{ g(t, \varepsilon ) ^{ -1}}
(\partial g(t, \varepsilon ) /
\partial t) $ and $\eta (t,\varepsilon )
= TL _{ g(t,\varepsilon ) ^{ -1}}
(\partial g(t,\varepsilon ) /
\partial \varepsilon ).$
\end{lem}

Thus, if {\bf i} holds, we define $\eta(t) = g(t)^{-1} \delta
g(t)$ for a variation $\delta g(t)$ with fixed endpoints. Then
if we let $\delta \xi = g(t)^{-1} \dot g(t)$, we have by the
above proposition
$\delta \xi = \dot{\eta } + [\xi ,\eta]$.  In addition, the
variation of $a(t) = g(t)^{-1} a_0$ is
$\delta a(t) = -\eta(t) a(t)$. Conversely, if  $\delta \xi =
\dot{\eta } + [\xi ,\eta]$ with $\eta(t)$ vanishing at the
endpoints, we define $\delta g(t) = g(t) \eta(t)$ and the above
proposition guarantees then that this $\delta g(t)$ is the
general variation of $g(t)$ vanishing at the endpoints. From
$\delta a(t) = -\eta(t)a(t)$ it follows that the variation of
$g(t)a(t) = a_0$ vanishes, which is consistent with the
dependence of $L_{a_0}$ only on $g(t), \dot g(t)$.
\quad $\blacksquare$

\paragraph{Cautionary Remarks.} Let us explicitly show that {\it
these Euler--Poincar\'e equations (\ref{eulerpoincare}) are not
the Euler--Poincar\'e equations for the semidirect product Lie
algebra\/} $\mathfrak{g}\,\circledS\, V^\ast$. Indeed, by
(\ref{eulerpoincare23}) the basic Euler--Poincar\'e equations
\[
\frac{d}{dt}\frac{\delta l}{\delta (\xi, a)} =
\operatorname{ad}_{(\xi, a)}^{\ast} \frac{ \delta l }{ \delta
(\xi,a)}\,, \quad (\xi, a) \in
\mathfrak{g}\,\circledS\, V^\ast
\]
for $l:\mathfrak{g}\,\circledS\, V^\ast \rightarrow \mathbb{R} $ become
\[
\frac{d}{dt} \frac{\delta l}{\delta \xi} =
 \operatorname{ad}_{\xi}^{\ast} \frac{ \delta l }{ \delta \xi}
+ \frac{\delta l}{\delta a} \diamond a, \quad
\frac{d}{dt}\frac{\delta l}{\delta a} = -\xi\frac{\delta l}{\delta a}\,,
\]
which is a {\em different} system from that given by the
Euler--Poincar\'e equation (\ref{eulerpoincare}) and
$\dot a = -\xi a$, even though the first equations of both systems are
identical.

\paragraph{The Legendre Transformation.} As we explained earlier,
one normally thinks of passing from Euler--Poincar\'e equations
on a Lie algebra
$\mathfrak{g}$ to Lie--Poisson
equations on the dual $\mathfrak{g}^\ast$
by means of the Legendre
transformation. In our case, we start with a
Lagrangian on $ \mathfrak{g}
\times V ^\ast $ and perform a partial
Legendre transformation in the variable
$ \xi $ only, by writing
\begin{equation}\label{legendre}
\mu = \frac{\delta l}{\delta \xi}\,, \quad
h(\mu, a) = \langle \mu, \xi\rangle - l(\xi, a).
\end{equation}
Since
\[
\frac{\delta h}{\delta \mu} = \xi +\left \langle \mu, \frac{\delta
\xi}{\delta \mu} \right \rangle - \left \langle \frac{\delta
l}{\delta \xi}\,, \, \frac{\delta \xi}{\delta \mu} \right \rangle\,
= \,\xi\,,
\]
and $\delta h / \delta a = -\delta l / \delta a$,
we see that (\ref{eulerpoincare}) and $\dot a(t) = -\xi(t) a(t)$
imply (\ref {leftham}) for the {\it minus\/} Lie--Poisson bracket
(that is, the sign + in (\ref {leftham})). If this Legendre
transformation is invertible, then we can also pass from the the
minus Lie--Poisson equations (\ref {leftham}) to the
Euler--Poincar\'e equations (\ref{eulerpoincare}) together with the
equations $\dot a(t) = -\xi(t) a(t)$.
\medskip

\paragraph{Right Representation and Right Invariant Lagrangian.}
There are four versions of the preceding theorem, the given left-left
version, a left-right, a right-left and a right-right version.
For us, the most important ones are the left-left and the
right-right versions. We state the remaining two in the appendix.

 Here we make the following assumptions:
\begin{itemize}
\item There is a {\it right\/} representation of Lie group $G$ on
the vector space $V$ and $G$ acts in the natural way on the {\it
right\/} on $TG \times V^\ast$: $(v_g, a)h = (v_gh, ah)$.
\item Assume that the function $ L : T G \times V ^\ast
\rightarrow \mathbb{R}$ is right $G$--invariant.
\item In particular, if $a_0 \in V^\ast$, define the
Lagrangian $L_{a_0} : TG \rightarrow \mathbb{R}$ by
$L_{a_0}(v_g) = L(v_g, a_0)$. Then $L_{a_0}$ is right
invariant under the lift to $TG$ of the right action of
$G_{a_0}$ on $G$, where $G_{a_0}$ is the isotropy group of $a_0$.
\item  Right $G$--invariance of $L$ permits us to define
$l: {\mathfrak{g}} \times V^\ast \rightarrow \mathbb{R}$ by
\[
l(v_gg^{-1}, a_0g^{-1}) = L(v_g, a_0).
\]
Conversely,  this relation defines for any
$l: {\mathfrak{g}} \times V^\ast \rightarrow
\mathbb{R} $ a right $G$--invariant function
$ L : T G \times V ^\ast
\rightarrow \mathbb{R} $.
\item For a curve $g(t) \in G, $ let
$\xi (t) := \dot{g}(t) g(t)^{-1}$ and define the curve
$a(t)$ as the unique solution of the linear differential equation
with time dependent coefficients $\dot a(t) = -a(t)\xi(t)$
with initial condition $a(0) = a_0$. The solution can be
written as $a(t) = a_0g(t)^{-1}$.
\end{itemize}

\begin{thm} \label{rarl}
The following are equivalent:
\begin{enumerate}
\item [{\bf i} ] Hamilton's variational principle
\begin{equation} \label{hamiltonprincipleright1}
\delta \int _{t_1} ^{t_2} L_{a_0}(g(t), \dot{g} (t)) dt = 0
\end{equation}
holds, for variations $\delta g(t)$
of $ g (t) $ vanishing at the endpoints.
\item [{\bf ii}  ] $g(t)$ satisfies the Euler--Lagrange
equations for $L_{a_0}$ on $G$.
\item [{\bf iii} ]  The constrained variational principle
\begin{equation} \label{variationalprincipleright1}
\delta \int _{t_1} ^{t_2}  l(\xi(t), a(t)) dt = 0
\end{equation}
holds on $\mathfrak{g} \times V ^\ast $, using variations of the form
\begin{equation} \label{variationsright1}
\delta \xi = \dot{\eta } - [\xi , \eta ], \quad
\delta a =  -a\eta ,
\end{equation}
where $\eta(t) \in \mathfrak{g}$ vanishes at the
endpoints.
\item [{\bf iv}] The Euler--Poincar\'{e} equations hold on
$\mathfrak{g} \times V^\ast$
\begin{equation} \label{eulerpoincareright1}
\frac{d}{dt} \frac{\delta l}{\delta \xi} = -
 \operatorname{ad}_{\xi}^{\ast} \frac{ \delta l }{ \delta \xi}
+ \frac{\delta l}{\delta a} \diamond a.
\end{equation}
\end{enumerate}
\end{thm}

The same partial Legendre transformation (\ref{legendre})
as before maps the Euler--Poincar\'e equations
(\ref{eulerpoincareright1}), together with the
equations $ \dot{a} = - a \xi $ for $a$ to the plus Lie--Poisson
equations  (\ref{rightsemi1.eqn}) and
(\ref{rightsemi2.eqn}) (that is, one chooses the overall minus sign
in these equations).
\paragraph{Generalizations.} The Euler--Poincar\'e equations are a
special case of the reduced Euler-Lagrange equations (see Marsden and
Scheurle [1993b] and Cendra, Marsden and Ratiu [1997]). This is shown
explicitly in  Cendra, Holm, Marsden and Ratiu [1997]. There is,
however, an easy generalization that is needed in some of the
examples we will consider. Namely, if $ L: TG \times V ^\ast \times
TQ $ and if the group $G$ acts in a trivial way on $ TQ $, then one
can carry out the reduction in the same way as above, carrying along
the Euler-Lagrange equations for the factor $ Q $ at each step. The
resulting reduced equations then are the Euler--Poincar\'e equations
above for the $ \mathfrak{g} $ factor, together the Euler-Lagrange
equations for the $ q \in Q $ factor. The system is coupled through
the dependence of $L$ on all variables. (For a full statement, see
Cendra, Holm, Hoyle and Marsden [1997], who use this extension to treat
the Euler--Poincar\'e formulation of the Maxwell-Vlasov equations for
plasma physics.)

\section{The Kelvin-Noether Theorem} \label{sec-KN-Theorem}
In this section, we explain a
version of the Noether theorem that holds for solutions of the
Euler--Poincar\'e equations. Our formulation is motivated and designed
for ideal continuum theories (and hence the name Kelvin-Noether), but it
may also of interest for finite dimensional mechanical systems.
Of course it is well known (going back at least to the pioneering work
of Arnold [1966a]) that the Kelvin circulation theorem for ideal flow is
closely related to the Noether theorem applied to continua using the
particle relabelling symmetry group.

There is a version of the theorem that holds for each of the choices
of conventions, but we shall pick the left-left conventions to
illustrate the result.

\paragraph{The Kelvin-Noether Quantity.} We start with a Lagrangian
$L _{a _0}$ depending on a parameter $a _0 \in V ^\ast$ as above. We
introduce a manifold ${\mathcal C}$ on which $G$ acts (we assume this
is also a left action) and suppose we have an equivariant map
${\mathcal K} : {\mathcal C} \times V ^\ast
\rightarrow \mathfrak{g} ^{\ast \ast} $.

As we shall see, in the case of continuum theories, the space ${\mathcal
C}$ will be a loop space and $\left\langle {\mathcal K} (c, a), \mu
\right\rangle$ for $c \in {\mathcal C}$ and $\mu \in
\mathfrak{g}^\ast$ will be  a circulation. This class of examples
also shows why we {\it do not} want to identify the double dual
$\mathfrak{g} ^{\ast \ast}$ with $\mathfrak{g}$.

Define the {\bfi Kelvin-Noether quantity}
$I : {\mathcal C} \times \mathfrak{g} \times V ^\ast
\rightarrow \mathbb{R}$ by
\begin{equation}\label{KelvinNoether}
I(c, \xi, a) = \left\langle{\mathcal K} (c, a), \frac{ \delta
l}{\delta \xi}( \xi , a)
\right\rangle.
\end{equation}

We are now ready to state the main theorem of this section.

\begin{thm}[Kelvin-Noether.] \label{KelvinNoetherthm}Fixing $c_0 \in
{\mathcal C}$, let $\xi (t), a(t)$ satisfy the
Euler--Poincar\'e equations and define $g(t)$ to be the solution of
$\dot{g}(t) = g(t) \xi(t)$ and, say, $ g (0) = e$. Let
$c(t) = g(t)^{-1} c_0$ and $I(t) = I(c(t), \xi(t), a(t))$. Then
\begin{equation}
\frac{d}{dt} I(t) = \left\langle {\mathcal K}(c(t), a (t) ),
                     \frac{\delta l}{\delta a} \diamond a
\right\rangle.
\end{equation}
\end{thm}

\noindent{\bf Proof.\,} First of all, write $ a (t) = g(t) ^{-1} a _0$
as we did
previously and use equivariance to write $I(t)$ as follows:
\[
   \left\langle {\mathcal K} (c(t), a(t) ) ,
          \frac{\delta l}{\delta \xi} (\xi(t), a(t)) \right\rangle
   =  \left\langle {\mathcal K} ( c_0, a _0 ), g(t)
         \left[ \frac{\delta l}{\delta \xi} (\xi(t),a(t)) \right]
\right\rangle\,.
\]
The $g ^{-1}$ pulls over to the right side as $g$ (and not $g^{-1}$)
because of our conventions of always using left representations.  We
now differentiate the right hand side of this equation. To do so, we
use the following well known formula for differentiating the
coadjoint action (see Marsden and Ratiu [1994], page 276):
\[
\frac{d}{dt} [g(t) \mu (t)]
=  g (t) \left[ - \operatorname{ad}_{\xi (t)}^\ast  \mu (t)
     +   \frac{d}{dt} \mu (t) \right] ,
\]
where, as usual,
\[ \xi (t) = g (t) ^{-1} \dot{g} (t) .
\]
Using this coadjoint action formula and the
Euler--Poincar\'e equations, we obtain
\begin{eqnarray*}
      \frac{d}{dt} I
  & = &  \frac{d}{dt} \left\langle {\mathcal K} ( c_0, a_0 ), g(t)
           \left[ \frac{\delta l}{\delta \xi} (\xi(t),a(t)) \right]
\right\rangle \\
  & = & \left\langle {\mathcal K} ( c _0, a_0 ) , \frac{d}{dt}
\left\{ g(t)
           \left[ \frac{\delta l}{\delta \xi} (\xi(t),a(t)) \right]
\right\} \right\rangle \\
  & = & \left\langle
      {\mathcal K} (c_0, a_0), g(t) \left[ - \operatorname{ad}^\ast _\xi
       \frac{\delta l}{\delta \xi}
      + \operatorname{ad} _\xi ^\ast \frac{\delta l}{\delta \xi}
      + \frac{\delta l}{\delta a} \diamond a \right]
\right\rangle \\
& = & \left\langle {\mathcal K} (c_0, a_0 ) , g(t)  \left[
\frac{\delta l}{\delta a} \diamond a\right] \right\rangle \\
& = & \left\langle {g(t) ^{-1} \mathcal K} ( c _0, a_0 ) , \left[
\frac{\delta l}{\delta a} \diamond a\right] \right\rangle \\
& = & \left\langle {\mathcal K} ( c (t), a (t)  ) ,  \left[
\frac{\delta l}{\delta a} \diamond a\right] \right\rangle\,.
\end{eqnarray*}
where, in the last steps, we used the definitions of the coadjoint action,
as well as the Euler--Poincar\'e equation (\ref{eulerpoincare})
and the equivariance of the map $ {\mathcal K} $.
\quad
$\blacksquare$
\medskip

\begin{cor} For the basic Euler--Poincar\'e equations, the Kelvin
quantity
$I (t) $, defined the same way as above but with
$ I : {\mathcal C} \times \mathfrak{g} \rightarrow\mathbb{R}$,
is conserved.
\end{cor}

For a review of the standard Noether theorem results
for energy and momentum conservation in the context of the
general theory, see, e.g., Marsden and Ratiu [1994].

\section{The Heavy Top} \label{sec-hvytop}
In this section we shall use Theorem \ref{lall} to derive
the classical Euler--Poisson equations for the heavy top.
Our purpose is merely to illustrate the theorem with a
concrete example.

\paragraph{The Heavy Top Lagrangian.} The heavy top kinetic energy is
given by the left invariant metric on $SO(3)$ whose value at the
identity is
$\langle \boldsymbol{\Omega}_1, \boldsymbol{\Omega}_2 \rangle =
\mathbb{I}\boldsymbol{\Omega}_1 \cdot \boldsymbol{\Omega} _2$,
where
$\boldsymbol{\Omega}_1 , \boldsymbol{\Omega}_2 \in \mathbb{R}^3$ are
thought of as elements of
$\mathfrak {so}(3)$, the Lie algebra of
$SO(3)$, via the isomorphism $\boldsymbol{\Omega} \in \mathbb{R}^3
\mapsto
\hat{\boldsymbol{\Omega}} \in \mathfrak {so}(3)$, $\hat
{\boldsymbol{\Omega}}\mathbf{v}:=\boldsymbol{\Omega}\times
\mathbf{v}$, and where
$\mathbb{I}$ is the (time independent) moment of inertia tensor in
body coordinates, usually taken as a diagonal matrix by choosing the
body coordinate system to be a principal axes body frame. This kinetic
energy is thus left invariant under the full group $SO(3)$. The
potential energy is given by the work done in lifting the weight of
the body to the height of its center of mass, with the direction of
gravity pointing downwards. If $M$ denotes
the total mass of the top, $g$ the magnitude of the gravitational
acceleration, $\boldsymbol{\chi}$ the unit vector of the oriented line
segment pointing from the fixed point about which the top
rotates (the origin of a spatial coordinate system) to the
center of mass of the body, and $\ell$ its length,
then the potential energy is given by $-{Mg}\ell\, \mathbf{R}^{-1}
\mathbf{e}_3
\cdot \boldsymbol{\chi}$, where $\mathbf{e}_3$ is the axis of the
spatial coordinate system parallel to the direction of gravity
but pointing upwards. This potential energy breaks the full
$SO(3)$ symmetry and is invariant only under the rotations
$S^1$ about the $\mathbf{e}_3$--axis.

However, for the application of Theorem \ref{lall} we are supposed to
think of the Lagrangian of the heavy top as a function on
$TSO(3)\times\mathbb{R}^3\rightarrow \mathbb{R}$. That is, we need to
think of the potential energy as a function of $(u_\mathbf{R},
\mathbf{v}) \in TSO(3)\times \mathbb{R}^3$. This means that we need to
replace the vector giving the direction of gravity $\mathbf{e}_3$ by an
arbitrary vector $\mathbf{v} \in \mathbb{R}^3$, so that the
potential equals
\[
U(u_\mathbf{R}, \mathbf{v}) = {Mg}\ell\, \mathbf{R}^{-1} \mathbf{v}
\cdot
\boldsymbol{\chi}.
\]
Thought of this way, the potential is $SO(3)$--invariant.
Indeed, if $\mathbf{R}' \in SO(3)$ is arbitrary, then
\begin{eqnarray*}
U(\mathbf{R}'u_\mathbf{R}, \mathbf{R}'\mathbf{v}) & = &
{Mg}\ell\, (\mathbf{R}'\mathbf{R})^{-1} \mathbf{R}'\mathbf{v} \cdot
\boldsymbol{\chi} \\ & = & {Mg}\ell\, \mathbf{R}^{-1} \mathbf{v} \cdot
\boldsymbol{\chi} \\ & = & U(u_\mathbf{R}, \mathbf{v})
\end{eqnarray*}
and the hypotheses of Theorem \ref{lall} are satisfied.
Thus, the heavy top equations of motion in the body representation are
given by the Euler--Poincar\'e equations
(\ref {eulerpoincare}) for the Lagrangian $l:
\mathfrak {so}(3) \times \mathbb{R}^3 \rightarrow
\mathbb{R} $.

\paragraph{The Reduced Lagrangian.} To compute the explicit expression of
$l$, denote by $\boldsymbol{\Omega}$ the angular velocity and by
$\boldsymbol{\Pi} = \mathbb{I}\boldsymbol{\Omega}$ the angular
momentum in the body representation. Let $\boldsymbol{\Gamma} =
\mathbf{R}^{-1}\mathbf{v}$; if
$\mathbf{v} =
\mathbf{e}_3$, the unit vector pointing upwards on the vertical
spatial axis, then $\boldsymbol{\Gamma}$ is this unit vector viewed
by an observer moving with the body. The Lagrangian
$l:\mathfrak{so}(3)
\times \mathbb{R}  ^3 \rightarrow \mathbb{R}$ is thus given by
\begin{eqnarray*}
l(\boldsymbol{\Omega}, \boldsymbol{\Gamma})
& = &  L(\mathbf{R}^{-1}u_\mathbf{R}, \mathbf{R}^{-1}\mathbf{v}) \\
& = &  \frac{1}{2} \boldsymbol{\Pi} \cdot \boldsymbol{\Omega}
+ U(\mathbf{R}^{-1}u_\mathbf{R}, \mathbf{R}^{-1}\mathbf{v})\\  & = &
\frac{1}{2} \boldsymbol{\Pi} \cdot \boldsymbol{\Omega}
+ {Mg}\ell\, \boldsymbol{\Gamma} \cdot \boldsymbol{\chi}\,.
\end{eqnarray*}
\paragraph{The Euler--Poincar\'e Equations.} It is now straightforward
to compute the Euler--Poincar\'e equations. First note that
\[
\frac{\delta l}{\delta \boldsymbol{\Omega}} = \boldsymbol{\Pi}, \quad
\frac{\delta l}{\delta \boldsymbol{\Gamma}} =
{Mg}\ell\, \boldsymbol{\chi}\,.
\]
Since
\[
 \operatorname{ad}_{\boldsymbol{\Omega}}^{\ast} \boldsymbol{\Pi}
= \boldsymbol{\Pi} \times \boldsymbol{\Omega}\,, \quad \mathbf{v}
\diamond
\boldsymbol{\Gamma} = -\boldsymbol{\Gamma} \times \mathbf{v}\,,
\]
and
\[ \boldsymbol{\Omega}\boldsymbol{\Gamma}
= -\boldsymbol{\Gamma} \times \boldsymbol{\Omega}\,,
\]
the Euler--Poincar\'e equations are
\[
\dot{\boldsymbol{\Pi}} = \boldsymbol{\Pi} \times \boldsymbol{\Omega} +
{Mg}\ell\, \boldsymbol{\Gamma}\times\boldsymbol{\chi}\,,
\]
which are coupled to the $\boldsymbol{\Gamma}$ evolution
\[
\dot{\boldsymbol{\Gamma}} = \boldsymbol{\Gamma} \times
\boldsymbol{\Omega}\,.
\]
This system of two vector equations comprises the classical
Euler--Poisson equations, which describe the motion of the
heavy top in the body representation.

\paragraph{The Kelvin-Noether theorem} Let
${\mathcal C} = \mathfrak{g}$ and let
$ {\mathcal K}: {\mathcal C} \times V ^\ast  \rightarrow
\mathfrak{g}^{\ast
\ast}
\cong \mathfrak{g}$ be the map $ (\mathbf{W}, \boldsymbol{\Gamma}
)\mapsto \mathbf{W}$. Then the Kelvin-Noether theorem gives the
statement
\[
\frac{d}{dt} \left\langle \mathbf{W}, \boldsymbol{\Pi} \right\rangle
= {Mg}\ell \left\langle \mathbf{W}, \boldsymbol{\Gamma} \times
\boldsymbol{\chi}
\right\rangle
\]
where $ \mathbf{W} (t) = \mathbf{R} (t) ^{-1} \mathbf{w} $; in other
words,
$\mathbf{W}(t)$ is the body representation of a space fixed vector.
This statement is easily verified directly. Also, note that $
\left\langle \mathbf{W}, \boldsymbol{\Pi} \right\rangle =
\left\langle \mathbf{w}, \boldsymbol{\pi} \right\rangle $, with
$\boldsymbol{\pi} =
\mathbf{R}(t)\boldsymbol{\Pi}$, so the Kelvin-Noether theorem may be
viewed as a statement about the rate of change of the momentum map of
the system (the spatial angular momentum) relative to the full
group of rotations, not just those about the vertical axis.

\section{The Euler--Poincar\'e
Equations \\ in Continuum Mechanics} \label{sec-EPPC}

In this section we will apply the Euler--Poincar\'e
equations in the case of continuum mechanical systems. We let
$\mathcal{D}$ be a bounded domain in $\mathbb{R} ^n$ with
smooth boundary $\partial \mathcal{D}$ (or, more generally,
a smooth compact manifold with boundary and given volume form or
density). We let $\operatorname{Diff}(\mathcal{D})$ denote the
diffeomorphism group of $\mathcal{D}$  of some given Sobolev class.
If the domain $\mathcal{D}$ is not compact, then various
decay hypotheses at infinity need to be imposed. Under such
conditions, $\operatorname{Diff}(\mathcal{D})$ is a smooth infinite
dimensional manifold and a topological group relative to the
induced manifold  topology. Right translation is smooth but left
translation and inversion are only continuous. Thus,
$\operatorname{Diff}(\mathcal{D})$ is not actually a Lie
group and the previous theory does not apply, strictly speaking.
Nevertheless, if one uses right translations and right
representations, the Euler--Poincar\'e equations of Theorem
\ref{rarl} do make sense, as a simple verification shows. We shall
illustrate below such computations, by verifying several key facts in
the proof.

Let $\mathfrak{X}(\mathcal{D})$ denote the space of
vector fields on $\mathcal{D}$ of some fixed differentiability class.
Formally, this is the {\it right\/} Lie algebra of
$\operatorname{Diff}(\mathcal{D})$, that is, its standard {\it left\/}
Lie algebra bracket is {\it minus\/} the usual Lie bracket for vector
fields. To distinguish between these brackets, we shall reserve in what
follows the notation $[\mathbf{u},\,\mathbf{v}]$ for the standard
Jacobi-Lie bracket of the vector fields $\mathbf{u},\,\mathbf{v} \in
\mathfrak{X}({\mathcal{D}})$ whereas the notation $
\operatorname{ad}_\mathbf{u} \mathbf{v} := -[\mathbf{u},\,\mathbf{v}]$
denotes the adjoint action of the {\it left\/} Lie algebra on itself.

We also let $\mathfrak{X}(\mathcal{D})^\ast$ denote the geometric dual
space of $\mathfrak{X}(\mathcal{D})$, that is,
$\mathfrak{X}(\mathcal{D})^\ast := \Omega^1({\mathcal{D}}) \otimes
{\rm Den}({\mathcal{D}})$, the space of one--form densities on
$\mathcal{D}$. If $\alpha \otimes m \in
\Omega^1({\mathcal{D}}) \otimes {\rm Den}({\mathcal{D}})$, the pairing
of  $\alpha \otimes m$ with $\mathbf{u} \in \mathfrak{X}(\mathcal{D})$ is
given by
\begin{equation}\label{continuumpairing}
\langle \alpha \otimes m, \mathbf{u} \rangle
= \int_{\mathcal{D}} \alpha\cdot \mathbf{u}\, m
\end{equation}
where $\alpha\cdot \mathbf{u}$ is the standard contraction of a
one--form with a vector field. For $\mathbf{u} \in
\mathfrak{X}(\mathcal{D})\,$ and $\alpha\otimes m \in
\mathfrak{X}(\mathcal{D})^\ast$, the dual of the adjoint
representation is defined by
\[
 \langle  \operatorname{ad}^\ast_\mathbf{u}(\alpha\otimes m), \mathbf{v}
\rangle
= -\int_{\mathcal{D}}\alpha \cdot [\mathbf{u}, \mathbf{v}]\;m
\]
and its expression is
\begin{equation}\label{continuumcoadjoint}
 \operatorname{ad}^\ast_\mathbf{u}(\alpha\otimes m) = (\pounds_\mathbf{u}
\alpha + ( \operatorname{div}_m\mathbf{u})\alpha)\otimes m
= \pounds_\mathbf{u}(\alpha\otimes m)\,,
\end{equation}
where  ${\rm div}_m\mathbf{u}$ is the divergence of $\mathbf{u}$ relative
to the measure $m$, that is, $\pounds_\mathbf{u}m = ({\rm
div}_m\mathbf{u})m$. Hence if $\mathbf{u} = u^i \partial/\partial x^i,\,
\alpha  = \alpha_i dx^i$, the one--form factor in the preceding
formula for $ \operatorname{ad}^\ast_\mathbf{u}(\alpha\otimes m)$ has the
coordinate expression
\begin{equation}\label{continuumcoadjoint-coord}
 \left ( u^j \frac{\partial
\alpha_i}{\partial
x^j} + \alpha_j \frac{\partial u^j}{\partial x^i} +
( \operatorname{div}_m\mathbf{u})\alpha_i \right )dx^i\,= \,
\left (\frac{\partial}{\partial x^j}(u^j\alpha_i) +
\alpha_j \frac{\partial u^j}{\partial x^i}\right ) dx^i\;,
\end{equation}
the last equality assuming that the divergence is taken
relative to the standard measure $m = d^n\mathbf{x}$ in $\mathbb{R} ^n$.
(On a Riemannian manifold the metric divergence needs to be used.)

Throughout the rest of the paper we shall use the following
conventions and terminology for the standard quantities in
continuum mechanics. Elements of $\mathcal{D}$ representing the
material particles of the system are denoted by $X$; their
coordinates $X^A, A=1,...,n$ may thus be regarded as the particle
labels. A {\bfi configuration}, which we typically denote by
$\eta$, is an element of $\operatorname{Diff}(\mathcal{D})$.
A {\bfi motion\/} $\eta_t$ is a path in $\operatorname{Diff}(\mathcal{D})$.
The {\bfi Lagrangian} or {\bfi material velocity\/} ${\bf V}(X,t)$
of the continuum along the motion $\eta_t$
is defined by taking the time derivative of the motion
keeping the particle labels (the reference particles) $X$ fixed:
\[
{\bf V}(X, t) := \frac{d\eta_t(X)}{dt}:=
\left.\frac{\partial}{\partial t}\right|_{X}\eta_t(X),
\]
the second equality being a convenient shorthand notation of the
time derivative for fixed $X$.

Consistent with this definition of velocity, the tangent space to
$\operatorname{Diff}(\mathcal{D})$ at $\eta \in
\operatorname{Diff}(\mathcal{D})$ is given by
\[
T_\eta \operatorname{Diff}(\mathcal{D})
= \{ {\bf V}_\eta: {\mathcal{D}} \rightarrow T {\mathcal{D}}  \mid {\bf
V}_\eta(X)
\in T_{\eta(X)}\mathcal{D} \}.
\]
Elements of $T_\eta \operatorname{Diff}(\mathcal{D})$ are usually
thought of as vector fields on $\mathcal{D}$ covering $\eta$. The
tangent lift of right and left translations on
$T\operatorname{Diff}(\mathcal{D})$ by $\varphi \in
\operatorname{Diff}(\mathcal{D})$
have the expressions
\[
{\bf V}_\eta\varphi := T_\eta R_\varphi ({\bf V}_\eta)
= {\bf V}_\eta \circ \varphi\,\qquad {\rm and} \qquad
\varphi{\bf V}_\eta := T_\eta L_\varphi ({\bf V}_\eta)
= T\varphi \circ {\bf V}_\eta \,.
\]

During a motion
$\eta_t$, the particle labeled by $X$ describes a path in
$\mathcal{D}$ whose points $x(X, t):= \eta_t(X)$ are
called the {\bfi Eulerian} or {\bfi spatial points\/} of this path.
The derivative $\mathbf{v}(x, t)$ of this path, keeping the Eulerian
point $x$ fixed, is called the {\bfi Eulerian} or {\bfi spatial
velocity\/} of the system:
\[
\mathbf{v}(x, t):= {\bf V}(X, t) :=
\left.\frac{\partial}{\partial t}\right|_x\eta_t(X).
\]
Thus the Eulerian velocity $\mathbf{v}$ is a time dependent vector
field on $\mathcal{D}$: $\mathbf{v}_t \in \mathfrak{X}(\mathcal{D})$,
where $\mathbf{v}_t(x) := \mathbf{v}(x, t)$. We also have
the fundamental relationship
\[
{\bf V}_t = \mathbf{v}_t \circ \eta_t\,,
\]
where ${\bf V}_t(X):= {\bf V}(X, t)$.
\medskip

The representation space $V^\ast$ of $\operatorname{Diff}(\mathcal{D})$
in continuum mechanics is often some subspace of $\mathfrak{T}
(\mathcal{D})\otimes {\rm Den}({\mathcal{D}})$, the tensor field
densities  on $\mathcal{D}$ and the representation is given by pull
back. It is thus a {\it right\/} representation of
$\operatorname{Diff}(\mathcal{D})$ on
$\mathfrak{T}(\mathcal{D})\otimes {\rm Den}({\mathcal{D}})$. The right
action of the Lie algebra
$\mathfrak{X}({\mathcal{D}})$ on
$V^\ast$ is given by $ a\mathbf{v} := \pounds_\mathbf{v} a$, the Lie
derivative of the tensor field density $a$ along the vector
field $\mathbf{v}$.

The Lagrangian of a continuum mechanical system is a function $L:
T\operatorname{Diff}(\mathcal{D}) \times V^\ast \rightarrow \mathbb{R} $
which is right invariant relative to the tangent lift of right
translation of $\operatorname{Diff}(\mathcal{D})$ on itself and pull
back on the tensor field densities.

Thus, the Lagrangian $L$ induces
a function $l: \mathfrak{X}(\mathcal{D}) \times V^\ast
\rightarrow \mathbb{R} $ given by
\[
l(\mathbf{v}, a) = L(\mathbf{v}\circ \eta, \eta^\ast a),\]
where $  \mathbf{v} \in \mathfrak{X}({\mathcal{D}})$ and $  a \in V^\ast
\subset {\mathfrak{T}}({\mathcal{D}})\otimes {\rm Den}({\mathcal{D}})$,
and where $\eta^\ast a$ denotes the pull back of $a$ by the
diffeomorphism $\eta$ and $\mathbf{v}$ is the Eulerian velocity. The
evolution of $a$ is given by the equation
\[
\dot a = -{\pounds}_\mathbf{v}\, a.
\]
The solution of this equation, given the
initial condition $a_0$, is $a(t) = \varphi_{t\ast} a_0$,
where the lower star denotes the push forward operation and
$\varphi_t$ is the flow of $\mathbf{v}$.

{\bfi Advected} Eulerian
quantities are defined in continuum mechanics to be those
variables which are Lie transported by the flow of the Eulerian
velocity field. Using this standard terminology, the above
equation states that the tensor field density $a$ (which may
include mass density and other Eulerian quantities) is advected.

As remarked, $V^\ast \subset {\mathfrak{T}}({\mathcal{D}})\otimes
{\rm Den}({\mathcal{D}})$.  On a general manifold, tensors of a given
type have natural duals. For example, symmetric covariant tensors are
dual to symmetric contravariant tensor densities, the pairing being
given by the integration of the natural contraction of these tensors.
Likewise, $k$--forms are naturally dual to
$(n-k)$--forms, the pairing being given by taking the integral of
their wedge product.

The operation $\diamond$ between elements of $V$ and
$V^\ast$  producing an element of  $\mathfrak{X}({\mathcal{D}})^\ast$
introduced in section 2 becomes
\begin{equation}\label{continuumdiamond}
\langle v \diamond a, \mathbf{u}\rangle
= -\int_{\mathcal{D}} v \cdot \pounds_\mathbf{u}\,a\;,
\end{equation}
where $v\cdot \pounds_\mathbf{u}\,a$ denotes the contraction, as
described above, of elements of $V$ and elements of $V^\ast$. (These
operations do {\it not} depend on a Riemannian structure.)

For a path $\eta_t \in \operatorname{Diff}(\mathcal{D})$ let
$\mathbf{v}(x, t)$
be its Eulerian velocity and consider as in the hypotheses of
Theorem \ref{rarl} the curve $a(t)$ with initial condition $a_0$
given by the equation
\begin{equation}
\dot a + \pounds_\mathbf{v} a = 0.
\label{continuityequation}
\end{equation}
Let $L_{a_0}({\bf V}) := L({\bf V}, a_0)$.
We can now state Theorem \ref{rarl} in this particular, but
very useful, setting.

\begin{thm}[Euler--Poincar\'{e} Theorem for Continua.]
\label{EPforcontinua}
Consider a path $\eta_t$ in  $\operatorname{Diff}(\mathcal{D})$ with
Lagrangian velocity ${\bf V}$ and Eulerian velocity $\mathbf{v}$. The
following are equivalent:

\begin{enumerate}
\item [{\bf i}] Hamilton's variational principle
\begin{equation} \label{continuumVP}
\delta \int_{t_1}^{t_2} L\left(X, {\bf V}_t (X),
a_0(X)\right)\,dt=0
\end{equation}
holds, for variations $\delta\eta_t$ vanishing at the endpoints.
\item [{\bf ii}] $\eta_t$ satisfies the Euler--Lagrange
equations for $L_{a_0}$ on
$\operatorname{Diff}(\mathcal{D})$.\footnote{We do not write these
equations explicitly since to do so would require either a
coordinatization of the diffeomorphism group, which is not easy to give
explictly, or requires more structure, such as an affine connection on
this group. Certainly writing the equations formally, imagining that
$\eta$ and $\dot{\eta}$ form valid coordinates in which the
Euler--Lagrange equations hold is not correct.}
\item [{\bf iii}] The constrained variational principle in
Eulerian coordinates
\begin{equation}\label{continuumconstrainedVP}
   \delta \int_{t_1}^{t_2} l(\mathbf{v},a)\ dt=0
\end{equation}
holds on $\mathfrak{X}(\mathcal{D}) \times V^\ast$, using
variations of the form
\begin{equation}\label{continuumvariations}
   \delta \mathbf{v} = \frac{\partial \mathbf{u}}{\partial t}
                   +[\mathbf{v},\mathbf{u}], \qquad
   \delta a = - \pounds_\mathbf{u}\,a,
\end{equation}
where $\mathbf{u}_t = \delta\eta_t \circ \eta_t^{-1}$ vanishes at
the endpoints.
\item [{\bf iv}] The Euler--Poincar\'{e} equations for continua
\begin{equation}\label{continuumEP}
   \frac{\partial }{\partial t}\frac{\delta l}{\delta \mathbf{v}}
   = -\, \operatorname{ad}^{\ast}_\mathbf{v}\frac{\delta l}
        {\delta \mathbf{v}}
   +\frac{\delta l}{\delta a}\diamond a
   =-\pounds_\mathbf{v} \frac{\delta l}{\delta \mathbf{v}}
    +\frac{\delta l}{\delta a}\diamond a\,,
\end{equation}
hold, where the $\diamond$ operation given by
(\ref{continuumcoadjoint}) needs to be determined on a
case by case basis, depending on the nature of the tensor $a$.
(Remember that $\delta l/\delta \mathbf{v}$ is a one--form density.)
\end{enumerate}
\end{thm}

\paragraph{Remarks.}

\begin{enumerate}
\item Of course, this theorem can be proven directly
by imitating the proof of Theorem \ref{lall} with appropriate
modifications for right representations and right actions. For
those used to the more concrete language of continuum mechanics as
opposed to that of Lie algebras, the following string of equalities
shows that {\bf iii} is equivalent to {\bf iv}:
\begin{eqnarray}\label{continuumEPderivation}
0 &=&\delta \int_{t_1}^{t_2} l(\mathbf{v}, a) dt
   =\int_{t_1}^{t_2}\left(\frac{\delta l}{\delta \mathbf{v}}\cdot
\delta\mathbf{v} +\frac{\delta l}{\delta a}\cdot \delta a\right)dt
\nonumber \\
  &=&\int_{t_1}^{t_2} \left[\frac{\delta l}{\delta \mathbf{v}}
  \cdot \left(\frac{\partial\mathbf{u}}{\partial t}
    -\operatorname{ad}_\mathbf{v}\,\mathbf{u}\right)
    -\frac{\delta l}{\delta a}\cdot \pounds_\mathbf{u}\, a \right]dt
\nonumber \\
   &=&\int_{t_1}^{t_2} \mathbf{u}\cdot
\left[-\,\frac{\partial}{\partial t}\frac{\delta l}{\delta\mathbf{v}}
   - \operatorname{ad}^*_\mathbf{v}\frac {\delta l}{\delta\mathbf{v}}
   +\frac{\delta l}{\delta a} \diamond a\right]dt\,.
\end{eqnarray}

\item Similarly, one can deduce by hand the form
(\ref{continuumvariations}) of the variations in the
constrained variational principle
(\ref{continuumconstrainedVP}) by a direct calculation. This
proceeds as follows. One writes the basic relation between
the spatial and material velocities, namely
$ \mathbf{v} (x,t) = \dot{\eta} (\eta_t ^{-1} (x),t) $. One then
takes the variation of this equation with respect to $\eta$
and uses the definition $ \mathbf{u} (x,t) = \delta \eta
(\eta_t ^{-1} (x),t) $ together with a calculation of its
time derivative. Of course, one can also do this calculation
using the inverse map
$\eta _t ^{-1} $ instead of $\eta$ as the basic variable, see
Holm, Marsden, and Ratiu [1986], Holm [1996a,b].

\item As we mentioned in the context of perfect fluids, the
preceding sort of calculation for $\delta \mathbf{v}$ in fluid
mechanics and the interpretation of this restriction on the
form of the variations as the so-called Lin constraints is
due to Bretherton [1970].

\item The coordinate expressions for $({\delta l/ \delta a})
\diamond a$ required to complete the equations of motion are given in
the next section for several choices of $a_0(X)$ in three
dimensions. Namely, we shall discuss the choices
corresponding to scalars, one-forms, two-forms, densities in
three dimensions, and symmetric tensors. In the equations of
motion, all of these quantities will be advected.

\item As with the general theory, variations of the action in
the advected tensor quantities contribute to the equations of
motion which follow from Hamilton's principle. At the level
of the action $l$ for the Euler--Poincar\'{e} equations, the
Legendre transform in the variable $\mathbf{v}$ alone is often
nonsingular, and when it is, it produces the Hamiltonian
formulation of Eulerian fluid motions with a Lie-Poisson
bracket defined on the dual of the semidirect product algebra
of vector fields acting amongst themselves by Lie bracket and
on tensor fields and differential forms by the Lie
derivative. This is a special instance of the more general
facts for Lie algebras that were discussed earlier.

\item As mentioned earlier, in the absence of the tensor
fields $a$ and when $l$ is the kinetic energy metric, the basic
Euler--Poincar\'e equations are the {\em geodesic spray
equations} for geodesic motion on the diffeomorphism group
with respect to that metric. See, e.g., Arnold [1966a],
Ovsienko and Khesin [1987], Zeitlin and Kambe [1993], Zeitlin
and Pasmanter [1994], Ono [1995a, 1995b] and Kouranbaeva
[1997] for details in particular applications of ideal
continuum mechanics.

\end{enumerate}

\paragraph{Remarks on the inverse map and the tensor fields
$a$ for fluids.} In the case of fluids in the Lagrangian
picture, the flow of the fluid is a diffeomorphism which
takes a fluid parcel along a path from its initial position
$X$, in a ``reference configuration" to its current position
$x$ in the ``container", i.e., in the Eulerian domain of
flow. As we have described, this forward map is denoted by
$\eta : X \mapsto x$. The inverse map $\eta^{-1}: x \mapsto
X$ provides the map assigning the Lagrangian labels to a
given spatial point. Interpreted as passive scalars, these
Lagrangian labels are simply advected with the fluid flow,
$\dot X=0$. In the Lagrangian picture, a tensor density in
the reference configuration $a_0(X)$ (satisfying $\dot
a_0(X)=0$) consists of {\it invariant} tensor functions of
the initial reference  positions and their differentials.
These tensor functions are parameters of the initial fluid
reference configuration (e.g., the  initial density
distribution, which is an invariant $n$-form).

When viewed in the Eulerian picture as
\[
   a_t (x):= (\eta _{t \ast}  a_0)(x)
      = (\eta^{-1 \ast}_t  a_0)(x),
\]
{\rm i.e.\/},
\[
     a_0 (X):= (\eta ^\ast _{t}  a_t)(X)
             = (\eta^{-1}_{t \ast} a_0)(X),
\]
the time invariant tensor density $a_0(X)$ in
the reference configuration acquires advective dynamics in the
Eulerian picture, namely
\[ \dot a_0(X) = \left(\frac{\partial}{\partial t} +
\pounds_\mathbf{v}\right)\,a(x,t)=0,\]  where $\pounds_\mathbf{v}$
denotes Lie derivative with respect to the Eulerian velocity field
$\mathbf{v}(x,t)$. This relation results directly from the
well known Lie derivative formula for tensor fields. (See, for
example, Abraham, Marsden and Ratiu [1988].)

Mapping the time invariant quantity $ a _0 $ (a tensor density
function of $X$) to the time varying quantity $ a _t $ (a tensor
density function of $x$) as explained above is a special case of
the general way we advect quantities in $ V ^\ast $ in the general
theory. Specifically, we can view this advection of $ a _t $ as
being the fluid analogue of the advection of the unit vector along
the direction of gravity (a spatially fixed quantity) by means of the
body rotation vector in the heavy top example.

Consistent with the fact that the heavy top is a {\it left invariant}
system while continuum theories are {\it right invariant}, the
advected tensor density $a_t$ is a spatial quantity, while the
advected direction of gravity is a body quantity. If we were to take
the inverse map $\eta^{-1}$ as the basic group variable, rather than
the map $\eta $, then continuum theories would also become left
invariant.

\paragraph{The continuity equation for the mass density.}
We will need to impose an additional assumption on our continuum
theory. Namely, we assume that amongst the tensor densities being
advected, there is a special one, namely the mass density. This of
course is a tensor density that occurs in all continuum theories. We
denote this density by $\rho d^n x$ and it is advected according to
the standard principles discussed above. Thus, $\rho$ satisfies the
usual continuity equation:
\[
    \frac{\partial }{\partial t} \rho
    + {\rm div} (\rho \mathbf{v} ) = 0.
\]
In the Lagrangian picture we have $\rho d^n x=\rho_0(X) d^n X$, where
$\rho_0(X)$ is the mass density in the reference configuration. It
will also be convenient in the continuum examples below to define
Lagrangian {\it mass} coordinates $\ell(X)$ satisfying $\rho d^n
x=d^n \ell$ with $\dot \ell = 0$. (When using Lagrangian mass
coordinates, we shall denote the density $\rho$ as $D$.) We assume
that $\rho$ (or $D$) is strictly positive.

\paragraph{The Kelvin-Noether Circulation Theorem.}
Let $\mathcal C$  be the space of continuous loops $\gamma : S^1
\rightarrow {\mathcal{D}}$ in
$\mathcal{D}$ and let the group
$\operatorname{Diff}(\mathcal{D})$ act on
$\mathcal C$ on the left by
$(\eta, \gamma)\in \operatorname{Diff}(\mathcal{D}) \times
{\mathcal C} \mapsto \eta\gamma \in {\mathcal C}$, where
$\eta\gamma = \eta\circ\gamma$.

Next we shall define the {\bfi circulation map\/}
$\mathcal K: {\mathcal C} \times V ^\ast
\rightarrow \mathfrak{X}(\mathcal{D})^{\ast\ast}$.
Given a one form density $\alpha \in \mathfrak{X} ^\ast $ we can
form a one form (no longer a density) by dividing it by the mass
density $\rho$; we denote the result just by $\alpha  / \rho$.
We let $ {\mathcal K} $ then be defined by
\begin{equation}
 \left\langle {\mathcal K} (\gamma , a) , \alpha
\right\rangle  = \oint _\gamma \frac{ \alpha }{ \rho } \, .
\label{circ.map}
\end{equation}
The expression in this definition is called the {\bfi circulation\/}
of the one--form $\alpha/\rho$ around the loop $\gamma$.

This map is equivariant in the sense that
\[ \left\langle {\mathcal K} ( \eta \circ \gamma , \eta _\ast a) ,
\eta _\ast \alpha \right\rangle =
\left\langle {\mathcal K} (\gamma, a), \alpha \right\rangle
\]
for any $ \eta \in \operatorname{Diff} ( {\mathcal{D}} ) $. This is
proved using the definitions and the change of variables formula.

Given the Lagrangian $l:\mathfrak{X}(\mathcal{D}) \times V^\ast
\rightarrow \mathbb{R}$, the Kelvin--Noether quantity is given
by (\ref{KelvinNoether}) which in this case becomes
\[
I(\gamma, \mathbf{v}, a) = \oint_\gamma \frac{1}{\rho}\frac{\delta
l}{\delta \mathbf{v}}\;.
\]
With these definitions of $\mathcal K$ and $I$, the statement of
Theorem \ref{KelvinNoetherthm} becomes the classical Kelvin
circulation theorem.

\begin{thm}[Kelvin Circulation Theorem.]\label{KelThmforcontinua}
Assume that $\mathbf{v}(x, t)$
satisfies the Euler--Poincar\'e equations for continua:
\[
\frac{\partial }{\partial t}\left(\frac{\delta
l}{\delta \mathbf{v}}\right)
   = -\pounds_\mathbf{v} \left(\frac{\delta l}{\delta \mathbf{v}}\right)
    +\frac{\delta l}{\delta a}\diamond a
\]
and $a$ satisfies the advection relation
\[\frac{\partial a}{\partial t} +
\pounds_\mathbf{v} a = 0.
\]
Let $\eta_t$ be the flow of the Eulerian
velocity field $\mathbf{v}$, that is, $\mathbf{v}_t  =
(d\eta_t/dt)\circ \eta_t^{-1}$. Define
$\gamma_t  := \eta_t\circ \gamma_0$ and $I(t) := I(\gamma_t ,
\mathbf{v}_t, a_t)$. Then
\[
\frac{d}{dt}I(t) = \oint_{\gamma_t }
\frac{1}{\rho}\frac{\delta l}{\delta a}\diamond a\;.
\]
\end{thm}
In this statement, we use a subscript $t$ to emphasise that the
operations are done at a particular $t$ and to avoid having to write
the other arguments, as in $ a _t (x) = a(x,t)$; we omit the
arguments from the notation when convenient. Due to the importance
of this theorem we shall give here a separate proof tailored for the
case of continuum mechanical systems.
\medskip

\noindent{\bf Proof.\,} First we change variables in the expression
for $I(t)$:
\[
I(t) = \oint_{\gamma_t }\frac{1}{\rho _t }\frac{\delta l}{\delta \mathbf{v}}
=\oint_{\gamma_0} \eta_t^\ast\left[\frac{1}{\rho_t}\frac{\delta l}
{\delta \mathbf{v}}\right] = \oint_{\gamma_0}
\frac{ 1 }{ \rho _0 } \eta_t^\ast\left[\frac{\delta
l} {\delta \mathbf{v}}\right].
\]
Next, we use the Lie derivative formula
\[
\frac{d}{dt}\left(\eta_t^*\alpha_t\right) =
\eta_t^*\left(\frac{\partial}{\partial t}\alpha_t + \pounds_\mathbf{v}
\alpha_t \right)\;,
\]
for an arbitrary one--form density $\alpha_t$.
This formula gives
\begin{eqnarray*}
      \frac{d}{dt} I(t)
  & = & \frac{d}{dt}  \oint_{\gamma_0}
\frac{ 1 }{ \rho _0 } \eta_t^\ast\left[\frac{\delta
l} {\delta \mathbf{v}}\right] \\
  & = & \oint_{\gamma_0} \frac{1}{\rho _0} \frac{d}{dt}
\left( \eta_t^\ast\left[\frac{\delta l} {\delta \mathbf{v}}\right]\right)  \\
  & = & \oint_{\gamma_0} \frac{1}{\rho _0} \eta_t^*\left[
\frac{\partial}{\partial t}
\left(\frac{\delta l} {\delta \mathbf{v}}\right) +
\pounds_\mathbf{v}\left(\frac{\delta l} {\delta \mathbf{v}} \right)\right].
\end{eqnarray*}
By the Euler--Poincar\'e equations, this becomes
\[
      \frac{d}{dt} I(t)
  =   \oint_{\gamma_0} \frac{1}{\rho _0} \eta_t^*\left[
\frac{\delta  l}{\delta  a}
\diamond a \right] = \oint_{\gamma_t} \frac{1}{\rho _t} \left[
\frac{\delta  l}{\delta  a}
\diamond a \right],
\]
again by the change of variables formula.\quad $\blacksquare$

\begin{cor}[Kelvin-Noether form.]
Since the last expression holds for every loop
$\gamma_t$, we may write it as
\begin{equation}\label{KThfm}
\left(\frac{\partial}{\partial t} + \pounds_\mathbf{v} \right)
\frac{1}{\rho} \frac{\delta l}{\delta \mathbf{v}}
= \frac{1}{\rho} \frac{\delta  l}{\delta  a} \diamond a\,.
\end{equation}
\end{cor}
This is the {\bfi Kelvin-Noether form} of the Euler--Poincar\'e equations
for ideal continuum dynamics.

\section{Applications of the Euler--Poincar\'{e} Theorem to Continua}
\label{sec-AppEPCont}

\paragraph{Variational formulae in three dimensional Euclidean
coordinates.} We compute explicit formulae for the variations $\delta a$
in the cases that the set of tensor fields $a$ consists of elements with
the following coordinate functions in a Euclidean basis on $\mathbb{R}^3$,
\begin{equation}
a\in\{b,\mathbf{A}\cdot d\mathbf{x},\mathbf{B}\cdot d\mathbf{S},D\,d^3x,
S_{ab}\,dx^a\otimes dx^b\}\,.
\label{Eul-ad-qts}
\end{equation}
These are the tensor fields that typically occur in ideal continuum
dynamics. Here, in three dimensional vector notation, we choose
$\mathbf{B}={\rm curl}\,\mathbf{A}$ and
$d(\mathbf{A}\cdot d\mathbf{x})=\mathbf{B}\cdot d\mathbf{S}$.
In Euclidean
coordinates on $\mathbb{R}^3$, this is $d(A_kdx^k)=A_{k,j}dx^j\wedge dx^k
=\frac{1}{2} \epsilon_{ijk}B^idx^j\wedge dx^k$, where $\epsilon_{ijk}$
is the completely antisymmetric tensor density on $\mathbb{R}^3$ with
$\epsilon_{123}=+1$. (The two form
$\mathbf{B}{\cdot}d\mathbf{S}=d(\mathbf{A}\cdot d\mathbf{x})$
is the physically interesting
special case of $B_{kj}dx^j{\wedge}dx^k$, in which
$B_{kj}=A_{k,j}$, so that $\nabla\cdot\mathbf{B}=0$.)

We have seen that invariance of the set
$a$ in the Lagrangian picture under the dynamics of
$\mathbf{v}$ implies in the Eulerian picture that
$(\frac{\partial}{\partial t} + \pounds_\mathbf{v})\,a=0$,
where $\pounds_\mathbf{v}$ denotes Lie derivative with respect to the
velocity vector field $\mathbf{v}$. According to Theorem
\ref{EPforcontinua}, equation (\ref{continuumEP}), the variations of the
tensor functions $a$ at fixed
$\mathbf{x}$ and $t$ are also given by Lie derivatives, namely
$\delta a = - \pounds_\mathbf{u}\,a$, or
\begin{eqnarray}
\delta b
&=& -\pounds_\mathbf{u}\ b = -\mathbf{u}\cdot\nabla\,b\,,
\nonumber \\
\delta \mathbf{A}\cdot d\mathbf{x}
&=& -\pounds_\mathbf{u}\,(\mathbf{A}\cdot d\mathbf{x})
=-\left((\mathbf{u}\cdot\nabla)\mathbf{A}+A_j\nabla u^j\right)
\cdot d\mathbf{x}
\nonumber \\
&=& \left(\mathbf{u}\times{\rm curl}\,\mathbf{A}
-\nabla(\mathbf{u}\cdot\mathbf{A})\right)\cdot d\mathbf{x}\,,
\nonumber \\
\delta\mathbf{B}\cdot d\mathbf{S}
&=&-\pounds_\mathbf{u}\,(\mathbf{B}\cdot
d\mathbf{S})
= \left({\rm curl}\,(\mathbf{u}\times\mathbf{B})\right)\cdot d\mathbf{S}
\ =\ d (\delta \mathbf{A}\cdot d\mathbf{x})
\,,
\nonumber \\
\delta D\ d^3x&=&-\pounds_\mathbf{u}\,(D\,d^3x)
= -\nabla\cdot({D\bf u})\ d^3x\,,
\nonumber \\
\delta S_{ab}\,dx^a\otimes dx^b
&=&-\pounds_\mathbf{u}\,(S_{ab}\,dx^a\otimes dx^b)
\nonumber \\
&=& -(u^k S_{ab,k}+S_{kb}u^k_{,a}+S_{ka}u^k_{,b})\,dx^a\otimes dx^b\,.
\end{eqnarray}
Hence, Hamilton's principle with this dependence yields
\begin{eqnarray}
0 &=&\delta \int dt\ l(\mathbf{v}; b,\mathbf{A},\mathbf{B},D,S_{ab})
\nonumber \\
&=&\int dt\ \left[ \frac{\delta l}{\delta \mathbf{v}}\cdot
\delta\mathbf{v} +\frac{\delta l}{\delta b} \ \delta b
+\frac{ \delta l}{ \delta D}\ \delta D
+\frac{ \delta l}{ \delta \mathbf{A}}\cdot\delta\mathbf{A}
+\frac{ \delta l}{ \delta \mathbf{B}}\cdot\delta\mathbf{B}
+\frac{ \delta l}{  \delta S_{ab}}\ \delta S_{ab}
\right]
\nonumber \\
&=&  \int dt\ \left[  \frac{\delta l}{\delta \mathbf{v}}
\cdot \left( \frac{\partial\mathbf{u}}{\partial
t}-\operatorname{ad}_\mathbf{v}\,\mathbf{u}\right)  -\frac{ \delta l}{
\delta b}\ \mathbf{u}\cdot\nabla\,b -\frac{ \delta l}{ \delta D}\
\left(\nabla\cdot({D\bf u})\right) \right.
            \nonumber \\&&\qquad\quad
+\   \frac{ \delta l}{ \delta \mathbf{A}}\cdot
\left(\mathbf{u}\times{\rm curl}\,\mathbf{A}
-\nabla(\mathbf{u}\cdot\mathbf{A})\right)
+\frac{ \delta l}{ \delta \mathbf{B}}\cdot
\left({\rm curl}\,(\mathbf{u}\times\mathbf{B})\right)
\nonumber \\&&\qquad\quad
-\ \left. \frac{ \delta l}{\delta S_{ab}}
\left(u^k S_{ab,k}+S_{kb}u^k_{,a}+S_{ka}u^k_{,b}\right)
\right]
\nonumber \\
&=&\int dt\
\left[    \mathbf{u}\cdot
\left(
       -\frac{ \partial}{\partial t}\frac{ \delta l}{ \delta\mathbf{v}}
-\operatorname{ad}^*_\mathbf{v}\ \frac{ \delta l}{ \delta\mathbf{v}}
-\frac{ \delta l}{ \delta b}\ \nabla\,b
+D\ \nabla\frac{ \delta l}{ \delta D}
\right.
\right.
\nonumber \\&&\qquad\qquad\qquad
-\ \left.
    \frac{ \delta l}{ \delta \mathbf{A}}\times{\rm curl}\, \mathbf{A}
   +\mathbf{A}\ {\rm div}\,\frac{ \delta l}{ \delta \mathbf{A}}\
   +\mathbf{B}\times{\rm curl}\, \frac{ \delta l}{ \delta \mathbf{B}}
\right)
\label{EP:eqn-u2} \\ &&\qquad\qquad
\left. +\ u^k
\left.
\left(   -\frac{ \delta l}{\delta S_{ab}} S_{ab,k}
         +(\frac{ \delta l}{\delta S_{ab}}S_{kb})_{,a}
         +(\frac{ \delta l}{\delta S_{ab}}S_{ka})_{,b}
\right)
\right]
\right.
\nonumber \\
&=& \int dt\ \mathbf{u}\cdot
\left[
      -\frac{ \partial}{\partial t}\frac{ \delta l}{ \delta\mathbf{v}}
      -\operatorname{ad}^*_\mathbf{v}\ \frac{ \delta l}{ \delta\mathbf{v}}
      +\frac{\delta l}{\delta a} \diamond a\
\right]
\nonumber \\
&=& \int dt\ \mathbf{u}\cdot
  \left[
  - \left(
          \frac{\partial }{\partial t} +\pounds_\mathbf{v}
  \right)  \frac{\delta l}{\delta \mathbf{v}}
            + \frac{\delta l}{\delta a}\diamond a
  \right]  \,,
\nonumber
\end{eqnarray}
where we have consistently dropped boundary terms arising from
integrations by parts, by invoking natural boundary conditions.
Thus, for the set of tensor fields $a$ in equation (\ref{Eul-ad-qts}) we
have the following Euclidean components of
$\frac{\delta l}{\delta a}\diamond a$,
\begin{eqnarray} \label{diamond-a-eq}
\left.
\left(    \frac{\delta l}{\delta a}\diamond a
\right)_k
\right.
&=&
-\ \left.
\frac{ \delta l}{ \delta b}\ b_{,k}
\ +\ D\left(\frac{ \delta l}{ \delta D}\right)_{,k}
\right.
\nonumber \\&&
+\ \left.\left(
   -\ \frac{ \delta l}{ \delta \mathbf{A}}\times{\rm curl}\, \mathbf{A}
   +\mathbf{A}\ {\rm div}\,\frac{ \delta l}{ \delta \mathbf{A}}\
   +\mathbf{B}\times{\rm curl}\, \frac{ \delta l}{ \delta \mathbf{B}}
\right)_k
\right.
\nonumber \\ &&
-\ \left.
       \frac{ \delta l}{\delta S_{ab}} S_{ab,k}
        \ +\ \left(\frac{ \delta l}{\delta S_{ab}}S_{kb}\right)_{,a}
        \ +\ \left(\frac{ \delta l}{\delta S_{ab}}S_{ka}\right)_{,b}
\right.\,.
\end{eqnarray}

\paragraph{Stress tensor formulation.}
For example, if we assume a Lagrangian in the form
\begin{equation} \label{act-3d}
l(\mathbf{v}; b,\mathbf{A},\mathbf{B},D, S_{ab})
=\int d^{\,3}x\ {\cal L}(\mathbf{v},\nabla\mathbf{v},
b,\mathbf{A},\mathbf{B},D, S_{ab}),
\end{equation}
where ${\cal L}$ is a {\em given} function, then we may use equation
(\ref{diamond-a-eq}) to express the Euler--Poincar\'e equations for
continua (\ref{continuumEP}) in this case in the {\bfi momentum
conservation form},
\begin{equation} \label{mom-cons}
   \frac{\partial }{\partial t}\frac{\delta l}{\delta \mathbf{v}}
   = -\, \operatorname{ad}^{\ast}_\mathbf{v}\frac{\delta l}{\delta \mathbf{v}}
   +\frac{\delta l}{\delta a}\diamond a
\quad \Rightarrow \quad
\frac{\partial m_i}{\partial t} = -\ \frac{\partial }{\partial
x^j}T^j_i\,,
\end{equation}
with {\bfi momentum density components} $m_i$, $i=1,2,3$ defined by
\begin{equation} \label{mom-comp}
m_i \equiv \frac{\delta l}{\delta v^i}
= \frac{\partial{\cal L}}{\partial v^i}
- \frac{\partial}{\partial x^k}
\left(\frac{\partial{\cal L}}{\partial
v^i_{,k}}\right),
\end{equation}
and {\bfi stress tensor} $T^j_i$ given by
\begin{eqnarray} \label{stress-tens}
T^j_i &=&
\left.
m_i v^j - \frac{\partial{\cal L}}{\partial v^k_{,j}}v^k_{,i}
- \frac{\partial{\cal L}}{\partial A_j} A_i
+ \frac{\partial{\cal L}}{\partial B^i}B^j
- \frac{\partial{\cal L}}{\partial S_{jb}}S_{ib}
- \frac{\partial{\cal L}}{\partial S_{aj}}S_{ia}
\right.
\nonumber \\ &&
\left.
+\ \delta^j_i\left({\cal L}-D\frac{\partial{\cal L}}{\partial D}
-  B^k\frac{\partial{\cal L}}{\partial B^k}
\right)
\right.\,.
\end{eqnarray}
Here, in the calculation of $T^j_i$, we have used the coordinate
expression (\ref{continuumcoadjoint-coord})
for $\operatorname{ad}^\ast_\mathbf{u}(\alpha\otimes m)$.

\paragraph{Kelvin-Noether form.}
The Euclidean components of the Euler--Poincar\'e equations for ideal
continua may also be summarized in Kelvin-Noether form (\ref{KThfm}) for
advected tensor fields $a$ in the set (\ref{Eul-ad-qts}). We adopt
the notational convention of the circulation map $\mathcal K$ in equation
(\ref{circ.map}) that a one form density can be made into a one form (no
longer a density) by dividing it by the mass density $D$ to produce, e.g.,
the one form in Euclidean components $\frac{1}{D}\frac{\delta l}{\delta
v^i}dx^i$ from the one form density $\frac{\delta l}{\delta \mathbf{v}}$.
With a slight abuse of notation (but in accord with the usual physics
conventions) we write the former coordinate expression as
$\frac{1}{D}\frac{\delta l}{\delta \mathbf{v}}\cdot d\mathbf{x}$.
We also denote the
Lie-derivative relation for the continuity equation as
$({\partial}/{\partial t}+\pounds_\mathbf{v})Dd^3x = 0$. Then, the
Euclidean components of the Euler--Poincar\'e equations for continua in
(\ref{EP:eqn-u2}) are expressed in Kelvin-Noether form (\ref{KThfm}) as
\begin{multline}
\left(
      \frac{\partial}{\partial t}
      +  \pounds_\mathbf{v}
\right)
\left(
      \frac{1}{D}\frac{\delta l}{\delta
      \mathbf{v}}\cdot d\mathbf{x}
\right)
    \,+\,\frac{1}{D}\frac{ \delta l}
    {\delta b}\nabla b \cdot d\mathbf{x}
    \,-\,\nabla
\left(\frac{\delta l}{\delta D}
\right) \cdot d\mathbf{x}
 \\
     + \frac{ 1}{ D}
\left(
       \ \frac{ \delta l}{ \delta \mathbf{A}}\times{\rm curl}\ \mathbf{A}
-\mathbf{A}\
       {\rm div}\frac{ \delta l}{ \delta \mathbf{A}}\
\right)     \cdot d\mathbf{x}\, -\frac{ 1}{ D}
\left(    \mathbf{B}\times{\rm curl} \ \frac{ \delta l}{ \delta \mathbf{B}}
\right)   \cdot d\mathbf{x}
                   \\
     + \frac{ 1}{ D}
\left(      \frac{ \delta l}{\delta S_{ab}} S_{ab,k}
          -(\frac{ \delta l}{  \delta S_{ab}}S_{kb})_{,a}
          -(\frac{ \delta l}{  \delta S_{ab}}S_{ka})_{,b}
\right) dx^k  = 0\,,
\label{EP-Kthm}
\end{multline}
where the components of the variational derivatives of the Lagrangian $l$
are to be computed according to the usual physics conventions, i.e., as
components of Fr\'{e}chet derivatives as in equation (\ref{fdf}). In
physical applications, the advected Eulerian tensor fields $a$ in
(\ref{Eul-ad-qts}) represent the buoyancy $b$ (or specific entropy, for
the compressible case), magnetic vector potential $\mathbf{A}$, magnetic
field intensity $\mathbf{B}$, mass density $D$, and Cauchy-Green strain
tensor $S_{ab}$, respectively. Formula (\ref{EP-Kthm}) is the Kelvin-Noether
form of the equation of motion for ideal continua in Euclidean coordinates.
This Euclidean component formula is especially convenient for direct
calculations in fluid dynamics, to which we turn our attention next.

\paragraph{Eulerian motion equation for an ideal incompressible
fluid.}
In the Eulerian velocity representation we consider
fluid motion in an $n$-dimensional domain and define the reduced
action
$\mathfrak{S}_{\rm red}$ and reduced Lagrangian $l(\mathbf{v},D)$ by
\begin{equation}
\mathfrak{S}_{\rm red}=\int dt \, l
    = \int dt\int d^{\,n}x\ \Big[{\frac{1}{2}}D|\mathbf{v}|^2 -
p(D-1)\Big]\,.
\label{lag-v}
\end{equation}
This action produces the following variations at fixed $\mathbf{x}$ and $t$
\begin{equation}
{\frac{1}{D}}{\frac{{\delta} l}{{\delta} \mathbf{v}}}
= \mathbf{v} \,,
\quad
{\frac{{\delta} l}{{\delta} D}} = {\frac{1}{2}}|\mathbf{v}|^2 - p\,,
\quad
{\frac{{\delta} l}{{\delta} p}} = -\,(D-1)\,.
\label{vds1}
\end{equation}
Hence, from equation (\ref{EP-Kthm}) for Hamilton principles of this type
we find the Eulerian motion equation,
\begin{equation}
\left(\frac{ \partial}{\partial t} + \pounds_\mathbf{v}\right)
\left(\frac{ 1}{ D}\frac{ \delta l}{ \delta \mathbf{v}}
\cdot d\mathbf{x}\right)
\,-\,\nabla\left(\frac{ \delta l}{ \delta D}\right)\cdot d\mathbf{x} = 0\,,
\quad{\rm or}\quad
\frac{ \partial\mathbf{v}}{\partial t}+ (\mathbf{v}\cdot\nabla)\mathbf{v}
+ \nabla p = 0\,,
\label{Eul-mot}
\end{equation}
for ``natural" boundary conditions, $\hat{\bf n}\cdot\mathbf{v}=0$ on the
boundary, where $\hat{\bf n}$ is the boundary's outward unit normal
vector. This is the Eulerian motion equation for an incompressible fluid
in $n$ dimensions. The constraint $D=1$ (volume or mass preservation) is
imposed by varying the Lagrange multiplier $p$, the presure.
Incompressibility then follows from substituting
$D=1$ into the Lie-derivative relation for $D$, which closes the ideal
incompressible fluid system,
\begin{equation}
\left( \frac{ \partial}{\partial t}+ \pounds_\mathbf{v}\right)D d^3x=0,
\quad {\rm i.e.,} \quad
\frac{ \partial D}{\partial t}
= -\ \nabla\cdot(D\mathbf{v})\,.
\end{equation}
This relation, together with the constraint $D = 1$ gives incompressibility
of the flow, $\nabla\cdot\mathbf{v}=0$.

\paragraph{Remark on Lagrangian mass coordinates.} An alternative way
to treat Hamilton's principle for the action (\ref{lag-v}) is to
perform variations at fixed $\mathbf{x}$ and $t$ of the {\it inverse} maps
$\mathbf{x} \mapsto \boldsymbol{\ell}$,
described by the Lagrangian mass coordinate functions $\ell^A(\mathbf{x},t)$,
$A=1,2,\dots,n$, which determine $\mathbf{v}$ and $D$ by the formulae
(in which one sums on repeated indices)
\begin{equation}
{\frac{\partial \ell^A}{\partial t}}=-v^iD^A_i\,,
\quad D^A_i={\frac{\partial \ell^A}{\partial x^i}} \,,
\quad D=\mathrm{det}(D^A_i)\,.
\label{lag-def}
\end{equation}
As discussed above, the relation of mass coordinates $\ell$ to the usual
Lagrangian coordinates $X$ is given by a change of variables in the
fluid reference configuration to make $\rho_0(X)d^nX=d^n\ell$. Variation
of an action of the form $\mathfrak{S}_{\rm red}(\mathbf{v},D)$ with respect
to $\ell^A$ with $p$ imposing volume preservation then yields (Holm, Marsden,
and Ratiu [1986], Holm [1996a]),
\begin{eqnarray}
{\delta} \mathfrak{S}_{\rm red} &=& \int dt \int d^nx\ \Big\{ D (D^{-1})^i_A
{\delta} l^A
\Big[{\frac{d}{dt}} {\frac{1}{D}}{\frac{{\delta} l}{{\delta} v^i}}
+ {\frac{1}{D}}{\frac{{\delta} l}{{\delta} v^j}}v^j_{,i}
- \left({\frac{{\delta} l}{{\delta} D}}\right)_{,i}\,\Big]
\nonumber \\
&&\hspace{1in}
-\, {\delta} p(D-1)\Big\}\,,
\label{hpg}
\end{eqnarray}
where
$d/dt={\partial}/{\partial t} + (\mathbf{v}\cdot\nabla)$ is the material
derivative of Eulerian quantities and we again invoke natural boundary
conditions when integrating by parts.

Hence, the vanishing of the coefficient of ${\delta}\ell^A$ in
the variational formula (\ref{hpg}) recovers the Euler--Poincar\'e
equation (\ref{Eul-mot}) for the Eulerian fluid  velocity, $\mathbf{v}$, by
stationarity of the action (\ref{lag-v}) with respect to
variations of the Lagrangian mass coordinates $\ell^A(\mathbf{x},t)$.
Similar arguments based on stationary variations of the action with
respect to the Lagrangian mass coordinates $\ell^A$ at fixed $\mathbf{x},t$
will also recover the more general Euler--Poincar\'e equations
(\ref{EP-Kthm}) from actions which depend on the velocity $\mathbf{v}$ and
the advected quantities in equation (\ref{Eul-ad-qts}) through their
dependence on the $\ell^A(\mathbf{x},t)$.

\paragraph{Adiabatic compressible MHD.} In the case of adiabatic
compressible magnetohydrodynamics (MHD), the action in Hamilton's
principle is given by
\begin{equation}
\mathfrak{S}_{\rm red}=\int dt\, l = \int dt\, d^3x\ \left(\frac{ D}{2}\
|\mathbf{v}|^2 - De(D,b)-\frac{1}{2}|\mathbf{B}|^2\right),
\label{mhdact}
\end{equation}
where $e(D,b)$ is the fluid's specific internal energy, whose dependence
on the density $D$ and specific entropy $b$ is given as the ``equation of
state" and which for an isotropic medium satisfies the thermodynamic
first law in the form $de=-pd(1/D)+Tdb$ with pressure $p(D,b)$ and
temperature $T(D,b)$. The variation of $l$ in (\ref{mhdact}) is
\begin{equation}
\delta \mathfrak{S}_{\rm red} = \int dt\, d^3x\
D\mathbf{v}\cdot\delta\mathbf{v}-DT\delta b +\left(\frac{1}{2}
|\mathbf{v}|^2 - h\right)\delta D - \mathbf{B}\cdot\delta\mathbf{B}.
\end{equation}
The quantity $h=e+p/D$ denotes the specific enthalpy, which thus satisfies
$dh=(1/D)dp+Tdb$. The Euler--Poincar\'e formula in the Kelvin-Noether form
(\ref{EP-Kthm}) yields the MHD motion equation as
\begin{equation}
\left(\frac{ \partial}{\partial t}+ \pounds_\mathbf{v}\right)
\left({\mathbf{v}}\cdot d\mathbf{x}\right)
- Tdb
+ \frac{ 1}{ D}\mathbf{B}\times{\rm curl}\ \mathbf{B}\cdot d\mathbf{x}
- d\left(\frac{1}{2} |\mathbf{v}|^2 - h\right) = 0,
\end{equation}
or, in three dimensional vector form,
\begin{equation}
\frac{ \partial\mathbf{v}}{\partial t} + (\mathbf{v}\cdot\nabla)\mathbf{v}
+\frac{ 1}{ D}\nabla p
+\frac{ 1}{ D}\mathbf{B}\times{\rm curl}\ \mathbf{B} = 0.
\end{equation}
By definition, the advected variables $\{b, \mathbf{B}, D\}$ satisfy the
the following Lie-derivative relations which close the ideal MHD
system,
\begin{eqnarray}
\left(\frac{ \partial}{\partial t}+ \pounds_\mathbf{v}\right) b=0,
&{\rm or}&
\frac{ \partial b}{\partial t}
= -\ \mathbf{v}\cdot\nabla\,b\,,
\nonumber \\
\left(\frac{ \partial}{\partial t}+ \pounds_\mathbf{v}\right)\mathbf{B}\cdot
d\mathbf{S}=0, &{\rm or}&
\frac{ \partial \mathbf{B}}{\partial t}
=  {\rm curl}\,(\mathbf{v}\times\mathbf{B}),
\nonumber \\
\left(\frac{ \partial}{\partial t}+ \pounds_\mathbf{v}\right)D d^3x=0,
&{\rm or}&
\frac{ \partial D}{\partial t}
= -\ \nabla\cdot(D\mathbf{v}),
\end{eqnarray}
and the function $p(D,b)= D^2\partial e/\partial D$ is specified by giving
the equation of state of the fluid, $e=e(D,b)$. If the condition
$\nabla\cdot\mathbf{B}=0$ holds initially, then it holds for all
time; since this constraint is preserved by the dynamical equation
for $\mathbf{B}$.

\paragraph{Adiabatic magneto-elastodynamics.}

When nonlinear elasticity is also a factor in the MHD evolution, there
is an additional Lie-derivative relation,
\begin{equation}
\left( \frac{ \partial}{\partial t}+ \pounds_\mathbf{v}\right)\,
(S_{ab}\,dx^a\otimes dx^b)=0\,,
\end{equation}
leading to the dynamical equation for the advected
Cauchy-Green strain tensor
$S_{ab}$ (which measures nonlinear strain in {\it spatial} coordinates),
\begin{equation}
\frac{ \partial}{\partial t}\,S_{ab}=
-\left(v^k S_{ab,k}+S_{kb}v^k_{,a}+S_{ka}v^k_{,b}\right).
\end{equation}
In this case, additional stress terms appear in the motion
equation for $\mathbf{v}$ that arise from the dependence of the specific
internal energy $e(D,b,S_{ab})$ on the Cauchy-Green strain tensor $S_{ab}$
in the MHD action (\ref{mhdact}) when the elasticity of the medium is
involved. The stress tensor per unit mass $\sigma^{ab}$ is determined from
the equation of state of such an magneto-elastic medium by the
Doyle-Erickson formula
$\sigma^{ab}\equiv \partial e/\partial S_{ab}$. The Euler--Poincar\'e
equation (\ref{EP-Kthm}) for ideal magneto-elasticity is then to be
\begin{equation}
\frac{ \partial{v_i}}{\partial t}+ {v^j}{v_{i,j}}
+\frac{ 1}{ D}p_{,i}
+\frac{ 1}{ D}B^j(B_{j,i}-B_{i,j})
-(\sigma^{ab}S_{ib})_{,a}
-(\sigma^{ab}S_{ia})_{,b} = 0\,,
\end{equation}
where we have used the specific enthalpy relation for an elastic
medium, $dh-Tdb=D^{-1}dp+\sigma^{ab} dS_{ab}$. Thus, adiabatic
magneto-elastodynamics summons all of the advected quantities in equation
(\ref{Eul-ad-qts}) and makes use of the entire Euler--Poincar\'e equation
(\ref{EP-Kthm}).

\paragraph{Adiabatic compressible Maxwell fluid dynamics via the
Kaluza-Klein construction.}

An adiabatic Maxwell fluid (MF) with (nonrelativistic) Eulerian fluid
velocity $\mathbf{v}$, density $D$, specific entropy $b$ and pressure
$p(D,b)$ satisfies the following system of equations,
\begin{eqnarray}
&&
\frac{ \partial\mathbf{v}}{\partial t}+ (\mathbf{v}\cdot\nabla)\mathbf{v}
+\frac{ 1}{ D}\nabla p(D,b) = \frac{ q}{m}\,({\bf
E}+\mathbf{v}\times\mathbf{B}),
\nonumber \\&&
\frac{ \partial D}{\partial t} = -\ \nabla\cdot(D\mathbf{v}),
\qquad
\frac{ \partial b}{\partial t}=-\ (\mathbf{v}\cdot\nabla)b,
\nonumber \\&&
\nabla\cdot{\bf E} = \frac{ q}{m}D,
\qquad
\frac{ \partial {\bf E}}{ \partial t}
=-\ \frac{ q}{m}D\mathbf{v}+\nabla\times\mathbf{B},
\nonumber \\&&
\nabla\cdot\mathbf{B}=0,
\qquad
\frac{ \partial \mathbf{B}}{\partial t}=-\ \nabla\times{\bf E}.
\label{MFeqs}
\end{eqnarray}
This system consists of the motion equation for a charged fluid moving
under the combined effects of pressure gradient and Lorentz forces;
the continuity equation for the mass density $D$; advection of the
specific entropy, $b$; and Maxwell's equations for the
electromagnetic fields ${\bf E}$ and $\mathbf{B}$ in the moving fluid
medium, whose polarizability and magnetization are neglected for
simplicity. (For the physically more realistic treatment of moving
media with electromagnetic induction in a similar framework,
including relativistic effects, see Holm [1987].) The equations for
$D$ and $b$ are the familiar advection laws. The coupling constant
$q/m$ is the charge-to-mass ratio of the fluid particles, and the
electric and magnetic fields ${\bf E}$ and $\mathbf{B}$ are defined
in terms of the scalar and vector potentials $\Phi$ and $\mathbf{A}$
by
\begin{equation}
\mathbf{E}\equiv-\,\frac{ \partial \mathbf{A}}{ \partial t}-\nabla\Phi,
\qquad
\mathbf{B}\equiv\nabla\times\mathbf{A}\,.
\label{EMpot}
\end{equation}
In the MF equations (\ref{MFeqs}), charged fluid motion is the source
for the electromagnetic fields which act self-consistently upon the
fluid through the Lorentz force. We shall show that equations
(\ref{MFeqs}) are Euler--Poincar\'e equations for the gauge invariant
action of ``Kaluza-Klein" form given by
\begin{eqnarray}
\mathfrak{S}_{\rm red}=\int dt\, l &=& \int dt\, d^3x\
\left(
      \frac{1}{2} D |\mathbf{v}|^2 + \frac{1}{2} D (\mathbf{A}\cdot\mathbf{v} -
          \Phi+{\tilde\omega})^2 - De(D,b)
\right.
\nonumber \\&&\qquad\qquad
+\  \left. \frac{1}{2} \Big|\frac{ \partial \mathbf{A}}{ \partial
t}+\nabla\Phi\Big|^2 -\frac{1}{2} |\nabla\times\mathbf{A}|^2\right),
\label{KKact}
\end{eqnarray}
where ${\tilde\omega}=\partial\theta/\partial t+\mathbf{v}\cdot\nabla\theta$
for a gauge field $\theta$ and $e(D,b)$ is the fluid's specific internal
energy, which satisfies the first law of thermodynamics in the form
$de=-pd(1/D)+Tdb$ with pressure $p(D,b)$ and temperature $T(D,b)$.

This action principle fits into the general theory with the
electromagnetic field variables playing the role of additional
configuration variables which are not acted on by the particle
relabelling group. They obey the usual Euler-Lagrange equations,
coupled to the Euler--Poincar\'e variables through the Lagrangian. In
other words, the primitive unreduced Lagrangian in this case is of
the abstract form $ L : TG \times V ^\ast \times TQ \times TC \rightarrow
\mathbb{R}$ in which $G$, the fluid particle relabelling group, acts
trivially on the Maxwell field variables $Q$ and the gauge field
$\theta\in C$. Note that the Lagrangian in equation (\ref{KKact}) is
invariant under translations of $\theta$, as well as under the
electromagnetic gauge transformations,
\begin{equation} \label{em-gauge}
\mathbf{A}\to\mathbf{A}+\nabla\zeta, \quad
\Phi\to\Phi-\partial\zeta/\partial t, \quad
\theta\to\theta-\zeta,
\end{equation}
for an arbitary function $\zeta$ of $\mathbf{x}$ and $t$.

We now take variations of the action. The variation of
$\mathfrak{S}_{\rm red}$ in equation (\ref{KKact}) may be written using
the definitions of $\mathbf{E}$ and $\mathbf{B}$, and the abbreviated
notation $c\equiv\mathbf{A}\cdot\mathbf{v}-\Phi+{\tilde\omega}$, as
\begin{eqnarray}
\delta \mathfrak{S}_{\rm red} &=& \int dt\, d^3x\
\Big[ D\left(\mathbf{v}+c\mathbf{A}
+c\nabla\theta\right)\cdot\delta\mathbf{v}
+\left(\frac{1}{2}|\mathbf{v}|^2
+\frac{1}{2}c^2-e-\frac{p}{D}\right)\delta D
\nonumber \\&&\qquad
-\ DT\ \delta b
+\left(cD\mathbf{v}+\frac{ \partial {\bf E}}{ \partial t}
-\nabla\times\mathbf{B}\right)\cdot\delta\mathbf{A}
+(-cD+\nabla\cdot{\bf E})\ \delta\Phi
\nonumber \\&&\qquad
-\ \left(\frac{ \partial cD}{\partial t}
+\nabla\cdot cD\mathbf{v}\right)\ \delta\theta\Big]\,,
\label{MFvar}
\end{eqnarray}
where terms arising from integration by parts vanish for the natural
boundary conditions given by
\begin{equation}
\mathbf{v}\cdot\hat{\bf n}=0, \quad
{\bf E}\cdot\hat{\bf n}=0,
{\rm\ and\ }
\hat{\bf n}\times \mathbf{B}=0
{\rm\ on\ the\ boundary,}
\end{equation}
and for variations $\delta g(t)$
of $ g (t) $ vanishing at the endpoints.
Stationarity of the action $\mathfrak{S}_{\rm red}$ in (\ref{KKact})
under variation of the gauge field $\theta$ gives the conservation law,
\begin{equation}
\left(\frac{ \partial}{\partial t}+ \mathbf{v}\cdot\nabla\right)c=0
\quad\hbox{via the continuity equation,}\quad
\frac{ \partial D}{\partial t}+ \nabla\cdot (D\mathbf{v})=0.
\label{c=q/m}
\end{equation}
Hence, we may set $c=q/m$ in equation (\ref{MFvar}) and then acquire the two
Maxwell equations with sources from stationarity of the action
$\mathfrak{S}_{\rm red}$ under variations of $\mathbf{A}$ and $\Phi$.
Once the flow velocity $\mathbf{v}$ is known, the relation $c=q/m$
determines the gauge function $\theta$ by ``quadrature", from the
definitions of $c$ and ${\tilde\omega}$ as
\begin{equation} \label{theta-eqn}
{\tilde\omega}\equiv
\frac{\partial\theta}{\partial t}+\mathbf{v}\cdot\nabla\theta
= \frac{q}{m} +\Phi -\mathbf{A}\cdot\mathbf{v} \,.
\end{equation}

The remaining variations of $\mathfrak{S}_{\rm red}$ in
$\{\mathbf{v},D,b\}$ for the Euler--Poincar\'e dynamics collect into the
Kelvin-Noether form of equation (\ref{EP-Kthm}) as
\begin{equation}\label{KKthm}
\left(\frac{ \partial}{\partial t}+ \mathbf{v}\cdot\nabla\right)
\left(\frac{ 1}{ D}\frac{ \delta l}{ \delta \mathbf{v}}\right)
+\frac{ 1}{ D}\frac{ \delta l}{  \delta v^j}\nabla v^j
\,+\,\frac{1}{D}\frac{ \delta l}{\delta b}   \nabla b
-\nabla \frac{ \delta l}{ \delta D}=0\,.
\end{equation}
Specifically, we have
\begin{eqnarray} \label{KKthm-v}
\left(\frac{ \partial}{\partial t}
+ \mathbf{v}\cdot\nabla\right)
\left(\mathbf{v}+c\mathbf{A}+c\nabla\theta\right)
\!\!\!&+&\!\!\!
(v_j+cA_j+c\theta_{,j})\nabla v^j
\nonumber \\
\,-\ T\nabla b
\!\!\!&-&\!\!\!\nabla \left(\frac{1}{2}|\mathbf{v}|^2
+\frac{1}{2}c^2-e-\frac{p}{D}\right)
=0\,.
\end{eqnarray}
Using the fundamental vector identity of fluid dynamics
in three dimensions,
\begin{equation}
({\bf b}\cdot\nabla){\bf a} + a_j\nabla b^j
=-{\bf b}\times(\nabla\times {\bf a}) + \nabla({\bf a}\cdot{\bf b})\,,
\label{fvid}
\end{equation}
with, in this case, ${\bf b}=\mathbf{v}$ and
${\bf a}=D^{-1}{\delta l}/{\delta \mathbf{v}}$, casts the Euler--Poincar\'e
equation (\ref{KKthm}) into its equivalent ``curl" form,
\begin{equation}
\frac{\partial}{\partial t}
\left(\frac{ 1}{ D}\frac{ \delta l}{ \delta \mathbf{v}}\right)
- \mathbf{v}\times\nabla\times
\left(\frac{ 1}{ D}\frac{ \delta l}{ \delta \mathbf{v}}\right)
\,+\,\frac{1}{D}\frac{ \delta l}{\delta b}   \nabla b
+\nabla\Big(\mathbf{v}\cdot\frac{ 1}{ D}\frac{\delta l}{\delta
\mathbf{v}}-\frac{\delta l}{\delta D}\Big) = 0\,.
\label{curl-mot}
\end{equation}
Similarly, applying the same vector identity with ${\bf
b}=\mathbf{v}$ and ${\bf a}=c(\mathbf{A}+\nabla\theta)$ in the
Maxwell fluid motion equation (\ref{KKthm-v}) yields,
\begin{eqnarray}
\frac{ \partial\mathbf{v}}{\partial t}+ (\mathbf{v}\cdot\nabla)\mathbf{v}
+\frac{ 1}{ D}\nabla p
\!\!&=&\!\!
c\ \left(-\frac{ \partial \mathbf{A}}{ \partial t}-\nabla\Phi
+\mathbf{v}\times(\nabla\times\mathbf{A})\right)
\nonumber \\
\!\!&=&\!\!
\frac{ q}{m}\ \left({\bf E} + \mathbf{v}\times \mathbf{B} \right)\,,
\label{KKmot}
\end{eqnarray}
where we have used the thermodynamic first law. Thus we find the Maxwell
fluid motion law -- the first among the equations in (\ref{MFeqs}) -- after
setting $c=q/m$ and using the definitions of the electromagnetic fields
${\bf E}$ and $\mathbf{B}$ in terms of the potentials $\mathbf{A}$ and
$\Phi$.

\bigskip
\begin{thm}[Kelvin circulation theorem for the Maxwell fluid.]
\label{MFKelvinthm} By the MF motion equation (\ref{KKmot}) and the
thermodynamic first law, we have
\begin{eqnarray}
\frac{ dI}{  dt}&=&
\oint_{\gamma_t}Tdb\,,
\label{KelThmMF}
\end{eqnarray}
where the circulation integral is given by
\[
I\equiv\oint_{\gamma_t} ( \mathbf{v}+ \frac{q}{m} \mathbf{A})\cdot
d\mathbf{x},
\]
for a curve ${\gamma_t}$ which moves with the
fluid velocity $\mathbf{v}$.
\end{thm}

\noindent{\bf Proof.\,} The proof is the same as for Theorem
\ref{KelThmforcontinua}; although it is also immediately seen from
the motion equation (\ref{KKthm-v}) after substituting $D^{-1}\delta l/
\delta \mathbf{v}=\mathbf{v}+c(\mathbf{A}
+ \nabla\theta)$ and $D^{-1}\delta l/
\delta b=-T$. \quad $\blacksquare$

\begin{cor}[Potential vorticity convection for the Maxwell fluid.]
\label{PV-cor} Stokes' theorem, advection of specific entropy b
and the continuity equation together imply convection of potential
vorticity for the adiabatic Maxwell fluid,
\begin{equation}
{\frac{\partial q}{\partial t}}+\mathbf{v}\cdot \nabla q
\equiv\frac{ dq}{  dt}=0
\quad {\rm with} \quad
q \equiv \frac{ 1}{ D}\nabla b\cdot{\rm curl}( \mathbf{v} + \frac{q}{m}
\mathbf{A}).
\label{pv-cor}
\end{equation}
\end{cor}
\paragraph{Remark.} The equation $dq/dt=0$ for convection of
potential vorticity for a general Lagrangian with dependence
$l(\mathbf{v},b,D)$, with
\begin{equation} \label{q-eqn}
q=\frac{1}{D}\nabla{b}\cdot {\rm curl}
\Big(\frac{ 1}{ D} \frac{\delta l}{\delta\mathbf{v}}\Big)\,,
\end{equation}
may also be proven directly from the ``curl" form of the Kelvin-Noether
equation (\ref{curl-mot}) in three dimensions, by taking the scalar product
of its curl with $\nabla{b}$ and applying the continuity equation for $D$.

\paragraph{Alternative interpretations of the Maxwell fluid formulation.}
Note that the first line of the Euler--Poincar\'e motion equation (\ref{KKmot})
for Maxwell fluids {\it persists}, when the electromagnetic energy terms are
{\it dropped} from the MF Lagrangian in equation (\ref{KKact}), to give the
action
\begin{equation}
\mathfrak{S}_{\rm red} = \int dt\ l = \int dt\, d^3x\ \left(\frac{1}{2} D
|\mathbf{v}|^2 + \frac{1}{2} D (\mathbf{A}\cdot\mathbf{v} - \Phi +
{\tilde\omega})^2 -
De(D,b)\right)\,.
\label{KK-Faraday-act}
\end{equation}
The Euler--Poincar\'e equation which results from this action is
\begin{equation}
\frac{ \partial\mathbf{v}}{\partial t}+ (\mathbf{v}\cdot\nabla)\mathbf{v}
+\frac{ 1}{ D}\nabla p
= c\ \left(-\frac{ \partial \mathbf{A}}{ \partial t}-\nabla\Phi
+\mathbf{v}\times(\nabla\times\mathbf{A})\right)\,.
\label{Faraday-mot}
\end{equation}
The Kelvin Theorem \ref{MFKelvinthm} and its potential vorticity
Corollary \ref{PV-cor} also persist for the dynamics derived from this
truncated action. The ``Lorentz force" terms in equation
(\ref{Faraday-mot}) in terms of $\mathbf{A}$ and $\Phi$ arise {\it
purely} from the Kaluza-Klein coupling term --- the second term in
the integrand of the action  $\mathfrak{S}_{\rm red}$ in equation
(\ref{KK-Faraday-act}). These ``Lorentz forces" may be interpreted
physically as noninertial forces resulting from having moved into a frame
of reference with a prescribed velocity given by
$\mathbf{A}(\mathbf{x},t)$. The velocity $\mathbf{v}$ then represents
fluid flow relative to this noninertial frame. This situation reduces
to {\em Faraday driving} of the fluid (Faraday [1831]), when
$\mathbf{A}(\mathbf{x},t)$ corresponds to a rigid motion of the
fluid container. For a simple example, set $c=f_0$; $\nabla\Phi=0$; and
$\mathbf{A}=\frac{1}{2}\hat{\bf z} \times \mathbf{x}$.
Then $c\mathbf{v}\times(\nabla\times\mathbf{A}) = \mathbf{v}\times f_0
\hat{\bf z}$ gives the Coriolis force and $\mathbf{v}$ corresponds to
fluid velocity in a uniformly rotating reference frame with constant
angular velocity $f_0$. This is the typical situation in {\em geophysical
fluid dynamics}.

Alternatively, the right hand side of equation (\ref{Faraday-mot})
may be interpreted as a {\it vortex force} arising from a given wave
field at the surface of an incompressible fluid (for $D=1$), as in
Craik and Leibovich [1976] (see also Holm [1996b]). The
Craik-Leibovich equations are formally identical to equation
(\ref{Faraday-mot}) when $\mathbf{A}$ is identified as the prescribed
mean Stokes drift velocity due to the presence of the wave field.

Fluid motion equations of the same form as
(\ref{Faraday-mot}) also appear in the generalized Lagrangian-mean
(GLM) formulation of wave, mean-flow interaction theories (see Andrews
and McIntyre [1978a,b]), in which ${\tilde\omega}$ is the
Doppler-shifted frequency of a wave packet interacting with a
Lagrangian-mean flow of velocity $\mathbf{v}$, and $\mathbf{A}$ is
the prescribed {\it pseudomomentum per unit mass} of the wave. For a
discussion of {\it self-consistent} Lie-Poisson Hamiltonian theories
of wave, mean-flow interaction in a similar form, see Gjaja and Holm
[1996].

\paragraph{Geodesic motion and the Kaluza-Klein construction for
incompressible fluids.} Hamilton's principle for the action in the ``minimal
coupling" form
\begin{equation}
\mathfrak{S}' =\int dt\, l = \int dt\, d^3x\ \left( \frac{1}{2} D
|\mathbf{v}|^2
+ \frac{ q}{m}D\mathbf{A}\cdot\mathbf{v} - \frac{ q}{m}D \Phi - De(D,b)
\right)\,,
\label{KK-Faraday-act1}
\end{equation}
yields the {\it same} Euler--Poincar\'e equation (\ref{Faraday-mot}) as
results from Hamilton's principle for the Kaluza-Klein action in
equation (\ref{KK-Faraday-act}). Thus, we see that by introducing the
auxiliary gauge field, $\theta$, the Kaluza-Klein construction
renders the minimal coupling form of the action for fluid dynamics
quadratic in the velocity, while preserving its corresponding
Euler--Poincar\'e equation. The Kaluza-Klein construction for charged
particle mechanics is discussed in Marsden and Ratiu [1994]. The
historical references are Kaluza [1921], Klein [1926] and Thirry
[1948].

In the {\it incompressible} case, when the Kaluza-Klein action is taken to
be
\begin{equation}\label{incomp-KK-act}
\mathfrak{S}_{\rm red} = \int dt\ l = \int dt\, d^3x\ \left(\frac{1}{2} D
|\mathbf{v}|^2 + \frac{1}{2} D (\mathbf{A}\cdot\mathbf{v} - \Phi +
{\tilde\omega})^2 - p(D-1)\right)\,,
\end{equation}
for arbitrary prescribed functions $\mathbf{A}$ and $\Phi$, the resulting
Euler--Poincar\'e equation (i.e., equation (\ref{Faraday-mot}) with $D=1$)
represents geodesic motion on the group of volume-preserving
diffeomorphisms with respect to the conserved kinetic-energy metric given
by
\begin{equation} \label{incomp-geod-met}
\|\mathbf{v}\|^2 = \int d^3x\ \left(\frac{1}{2}
|\mathbf{v}|^2 + \frac{1}{2} (\mathbf{A}\cdot\mathbf{v} - \Phi +
{\tilde\omega})^2\right)\,,
\end{equation}
where $c\equiv\mathbf{A}\cdot\mathbf{v}-\Phi+{\tilde\omega}$ is an advected
quantity, $dc/dt=0$. This observation extends the geodesic property of
incompressible ideal fluid flows established in Arnold [1966a] to the case of
incompressible Maxwell fluid flows, as well as to the case of incompressible
ideal fluid flows in an arbitrarily moving reference frame. From
the Euler--Poincar\'e point of view, this extension enlarges the
particle relabelling group $G$ from the group of diffeomorphisms
to the group of automorphisms of the single particle
Kaluza-Klein bundle. The total system for the incompressible
Maxwell fluid flows is then geodesic motion on the product of
this automorphism group with the Maxwell fields themselves. We
believe that a similar extension may be involved in the results
of Ono [1995a, 1995b].

We should also remark that when one has equations in geodesic form,
one can make use of all the attendant geometry to obtain
additional interesting results. Examples of this applied to
questions of stability and conjugate points are given in the works of
Misiolek listed in the references.

\section{Approximate Model Fluid Equations which Preserve the
Euler--Poincar\'e Structure}\label{sec-mod-eqns}

The preceding section demonstrates the applicability of the
Euler--Poincar\'e theorem for ideal continua when the
equations of motion are given. Here we discuss approximate
fluid models which preserve the Euler--Poincar\'e structure,
and are obtained by making asymptotic expansions and other
approximations in Hamilton's principle for a given set of
model equations. As examples, in this section we first
discuss the derivation of the quasigeostrophic approximation
in geophysical fluid dynamics from an approximation of
Hamilton's principle for the rotating shallow water
equations. Next, we discuss the Boussinesq approximation for
dispersive water waves in one dimension. As an example of the
type of ``bonus" which may appear in making simplifying
approximations while preserving mathematical structure, we
derive the integrable Camassa-Holm equation (Camassa and Holm
[1993], Camassa, Holm and Hyman [1994], Alber et al. [1994, 1995, 1997]),
by making asymptotic approximations in the Hamilton's principle for the
Boussinesq equations. The Camassa-Holm equation in one dimension is a
completely integrable partial differential equation for dispersive water
waves that was actually discovered by making structure
preserving approximations of this type. This equation turns out to
describe geodesic motion on the group of diffeomorphisms of either the
real line or the periodic interval, with metric given by the $H^1$ norm
of the velocity. We also derive a multidimensional analogue of the
one-dimensional Camassa-Holm equation by invoking the
$n$-dimensional version of this geodesic property. There are
also other potential advantages of making structure
preserving approximations, e.g., for numerical integrations. However,
discussion of these other advantages is deferred to another place. (See
Marsden and Wendlandt [1997], Wendlandt and Marsden [1997] and Marsden,
Patrick and Shkoller [1997] for recent advances in this direction.)

\paragraph{Rotating shallow water dynamics as Euler--Poincar\'e
equations.} We first consider dynamics of rotating shallow
water (RSW) in a two dimensional domain with horizontal
coordinates $\mathbf{x}=(x_1,x_2)$. RSW motion is governed by
the following nondimensional equations for horizontal fluid
velocity $\mathbf{v}=(v_1,v_2)$ and depth $D$,
\begin{equation}
\epsilon\frac{ d}{ dt}\mathbf{v}+ f\hat{\mathbf{z}}\times\mathbf{v}
+ \nabla \psi = 0\, ,
\qquad
\frac{ \partial D}{ \partial t} + \nabla\cdot D\mathbf{v} = 0\, ,
\label{rsw}
\end{equation}
with notation
\begin{equation}
\frac{ d}{ dt}\equiv\left(\frac{ \partial}{\partial t}
+ \mathbf{v}\cdot\nabla\right)
\quad\hbox{and}\quad
\psi\equiv\bigg(\frac{ D-B}{\epsilon\mathcal{F} }\bigg)\, .
\label{Notation}
\end{equation}
These equations include variable Coriolis parameter $f=f(\mathbf{x})$ and
bottom topography $B=B(\mathbf{x})$.

The dimensionless scale factors appearing in the RSW equations
(\ref{rsw}) and (\ref{Notation}) are the Rossby number $\epsilon$
and the rotational Froude number ${\mathcal F} $, given
in terms of typical dimensional scales by
\begin{equation}
\epsilon = \frac{ \mathcal{V}_0}{  f_0 L} \ll1
\quad\hbox{and}\quad
{\mathcal F} =\frac{ f_0^2 L^2}{ g B_0} = O(1) \, .
\label{rn}
\end{equation}
The dimensional scales $(B_0,L,{\cal V}_0,f_0,g)$ denote equilibrium fluid
depth, horizontal length scale, horizontal fluid velocity, reference
Coriolis parameter, and gravitational acceleration, respectively.
Dimensionless quantities in equations (\ref{rsw}) are unadorned and are
related to their dimensional counterparts (primed), according to
\begin{eqnarray}
&&\mathbf{v}'={\cal V}_0\mathbf{v},\quad \mathbf{x}' = L\mathbf{x},
\quad t' = \left(\frac{ L}{  \mathcal{V}_0}\right)t,
\quad f' = f_0f ,
\nonumber \\
&&B' = B_0B, \quad D' = B_0 D,
\quad\hbox{and}\quad
D'-B' = B_0(D - B).
\label{scl}
\end{eqnarray}
Here, dimensional quantities are:
$\mathbf{v}'$, the horizontal fluid velocity; $D'$, the fluid depth;
$B'$, the equilibrium depth; and $D'-B'$, the free surface elevation.

For barotropic horizontal motions at length scales
$L$ in the ocean, say, for which ${\mathcal F} $ is order $O(1)$
-- as we shall assume -- the Rossby number $\epsilon$
is typically quite small ($\epsilon\ll1$) as indicated in equation
(\ref{rn}). Thus, $\epsilon\ll1$ is a natural parameter for making
asymptotic expansions. For example, we shall assume
$|\nabla f|=O(\epsilon)$ and
$|\nabla B|=O(\epsilon)$, so we may write $f=1+\epsilon f_1(\mathbf{x})$ and
$B=1+\epsilon B_1(\mathbf{x})$. In this scaling, the leading order
implications of equation (\ref{rsw}) are
$\mathbf{v}=\hat{\mathbf{z}}\times\nabla\psi$
and $\nabla \cdot\mathbf{v}=0$. This is geostrophic balance.

A simple calculation using equation (\ref{Eul-mot}) shows that the RSW
equations (\ref{rsw}) arise as Euler--Poincar\'e equations from Hamilton's
principle with action
$\mathfrak{S}_{\rm RSW}$,
\begin{equation}
\mathfrak{S}_{\rm RSW} = \int dt \, l_{\rm RSW}
= \int dt\int dx_1dx_2\,\left[ D\mathbf{v}\cdot\mathbf{R}(\mathbf{x})
- \frac{ (D-B)^2}{2\epsilon\mathcal{F}}
+ \frac{\epsilon}{2}D|\mathbf{v}|^2\right]\,,
\label{swlag}
\end{equation}
where
${\rm curl}\,\mathbf{R}(\mathbf{x})\equiv f(\mathbf{x})
\hat{\mathbf{z}}$ yields the prescribed Coriolis parameter.
The RSW equations (\ref{rsw}) themselves can be regarded as being derived
from asymptotics in Hamilton's principle for three dimensional
incompressible fluid motion, see Holm [1996a]. However, this viewpoint is
not pursued further here, as we proceed to describe the relation of RSW
to the quasigeostrophic approximation of geophysical fluid dynamics.

\paragraph{Quasigeostrophy.}

The quasigeostrophic (QG) approximation is a useful model in the
analysis of geophysical and astrophysical fluid dynamics, see, e.g.,
Pedlosky [1987]. Physically, QG theory applies when the motion is nearly
in geostrophic balance, i.e., when pressure gradients nearly balance the
Coriolis force in a rotating frame of reference, as occurs in meso- and
large-scale oceanic and atmospheric flows on Earth. Mathematically, the
simplest case is for a constant density fluid in a planar domain with
Euclidean coordinates $\mathbf{x}=(x_1,x_2)$. QG dynamics for this case is
expressed by the following nondimensional evolution equation for the
stream-function $\psi$ of the incompressible geostrophic fluid velocity
$\mathbf{v}=\hat{\mathbf{z}}\times\nabla\psi$,
\begin{equation}
{\frac{\partial(\Delta\psi - {\mathcal F} \psi)}{\partial t}}
+[\psi,\Delta\psi]
+\beta {\frac{\partial\psi}{\partial x_1}} = 0\,.
\label{qgb}
\end{equation}
Here $\Delta$ is the Laplacian operator in the plane, ${\mathcal F}$ denotes
rotational Froude number, $[a,b]\equiv{\partial (a,b)}/{\partial(x_1,x_2)}$ is
the Jacobi bracket (Jacobian) for functions $a$ and $b$ defined on
$\mathbb{R}^2$ and $\beta$ is the gradient of the Coriolis parameter, $f$,
taken as $f=1+\beta x_2$ in the $\beta$-plane approximation, with constant
$\beta$. (Neglecting $\beta $ gives the $f$-plane approximation of QG
dynamics.)
The QG equation (\ref{qgb}) may be derived from an asymptotic expansion of
the RSW equations (\ref{rsw}) by truncating at first order in the Rossby
number, cf. Pedlosky [1987]. Equation (\ref{qgb}) may be written
equivalently in terms of the potential vorticity,
$q$, as in equation (\ref{pv-cor}),
\begin{equation}
{\frac{\partial q}{\partial t}}+\mathbf{v}\cdot \nabla q=0,
\quad\hbox{where}\quad
q\equiv {\Delta }\psi -{\mathcal F} \psi +f
\quad\hbox{for QG.}
\label{qgb'}
\end{equation}
This form of QG dynamics expresses its basic property of potential
vorticity conservation on geostrophic fluid parcels.

The QG approximation to the RSW equations introduces
``quasigeostrophic particles" which move with geostrophic velocity
$\mathbf{v} =\hat{\mathbf{z}}\times\nabla\psi$ and,
thus, trace the geostrophic component of the RSW fluid flow. These QG
fluid trajectories are described as functions of Lagrangian mass
coordinates
$\boldsymbol{\ell}=(\ell_1,\ell_2)$ given by
$\mathbf{x}(\boldsymbol{\ell},t)$ in the domain of flow.

\paragraph{Hamilton's principle derivation of QG as Euler--Poincar\'e
equations.} As in Holm and Zeitlin [1997], we consider the following
action for QG written in the Eulerian velocity representation with the
integral operator $(1-{\mathcal F}\Delta^{-1})$,
\begin{equation}
 \mathfrak{S}_{\rm red}=\int dt\, l = \int dt\int dx_1dx_2\
\Big[\frac{\epsilon}{2} D\mathbf{v}\cdot (1-{\mathcal
F}\Delta^{-1})\mathbf{v} + D\mathbf{v}\cdot\mathbf{R}(\mathbf{x}) -
\psi(D-1)\Big]
\,.  \label{QG-lag-v}
\end{equation}
This choice can be found as an asymptotic approximation of the RSW action
$\mathfrak{S}_{\rm RSW}$ in equation (\ref{swlag}), in the limit of small wave
amplitudes of order $O(\epsilon^2)$ and constant mean depth to the same
order, when the wave elevation is determined from the fluid velocity by
inverting the geostrophic relation,
$\mathbf{v}=\hat{\mathbf{z}}\times\nabla\psi$.
The variational derivatives of the reduced Lagrangian $\mathfrak{S}_{\rm
red}$ at fixed $\mathbf{x}$ and $t$ are
\begin{eqnarray}
\frac{1}{D}\frac{\delta l}{\delta \mathbf{v}}
&=& \mathbf{R} +
\epsilon\Big[\mathbf{v} - \frac{\mathcal F}{2}\Delta^{-1}\mathbf{v}
- \frac{{\mathcal F} }{2D}\Delta^{-1}(D\mathbf{v})\Big]\,,
\nonumber \\
\frac{\delta l}{\delta D}
&=& \frac{\epsilon}{2}\mathbf{v}\cdot
(1-{\mathcal F} \Delta^{-1})\mathbf{v} +
\mathbf{v}\cdot\mathbf{R} - \psi\,,
\nonumber \\
\frac{\delta l}{\delta \psi} &=& -\,(D-1)\,,
\label{vds2}
\end{eqnarray}
where we have used the symmetry of the Laplacian operator and assumed no
contribution arises from the boundary when integrating by parts. For example,
we may take the domain to be periodic. Hence, the Euler--Poincar\'e
equation (\ref{Eul-mot}) for action principles of this type and the
fundamental vector identity (\ref{fvid}) combine to give the Eulerian QG
``motion equation",
\begin{eqnarray}
\epsilon\frac{\partial}{\partial t}(1-{\mathcal F} \Delta^{-1})\mathbf{v}
\!\!\!&-&\!\!\!
\mathbf{v}\times\mathrm{curl}\left(\epsilon(1-{\mathcal F}
\Delta^{-1})\mathbf{v} +
\mathbf{R}\right)
\nonumber\\
&&+\nabla\left(\psi+{\frac{\epsilon}{2}}\mathbf{v}
\cdot(1-{\mathcal F} \Delta^{-1})\mathbf{v}\right)=0\, ,
\label{QG-mot-eqn1}
\end{eqnarray}
upon substituting the constraint $D=1$, imposed by varying $\psi$. The
curl of this equation yields
\begin{equation}
{\frac{\partial{q}}{\partial t}}
+ \mathbf{v}\cdot\nabla{q} + {q}\nabla\cdot\mathbf{v} = 0\,,
\label{qg-vort-eqn}
\end{equation}
where the potential vorticity $q$ is given by
\begin{equation} \label{pv-QG}
{q}=\epsilon\hat{\mathbf{z}}\cdot\hbox{curl}\,
(1-{\mathcal F} \Delta^{-1})\mathbf{v} + f
= \epsilon(\Delta\psi-{\mathcal F} \psi) + f\,,
\end{equation}
with
\begin{equation} \label{f-def1}
f\equiv\hat{\mathbf{z}}\cdot{\rm curl}\mathbf{R}=1+\beta x_2,
\end{equation}
and $\beta$ is assumed to be of order $O(\epsilon)$. The constraint $D=1$
implies $\nabla\cdot\mathbf{v}=0$ (from the kinematic relation
$\partial D/\partial t+\nabla\cdot D\mathbf{v}=0$) and
when $\mathbf{v}=\hat{\mathbf{z}}\times\nabla\psi$ is substituted, the
equation for ${q}=\Delta\psi-{\mathcal F} \psi + f$ yields the QG
potential vorticity convection equation (\ref{qgb'}). Thus, the QG
approximation follows as the Euler--Poincar\'e equation for an asymptotic
expansion of the action for the RSW equations when the potential energy is
modelled by inverting the geostrophic relation. The {\it same} QG equation
follows upon recasting the action (\ref{QG-lag-v}) in the Kaluza-Klein form
(\ref{incomp-KK-act}) for incompressible fluids,
\begin{equation}
 \mathfrak{S}_{\rm red}=\int dt\, l = \int dt\, dx_1dx_2\
\Big[{\frac{\epsilon}{2}} D\mathbf{v}\cdot (1-{\mathcal F}
\Delta^{-1})\mathbf{v}
+ \frac{1}{2} D (\mathbf{R}\cdot\mathbf{v}
+ {\tilde\omega})^2 - \psi(D-1)\Big]\,,
\label{KK-QG-act}
\end{equation}
where ${\tilde\omega}$ is defined as ${\tilde\omega}
=d\theta/dt
=\partial\theta/\partial t + \mathbf{v}\cdot\nabla\theta$
for the gauge field $\theta$, as in the case of the Maxwell fluid. Thus,
the QG motion equation (\ref{QG-mot-eqn1}) with the beta-effect (included
in $\mathbf{R}$) describes geodesic motion on the group of area-preserving
diffeomorphisms with respect to the conserved kinetic-energy metric given
by
\begin{equation} \label{metric-QG}
\|\mathbf{v}\|^2 = \int dx_1dx_2\
\Big[{\frac{\epsilon}{2}} \mathbf{v}\cdot (1-{\mathcal F}
\Delta^{-1})\mathbf{v}
+ \frac{1}{2}  (\mathbf{R}\cdot\mathbf{v}
+ {\tilde\omega})^2 \Big]\,,
\end{equation}
where $c'\equiv\mathbf{R}\cdot\mathbf{v}+{\tilde\omega}$ is an advected
quantity, $dc'/dt=0$. This observation from the Euler--Poincar\'e
viewpoint confirms the geodesic interpretation of the QG equations for
motion in the $\beta$-plane established in Zeitlin and Pasmanter [1994].

\paragraph{1D Boussinesq dispersive shallow water equations.}
For one dimensional shallow water motion with prescribed mean depth $B(x)$
we choose the following action
\begin{equation}
 \mathfrak{S}_{\rm red}=\int dt\, l = \int dt\, dx\ \Big[\frac{1}{2}
Dv^2+\frac{
\alpha^2}{2}(Dv)_x^2 -\frac{ g}{2}(D-B(x))^2\Big],
\label{1dswact}
\end{equation}
in which $g$ and $\alpha^2$ are constants and subscript $x$ denotes
partial derivative. The second term, proportional to $\alpha^2$,
represents the kinetic energy due to vertical motion. The last term is the
potential energy. Recall that the surface elevation $h\equiv{D-B(x)}$
satisfies $\partial h/\partial t=-(Dv)_x$ for shallow water dynamics in
one dimension. Thus, the last two terms are analogous to the Lagrangian
$$
\frac{\alpha^2}{2} \Big(\frac{\partial h}{\partial t}\Big)^2
-\ \frac{g}{2}\, h^2\,.
$$
This is the Lagrangian for a {\it harmonic oscillator} whose displacement is
given by the surface elevation $h$
and whose natural frequency is $\sqrt{g}/\alpha$.  For the choice
of action in equation (\ref{1dswact}) and for boundary conditions such that
$v_x\to0$ as $|x|\to\infty$, the variation is
\begin{equation}
\delta \mathfrak{S}_{\rm red} = \int dt\, dx\
\Big[D(v-\alpha^2(Dv)_{xx})\delta v +\Big(\frac{
v^2}{2}-\alpha^2v(Dv)_{xx}-gh\Big)\delta D\Big].
\label{1dswvar}
\end{equation}
The Kelvin-Noether form of the Euler--Poincar\'e equations in
(\ref{Eul-mot}) then gives
\begin{equation}\!\!\!\!
0= \Big( \frac{\partial}{\partial t}+ \pounds_v\Big)
\Big( \frac{ 1}{ D}\frac{ \delta l}{\delta v}dx \Big)
-d \Big( \frac{ \delta l}{ \delta D}\Big)
=\Big[
\frac{ \partial}{\partial t}(v-\alpha^2(Dv)_{xx})
+vv_x+gh_x\Big]dx\,.
\label{1dsweq}
\end{equation}
Upon inserting the one-dimensional continuity equation into the
$\alpha^2$ term and rearranging slightly, we find the system of equations
\begin{equation}
\frac{ \partial v}{ \partial t}+vv_x+gh_x
+\alpha^2\frac{ \partial^2 h_x}{ \partial t^2}=0,
\quad
\frac{ \partial h}{ \partial t}+\left(hv+B(x)v\right)_x=0\,,
\label{1Bsseq}
\end{equation}
the second of which is just a restatement of continuity. According to
Whitham [1974] these equations were favored by Boussinesq, who first
formulated them by using the method of asymptotic expansions. Here we see that
the Boussinesq shallow water equations (\ref{1Bsseq}) are also
Euler--Poincar\'e equations on $\operatorname{Diff}(\mathbb{R})$ derived from
the action (\ref{1dswact}). The term proportional to $\alpha^2$ in equation
(\ref{1Bsseq}) arises from the kinetic energy due to vertical motion in the
action (\ref{1dswact}) and produces the wave dispersion responsible for
solitary wave solutions of these equations.

\paragraph{1D Camassa--Holm equation for peakons.}
In the limit that the potential energy $gh^2/2$ is negligible
compared to the kinetic energy (e.g., for weak gravity, or small surface
elevation), we may ignore the last term in the action (\ref{1dswact}) for the
Boussinesq equations in one dimension, set $D=B(x)$ in the other terms
and rescale to $\alpha^2=1$, conforming to the notation in Camassa and
Holm [1993]. We thereby obtain the following simplified expression for the
shallow water action in this regime,
\begin{equation}
\mathfrak{S}_{\rm red} = \int dt\, dx\,B(x)\, \left(\frac{1}{2}
v^2+\frac{1}{2} v_x^2\right)\,.
\label{1dchact}
\end{equation}
For this action, when we also assume $B(x)=1$ (for constant bottom
topography) Hamilton's principle implies simply
\begin{equation}
0=\delta \mathfrak{S}_{\rm red} = \int dt\, l = \int dt\, dx\
(v-v_{xx})\delta{v}\,,
\label{1dchvar}
\end{equation}
for vanishing boundary conditions for $v_x$ on the real line
as $|x|\to\infty$.
Hence, $\delta l/\delta v=v-v_{xx}$ and the basic Euler--Poincar\'e
equations for this case reduce to
\begin{equation}
\frac{ \partial}{\partial t}(v-v_{xx})=-\operatorname{ad}_v^*(v-v_{xx})
=-v(v-v_{xx})_x-2(v-v_{xx})v_x\,.
\label{1dcheq}
\end{equation}
This is the $\kappa=0$ case of the completely integrable partial
differential equation derived by Camassa and Holm [1993],
\begin{equation}
\frac{ \partial}{\partial t}(v-v_{xx})+2 \kappa v_x
=-3vv_x + 2v_xv_{xx}+vv_{xxx}\,.
\label{1dcheq-k}
\end{equation}
For $\kappa=0$, this equation admits `peakon' solutions. The peakons
are solitons which interact elastically and possess a peak, at which the
derivative $v_x$ reverses sign. The simplest case is the single peakon,
which is a solitary travelling wave solution given by $v(x,t)=c_0\,{\rm
exp}-|x-c_0t\,|$, with a constant wave speed $c_0$. The multi-peakon
solutions of the Camassa-Holm equation are obtainable from its associated Lax
pair and linear isospectral problem, as shown in Camassa and Holm [1993].

Being basic Euler--Poincar\'e, equation (\ref{1dcheq}) describes
geodesic motion. Camassa and Holm [1993] note that the integrable
dynamics of $N$ peakons interacting nonlinearly via equation
(\ref{1dcheq}) reduces to finite dimensional geodesic motion on a
manifold with $N$ corners. This geodesic property persists to infinite
dimensions and although the equation was originally intended to be an
approximation of shallow water motion, it turns out equation
(\ref{1dcheq}) is also the geodesic spray equation for motion on the
group of diffeomorphisms of the real line with metric given by the $H^1$
norm of $v$, see Kouranbaeva [1997]. The $\kappa\ne0$ case of the
Camassa-Holm equation (\ref{1dcheq-k}) may be obtained formally by
shifting $(v-v_{xx})$ by $\kappa$ in equation (\ref{1dcheq}) and
retaining homogeneous boundary conditions for $(v-v_{xx})$ as
$|x|\to\infty$. The corresponding statement about geodesic motion for
$\kappa\ne0$, however, is rather more technical than for $\kappa=0$ and
involves the Gel'fand-Fuchs co-cycle and the Bott-Virasoro group, see
Misiolek [1997] for details. See Alber et al. [1994, 1995, 1997] for
discussions of the periodic solutions of the Camassa-Holm equation and a
related integrable shallow water equation in the Dym hierarchy,
\begin{equation}
2 \kappa v_x
= \frac{ \partial}{\partial t}v_{xx} + 2v_xv_{xx}+vv_{xxx}\,.
\label{1dcheq-hi-k}
\end{equation}
This equation is the ``high wave number limit" of the Camassa-Holm
equation (\ref{1dcheq-k}).

\paragraph{Higher dimensional Camassa--Holm equation.}
As we have seen, the Camassa-Holm (CH) equation in one dimension
describes geodesic motion on the diffeomorphism group with respect to the
metric given by the $H^1$ norm of the Eulerian fluid velocity. Thus, a
candidate for its $n$-dimensional incompressible generalization should be
the Euler--Poincar\'e equation that follows from the Lagrangian given by
the $H^1$ norm of the fluid velocity in
$n$ dimensions, subject to volume preservation (for $n\ne1$),
\begin{equation}\label{CH-lag}
\mathfrak{S}_{\rm red} = \int dt \; l=\int dt\int_{\cal M} d^{\,n}x
\,\frac{D}{2} (v_i v^i + \alpha^2 v_i^{,j}v_{,j}^i) - p(D-1)\,,
\end{equation}
where we have restored the length-scale, or aspect-ratio parameter, $\alpha$.
Varying this action at fixed $ \mathbf{x}$ and $t$ gives
\begin{eqnarray}
\delta \mathfrak{S}_{\rm red} &=&\!\! \int dt\int_{\cal M} d^{\,n}x \,\Big[
(v_i v^i + \alpha^2 v_i^{,j}v_{,j}^i) \delta D - (D-1) \delta p
+ \left(Dv_i - \alpha^2 \partial_j (D v_i^{,j}) \right) \delta v^i\Big]
\nonumber \\
&&\hspace{.5in}
+\ \alpha^2 \int dt\oint_{\partial{\cal M}} d^{\,n-1}x \
\hat{n}_j(D v_i^{,j} \delta v^i)
\,,
\label{var.ndCH}
\end{eqnarray}
whose natural boundary conditions on $\partial{\cal M}$ are
\begin{equation}\label{boundary.equation}
\mathbf{v} \cdot \hat{\mathbf{n}} = 0
\quad {\rm and} \quad
(\hat{\mathbf{n}} \cdot \nabla)
\mathbf{v}\ \|\ \hat{\mathbf{n}},
\end{equation}
where $\|$ denotes ``parallel to" in the second boundary condition,
which of course is not imposed when $\alpha^2$ is absent. (Recall
that $\delta\mathbf{v}$ in equation (\ref{var.ndCH}) is arbitrary
except for being tangent on the boundary. This tangency, along with the
second condition in equation (\ref{boundary.equation}) is sufficient for the
boundary integral in equation (\ref{var.ndCH}) to vanish.) By equation
(\ref{EP-Kthm}) or (\ref{KKthm}), the Euler--Poincar\'e equation for
the action $\mathfrak{S}_{\rm red}$ in equation (\ref{CH-lag}) is
\begin{equation}
\left(\frac{ \partial}{\partial t}+ \mathbf{v}\cdot\nabla\right)
(\mathbf{v} - \alpha^2\Delta \mathbf{v})
+(v_j-\alpha^2\Delta v_j)\nabla v^j
-\nabla\left(\frac{1}{2}|\mathbf{v}|^2
+ \frac{\alpha^2}{2}|\nabla\mathbf{v}|^2 - p\right) = 0\,,
\label{nd:CHeqn}
\end{equation}
where
$({\nabla}\mathbf{v})^i_j=v^i_{,j}
  \equiv{\partial}v^i/{\partial}x^j$,
$|\nabla\mathbf{v}|^2\equiv v^i_{,j}v_i^{,j}= {\rm
tr}(\nabla\mathbf{v}\cdot\nabla\mathbf{v}^T)$ and superscript
$(\bullet)^T$ denotes transpose. We have also used the constraint
$D=1$, which as before implies incompressibility via the continuity
equation for $D$. Requiring the motion equation (\ref{nd:CHeqn}) to
preserve ${\rm div}\,\mathbf{v}=0$ implies a Poisson equation for
the pressure $p$ with a Neumann boundary condition, which is obtained just
as in the case of incompressible ideal fluid dynamics by taking the
normal component of the motion equation evaluated at the boundary.

\paragraph{Properties of the Camassa--Holm equation.}
Since the CH action $\mathfrak{S}_{\rm red}$ in (\ref{CH-lag}) is translation
invariant, the Noether theorem ensures the CH equation (\ref{nd:CHeqn})
conserves a momentum. In fact, by the stress tensor formulae
(\ref{mom-cons})--(\ref{stress-tens}), equation (\ref{nd:CHeqn}) may be
rewritten as
\begin{equation} \label{mom-cons-CH}
\frac{\partial m_i}{\partial t} = -\ \frac{\partial }{\partial
x^j}T^j_i\,.
\end{equation}
In this case, the momentum density $m_i$, $i=1,2,3$ defined in equation
(\ref{mom-comp}) is given by
\begin{equation} \label{mom-CH}
m_i \equiv \frac{\delta l}{\delta v^i}\Big|_{D=1}
=\ v_i-\alpha^2\Delta v_i\,,
\end{equation}
and the stress tensor $T^j_i$ defined in equation (\ref{stress-tens}) is
given by
\begin{equation} \label{stress-tens-CH}
T^j_i = (v_i-\alpha^2\Delta v_i) v^j
        - \alpha^2\, v^k_{,i} v^{,j}_k  + \delta^j_i p\,.
\end{equation}
Thus, equation (\ref{mom-cons-CH}) implies conservation of the total
momentum, $\mathbf{M}=\int_{\cal M} \mathbf{m}\, d^{\,3}x$, provided the
normal component of the stress tensor $T^j_i$ vanishes on the boundary.

Since the CH equation (\ref{nd:CHeqn}) is Euler--Poincar\'e, it also has a
corresponding {\bfi Kelvin-Noether circulation theorem}. Namely, cf. equation
(\ref{KelThm2}),
\begin{equation}
\frac{ d}{dt}\oint_{\gamma_t}(\mathbf{v} - \alpha^2 \Delta \mathbf{v})\cdot
d\mathbf{x} = 0\,,
\label{KelThm2}
\end{equation}
for any closed curve ${\gamma_t}$ that moves with the fluid velocity
$\mathbf{v}$. This expression for the Kelvin-Noether property of the CH
equation in 3D is reminiscent of corresponding expressions in wave, mean-flow
interaction theory. This correspondence suggests a physical interpretation
of the $\alpha^2$ term in the Kelvin-Noether circulation integral as a {\it
Lagrangian mean closure relation} for the pseudomomentum of the high
frequency (i.e., rapidly fluctuating, turbulent) components of the flow. In
this interpretation, $\alpha$ corresponds to the typical length scale at
which these high frequency components become important. See Holm, Foias and
Titi [1998] for more discussion of using the 3D CH equation (\ref{nd:CHeqn})
as the basis for a turbulence closure model.

In three dimensions, we may use the vector identity (\ref{fvid}) to
re-express the CH motion equation (\ref{nd:CHeqn}) in its
``curl" form, as
\begin{eqnarray}\label{3d:CHeqn}
\frac{\partial }{\partial t}(1-\alpha^2\Delta) \mathbf{v}
&-& \mathbf{v} \times \left(\nabla \times (1-\alpha^2\Delta)
\mathbf{v}\right)
\nonumber \\
&+ & \nabla \left(\mathbf{v} \cdot \left( 1 - \alpha^2\Delta\right)
\mathbf{v} - \frac{1}{2} |\mathbf{v}|^2
- \frac{\alpha^2}{2} |\nabla\mathbf{v}|^2 + p \right) = 0\,.
\end{eqnarray}
The inner product of $\mathbf{v}$ with this equation then implies
{\bfi conservation of energy},
\begin{equation}
E = \frac{1}{2} \int_{\cal M} d^{\,3}x \
    \left(\mathbf{v}\cdot(1-\alpha^2\Delta)\mathbf{v}\right)
  = \frac{1}{2} \int d^{\,3}x \
    \left(|\mathbf{v}|^2+\alpha^2|\nabla\mathbf{v}|^2\right)
\,,
\label{KE-CH}
\end{equation}
upon integrating by parts and using the boundary conditions
(\ref{boundary.equation}). Naturally, this energy is also conserved
in $n$ dimensions. In fact, Legendre transforming the action
(\ref{CH-lag}) gives the following {\bfi Hamiltonian} (still
expressed in terms of the velocity, instead of the momentum density
$\mathbf{m}= \delta l/ \delta \mathbf{v}$),
\begin{equation}
H = \int_{\cal M} d^{\,n}x\ \Big[\,\frac{D}{2}
\left(|\mathbf{v}|^2+\alpha^2|\nabla\mathbf{v}|^2\right)
 + p(D-1)\Big]
\,.  \label{CH-ham-v}
\end{equation}
Thus, when evaluated on the constraint manifold $D=1$, the Lagrangian
and the Hamiltonian for the CH equation coincide in $n$ dimensions. (This,
of course, is not unexpected for a stationary principle giving rise to
geodesic motion.)

The curl of the 3D Camassa-Holm motion equation (\ref{3d:CHeqn}) yields
\begin{equation} \label{vortex-stretching}
\frac{ \partial}{\partial t}\mathbf{q}
= \mathbf{q}\cdot\nabla\mathbf{v} - \mathbf{v}\cdot\nabla\mathbf{q}
\equiv [\mathbf{v},\mathbf{q}\,],
\quad \hbox{where} \quad
\mathbf{q}\equiv{\rm curl}
(\mathbf{v} - \alpha^2 \Delta \mathbf{v})\,,
\end{equation}
and we have used incompressibility and commutativity of the divergence and
Laplacian operators. Thus, $\mathbf{v}$ is the transport velocity for the
generalized vorticity $\mathbf{q}$ and the ``vortex stretching" term
$\mathbf{q}\cdot\nabla\mathbf{v}$ involves $\nabla\mathbf{v}$, whose $L^2$
norm is {\em controlled} by the conservation of energy in equation
(\ref{KE-CH}). Boundedness of this norm will be useful in future analytical
studies of the 3D Camassa-Holm equation; for example, in the investigation of
the Liapunov stability properties of its equilibrium solutions.

\paragraph{3D periodic CH motion.}
In a three dimensional periodic domain, the conserved energy $E$ in equation
(\ref{KE-CH}) may also be expressed as
\begin{equation}
E = \frac{1}{2} \int_{\cal M} d^{\,3}x \
    \left(|\mathbf{v}|^2
    + \alpha^2 |{\rm curl}\mathbf{v}|^2 \right)\,,
\label{enstrophic}
\end{equation}
upon integrating by parts and using ${\rm div}\,\mathbf{v}=0$.  Thus, in
the 3D periodic case, the CH energy $E$ may be interpreted as the sum of the
kinetic energy and the enstrophy (i.e., the $L^2$ norm of vorticity) of the
Euler fluid.

The inner product of the generalized vorticity $\mathbf{q}$ with the
motion equation (\ref{3d:CHeqn}) implies conservation of {\bfi helicity},
for three dimensional periodic motion. Namely, the quantity
\begin{equation} \label{helicity-def}
\Lambda \equiv \int_{\cal M} d^{\,3}x \
(1  - \alpha^2 \Delta )\mathbf{v}
\cdot{\rm curl}(1  - \alpha^2 \Delta )\mathbf{v}
\quad \hbox{(helicity)},
\end{equation}
is also a constant of motion for three dimensional periodic CH motion.

>From the CH vorticity equation (\ref{vortex-stretching}), we see
that steady 3D solutions of the CH equation  (denoted with
subscript $e$ for ``equilibrium") are characterized by the
vector-field commutation relation
$[\mathbf{v}_e,\mathbf{q}_e]=0$.  Thus, the velocity of a steady
CH flow $\mathbf{v}_e$ generates a volume preserving
diffeomorphism that leaves invariant its corresponding steady
generalized vorticity $\mathbf{q}_e$. For example, the {\bfi CH
Beltrami flows} for equation (\ref{3d:CHeqn}) are characterized
by $\mathbf{v}_e=\mu\mathbf{q}_e$, for a constant
$\mu$. The CH Beltrami flows verify the invariance property; hence, they
are steady. These steady solutions are also {\it
critical points} of the sum $E+\Lambda/2\mu$ of the energy $E$ in equation
(\ref{enstrophic}) and $1/2\mu$ times the conserved helicity in equation
(\ref{helicity-def}). Hence, they are {\it relative equilibrium solutions} of
the CH equation. The CH Beltrami flows are divergenceless vector
eigenfunctions of the product of the curl operator and the Helmholtz
operator, $(1-\alpha^2\Delta)$. They are the CH analogues of ``A-B-C flows"
for the ideal Euler fluid.

\paragraph{Constitutive properties of the CH ``fluid."}
Physically, conservation of the energy in equation (\ref{enstrophic}) means
that the CH fluid can exchange energy between its translational, and its
rotational and shear motions. One may ask, what constitutive relation
describes such a fluid?

One may verify directly that the 3D Camassa--Holm equation
(\ref{nd:CHeqn}) in Cartesian coordinates implies the following
formula for the geodesic spray form of the CH equations in 3D:
\begin{eqnarray} \label{nd:CHeqn-geodesic1}
(1-\alpha^2\Delta)
\left(\frac{ \partial}{\partial t} + \mathbf{v}\cdot\nabla\right)
v_i
\!\!\!&=&\!\!\!
\alpha^2(\Delta{v_j})(v^j_{,i}-v_i^{,j})
\nonumber \\
\!\!\!& - &\!\!\!\frac{\partial}{\partial{x^j}}
\left[\left(p - \frac{\alpha^2}{2} v^i_{,k}v_i^{,k}\right)
\delta^j_i
 +2\alpha^2v^j_{,k}v_i^{,k}\right].
\end{eqnarray}
In vector notation, this is
\begin{eqnarray} \label{nd:CHeqn-geodesic2}
(1-\alpha^2\Delta)
\left(\frac{ \partial}{\partial t} + \mathbf{v}\cdot\nabla\right)
\mathbf{v}
\!\!\!&=&\!\!\!
\alpha^2(\Delta\mathbf{v})\times{\rm{curl}\,\mathbf{v}}
\nonumber \\
\!\!\!& - &\!\!\!{\rm div}\,
\left[\left(p - \frac{\alpha^2}{2} |\nabla\mathbf{v}|^2\right)\mathbf{{\hat
I}} +2\alpha^2\nabla\mathbf{v}\cdot\nabla\mathbf{v}^T\right],
\end{eqnarray}
where $\mathbf{{\hat I}}$ is the unit tensor. Viewed this way, the CH fluid
acceleration $d\mathbf{v}/dt$ is nonlocal, nonlinear, and (as we know from
the general Euler--Poincar\'e theory) geodesic.  Holm, Foias and Titi
[1998] show, among other things, that the geodesic spray equation
(\ref{nd:CHeqn-geodesic2}) may be rearranged into a viscoelastic
constitutive relation of Jeffreys type.

\paragraph{Discussion.}
The essential idea of the CH equation is that its specific momentum (i.e.,
its momentum per unit mass) is transported by a velocity which is
smoothed by inverting the elliptic Helmholtz operator $(1-\alpha^2\Delta)$,
where $\alpha$ corresponds to the length scale at which this smoothing
becomes important, i.e., when it becomes of order $O(1)$.
When the smoothing operator $(1-\alpha^2\Delta)^{-1}$ is applied to the
transport velocity in Euler's equation to produce the CH equation, its
effect on length scales smaller than $\alpha$ is that steep gradients of the
specific momentum function tend not to steepen much further, and thin vortex
tubes tend not to get much thinner as they are transported. And, its effect
on length scales that are considerably larger than $\alpha$ is negligible.
Hence, the transport of vorticity in the CH equation is intermediate between
that for the Euler equations in 2D and 3D. As for Euler vorticity, the curl
of the CH specific momentum is {\em convected} as an {\em active} two form,
but its transport velocity is the {\em smoothed, or filtered}
CH specific momentum.

The effects of this smoothing or filtering
of the transport velocity in the
CH equation can be seen quite clearly from its Fourier spectral
representation in the periodic case. In this case, we define
$\mathbf{m}_\mathbf{k}$ as the
$\mathbf{k}$-th Fourier mode of the specific momentum
$\mathbf{m}\equiv(1-\alpha^2\Delta)\mathbf{v}$
for the CH equation; so that
$\mathbf{m}_\mathbf{k} \equiv (1+\alpha^2
|\mathbf{k}|^2)\mathbf{v}_\mathbf{k}$.
Then the Fourier spectral representation of
the CH equation for a periodic
three-dimensional domain is expressed as
\begin{equation} \label{CH-spectral}
\Pi\left(
\frac{d}{dt} \mathbf{m}_\mathbf{k}
- \sum_{\mathbf{p}+\mathbf{q}=\mathbf{k}}
\frac{ \mathbf{m}_\mathbf{p} }{1+\alpha^2|\mathbf{p}|^2 }
\times ( \mathbf{q}\times\mathbf{m}_\mathbf{q}) \right)
=0,
\end{equation}
where $\Pi$ is the Leray projection onto Fourier modes transverse to
$\mathbf{k}$. (As usual, the Leray projection ensures
incompressibility.) In this Fourier spectral representation of the
CH equation, one sees that the coupling to high modes is suppressed
by the denominator when $1 + \alpha^2|\mathbf{p}|^2\gg1$.
Consequently, when $|\mathbf{p}|\geq{O(1/\alpha)}$, the smoothing of
the transport velocity suppresses the development of higher modes
$|\mathbf{k}|{\ge}O(2|\mathbf{p}|)$. And, it {\em also} suppresses
the ``stochastic backscatter" from higher modes to lower ones,
$|\mathbf{k}|=O(1)$. Thus, thinking of ``interaction triangles"
among the modes, one sees that both the very acute and very obtuse
$\mathbf{p}+\mathbf{q}=\mathbf{k}$ triangles are suppressed, leaving
active only the nearly equilateral ones, when
$|\mathbf{p}|\geq{O(1/\alpha)}$. Hence, the CH smoothing of the
transport velocity suppresses {\em both} the forward and backward
cascades for wave numbers
$|\mathbf{p}|\geq{O(1/\alpha)}$, but leaves the Euler dynamics
essentially unchanged for smaller wave numbers. As we have seen, the
result is that the vortex stretching term in the dynamics of
$\mathbf{q}={\rm curl}\,\mathbf{m}$ is mollified in the CH model and
so the vortices at high wave numbers will tend to be ``shorter and
fatter'' than in the corresponding Euler case.

When the kinetic energy terms are neglected relative to the gradient
velocity terms, the CH action
$\mathfrak{S}_{\rm red}$ in (\ref{CH-lag}) becomes
\begin{equation}\label{CH-lag-hi-k}
\mathfrak{S}_{\rm red}^{\infty} = \int dt \; l=\int dt\int_{\cal M} d^{\,n}x
\left[\,\frac{D}{2} v^i_{,j} v_i^{,j} - p(D-1)\right]\,,
\end{equation}
whose Euler--Poincar\'e equation in 3D implies the following,
\begin{equation}\label{3d:CHeqn-hi-k}
\frac{\partial }{\partial t}\Delta \mathbf{v}
- \mathbf{v} \times \left(\nabla \times \Delta \mathbf{v}\right)
+ \nabla \left(\mathbf{v} \cdot  \Delta\mathbf{v}
+ \frac{1}{2} |\nabla\mathbf{v}|^2 - p \right) = 0\,.
\end{equation}
This equation is the ``high wave number limit" of the 3D Camassa-Holm
equation (\ref{3d:CHeqn}).
Scale invariance is restored in this limiting equation and its corresponding
group invariant (e.g., self-similar) solutions may be illuminating.

\paragraph{The Riemannian CH equations.} One can
formulate the CH equations on a general Riemannian manifold,
possibly with boundary. Although this will be the subject of future
papers, we will make some comments about some of the features (some
of them conjectural) here.

We start with a smooth, oriented, compact Riemannian manifold $M$,
possibly with a smooth boundary. We first define the group
${\rm Diff}_{\rm CH}$ to be the group of diffeomorphisms $\eta: M
\rightarrow M$ of class $H^s$, where $s > (n/2) + 2$ with the
boundary condition that the tangent map $ T \eta : TM \rightarrow
TM $ takes the outward normal direction to the boundary $ \partial
M $ at a point $x \in M$  to the outward normal direction at the
point $ \eta (x) $. We first {\bf conjecture} that {\it this group
is a smooth manifold and is a Lie group (in the same sense as in
Ebin and Marsden [1970]) with Lie algebra the set of vector fields
$v$ on $M$ which are tangent to the boundary of $M$ and that satisfy
the boundary condition}
\[
   \left\langle \nabla _n v , u \right\rangle = S (u,v)
\]
{\it for all vectors $u$ tangent to the boundary}. Here $ S (u,v)
$ is the second fundamental form of the boundary. This condition on
the boundary is the CH analogue to the condition of parallel to the
boundary in the case of the Euler equations. The condition comes
about by differentiation of the condition on $\eta$ in the
definition of ${\rm Diff}_{\rm CH} $ using a routine calculation.

Notice that the boundary conditions are different from those
previously (see equation (\ref{boundary.equation}). This appears,
however, to be needed for the group theoretic version of
the equations. Now we put a right invariant Lagrangian on
$ {\rm Diff}_{\rm CH} $ which, at the identity, is given by
\begin{equation} \label{CHenergy.equation}
L (v) =
\frac{1}{2} \int _M \left( \| v \| ^2 + \| \nabla v \| ^2 \right) d
\mu   - \frac{1}{2} \oint _{\partial M} S (u,u) d A
\end{equation}
where $ \nabla v $ is the covariant derivative of $v$, where $ d
\mu $ is the Riemannian volume element and where $ d A $ is the area
element of the boundary. We are using terminology
appropriate to the case in which $M$ is three dimensional, but of
course there is no restriction on the dimension of $M$. Also,
$ \| \nabla v \| ^2 $ denotes the norm in the sense of the full
Riemannian contraction of the tensor $ \nabla v $. The associated
Laplace operator is usually called the {\it rough Laplacian}.

At this point there are some choices one can make. One can use a
different $ H ^1 $ metric built out of thinking of $ v $ as a one
form and using the $ d $ and $ \delta $ operators and the
corresponding {\it Laplace deRham} operator. This leads to a
slightly different system in general, but one that has similar
analytical properties.

The {\bfi Riemannian CH equations} are, by definition, the
Euler--Poincar\'e equations for this group and this Lagrangian. The
boundary term in the Lagrangian is designed to make the boundary
conditions in the resulting equations come out agreeing with those
for the Lie algebra of the group $ {\rm Diff}_{\rm CH} $. Apart from
the boundary conditions, the resulting equations agree with the ones
we developed in Euclidean space, but in general one replaces the
Laplacian with the rough Laplacian. Note that since the Lagrangian
is quadratic in $v $, the equations on ${\rm Diff}_{\rm CH}$ are
geodesic equations (possibly with respect to an indefinite metric).

\begin{cnj} As in the case of the Euler
equations (Ebin and Marsden [1970]), the geodesic spray of the
Riemannian CH equations is smooth if $ s > (n/2) + 2 $.
\end{cnj}

This conjecture is based on a direct examination of the expression
for the spray of the Riemannian CH equations. If this is true,
then other analytic things, including results on the limit of zero
viscosity (or viscoelasticity) also hold. We also note that because
energy conservation involves a stronger norm than in the Euler
equations for ideal flow, one expects other analytic properties of
the Riemannian CH equation to be improved. This would include
results on stability and long time existence.

Another consequence of this would be that {\it the spray of the
incompressible Riemannian CH equations would also be
smooth}. This follows since the projection map is smooth and
the fact that the spray of the incompressible equations is given by
the composition of the spray of the compressible ones and the
tangent of the projection map. (These facts are proved in Ebin and
Marsden [1970]).

As we have mentioned, all of these things will be explored in detail
in other publications.

\paragraph{2D Camassa--Holm equation.}
In two dimensions, the curl of the Euler--Poincar\'e motion equation
(\ref{3d:CHeqn}) produces a scalar relation for potential vorticity
convection, namely,
\begin{equation}
{\frac{\partial q}{\partial t}}+\mathbf{v}\cdot \nabla q=0,
\quad\hbox{where}\quad
q\equiv (1-\alpha^2\Delta){\Delta }\psi
\quad\hbox{for 2dCH.}
\label{CH:PVadv}
\end{equation}
In terms of the stream function $\psi$, with
$\mathbf{v}=\hat{\mathbf{z}}\times\nabla\psi$,
the boundary conditions (\ref{boundary.equation}) in two dimensions
with
$\hat{\mathbf{s}}=
\hat{\mathbf{z}}\times\hat{\mathbf{n}}$ become
\begin{equation}
\psi = {\rm const}
\quad {\rm and} \quad
\hat{\mathbf{n}}\cdot
\nabla\nabla\psi\cdot\hat{\mathbf{n}}=0
\Rightarrow
\Delta\psi=0
\quad \hbox{on the boundary.}
\label{2dchbc}
\end{equation}
Potential vorticity convection (\ref{CH:PVadv}) for the 2dCH equation,
combined with incompressibility and the first boundary condition in
(\ref{boundary.equation}) imply conservation of the following
quantity,
\begin{equation}
C_\Phi \equiv \int d^2x\ \Phi (q)\,,
\end{equation}
for any suitably well-behaved function $\Phi$. (This is the
Casimir, in the Lie-Poisson bracket formulation.)
Substituting $\mathbf{v}=\hat{\mathbf{z}}\times\nabla\psi$
and using the divergence theorem yields an expression for the
kinetic energy Lagrangian for the 2dCH equation in terms of the stream
function, $\psi$. Namely,
\begin{eqnarray}
E &=& \frac{1}{2} \int d^{\,2}x \
    \left(\mathbf{v}\cdot(1-\alpha^2\Delta)\mathbf{v}\right)
= \frac{1}{2} \int d^{\,2}x \
    \left(\nabla \psi \cdot(1-\alpha^2\Delta)\nabla \psi \right)
\\ \label{KE-psi}
&=& \frac{1}{2} \sum_i\psi^{(i)}\oint_{\gamma^{(i)}} ds \
\frac{\partial}{\partial n}(1-\alpha^2\Delta) \psi^{(i)}
- \frac{1}{2} \int d^{\,2}x \
    \left(\psi(1-\alpha^2\Delta)\Delta \psi \right)\,,
\nonumber
\end{eqnarray}
where we sum over the connected components of the boundary
$\gamma^{(i)}$ and use $\psi^{(i)}$ constant on the $i$th component. Thus,
solutions of the CH equation (\ref{nd:CHeqn}) in two dimensions optimize the
integrated product of the stream function and potential vorticity,
constrained by the circulation of $\mathbf{v} - \alpha^2\Delta \mathbf{v}$
on each connected component of the boundary. The Lagrange multiplier for this
circulation is the corresponding boundary component's (constant) stream
function.

\paragraph{Steady CH solutions in two dimensions.}
Steady solutions of the 2dCH equation (\ref{CH:PVadv}) with
boundary conditions (\ref{2dchbc}) satisfy
\begin{equation}
\label{steady-2dCH}
\hat{\mathbf{z}} \cdot \nabla \psi_e \times \nabla q_e = J(q_e,\psi_e)=0,
\quad\hbox{for}\quad
q_e\equiv (1-\alpha^2\Delta){\Delta }\psi_e\,.
\end{equation}
Thus, steady CH solutions exist when there is a functional relation between
the potential vorticity $q_e$ and its associated stream function $\psi_e$.
For example, one could have $q_e=F(\psi_e)$ for function $F$. In
particular, the {\it linear} steady flows satisfy
\begin{equation}
q_e=\Delta (1-\alpha^2\Delta) \psi_e = -|\mathbf{k}|^2(1+\alpha^2
|\mathbf{k}|^2)\psi_e.
\label{FT-soln}
\end{equation}
These are sines and cosines with wave number $\mathbf{k}$ for periodic
boundary conditions. The corresponding fluid velocity for any of these
steady 2dCH solutions is found from
$\mathbf{v}_e=\hat{\mathbf{z}}\times\nabla\psi_e$.

As another example, the {\bfi point
potential-vortex solution} centered at $z'=x'+iy'$ in the infinite
$x,y$ plane has stream function
\begin{equation}
\psi(|z-z'|) = {\rm log}(|z-z'|) + K_0(|z-z'|/\alpha),
\label{pt-vort-soln}
\end{equation}
where $K_0(|z-z'|/\alpha)$ is the Bessel function of the second kind. This
stream function satisfies
\[ q = \Delta (1-\alpha^2\Delta) \psi= 2\pi\delta (|z-z'|), \]
where $\Delta$ is the Laplacian operator in the plane and $\delta
(|z-z'|)$ is the Dirac delta function. (The proof uses $\Delta {\rm
log}(|z-z'|)=2\pi\delta (|z-z'|)$ and
$(1-\alpha^2\Delta)K_0(|z-z'|/\alpha)=2\pi\delta (|z-z'|)$.)
The {\bfi circular potential-vortex patch solution},
$q = \Delta (1-\alpha^2\Delta) \psi = C = {\rm const}$
for $|z-z'|\le a$ and $q = 0$ for $|z-z'|>a$, has stream function
\begin{eqnarray}
\psi &=& C\left(\frac{|z-z'|^2}{4}+1\right) \hbox{ for } |z-z'|\le a,
\nonumber \\
\psi &=& {\rm log}(|z-z'|)
+ \alpha^2K_0(|z-z'|/\alpha) \hbox{ for }  |z-z'|\ge a\,,
\label{vort-patch-soln}
\end{eqnarray}
where the constant $C$ is chosen so the velocity is
continuous at $|z-z'|=a$. The interior of this solution is also a
uniformly rotating vortex patch. These special potential-vortex solutions
illustrate the ``screening" in the vortex interaction dynamics
for the 2dCH equation introduced by the Helmholtz operator
$(1-\alpha^2\Delta)$, which modifies the momentum density of the 2dCH flow
relative to the standard incompressible Euler equations in the plane. As
we have seen, the corresponding screening length $\alpha$ is an
additional parameter in the model.

\paragraph{Quasigeostrophic analogue of CH in two dimensions.} We extend the
nondimensional QG action principle in equation (\ref{QG-lag-v}) for QG
dynamics in a periodic two-dimensional domain to include CH $\alpha^2$
terms, as follows,
\begin{equation}\label{QG-ch-lag-v}
 \mathfrak{S}_{\rm red}=\int dt\, l = \int dt\int dx_1dx_2\
\Big[\frac{\epsilon}{2} D\mathbf{v}\cdot
(1-{\mathcal F} \Delta^{-1} - \alpha^2\Delta)\mathbf{v}
+ D\mathbf{v}\cdot\mathbf{R} - \psi(D-1)\Big]
\,.
\end{equation}
The corresponding Euler--Poincar\'e equation is to be compared to the QG
motion equation (\ref{QG-mot-eqn1}). We find,
\begin{eqnarray}\label{QG-mot-CH}
\epsilon\frac{\partial}{\partial t}
(1-{\mathcal F} \Delta^{-1}- \alpha^2\Delta)\mathbf{v}
\!\!\!&-&\!\!\!
\mathbf{v}\times\mathrm{curl}\left(\epsilon(1-{\mathcal F}
\Delta^{-1} - \alpha^2\Delta)\mathbf{v} +
\mathbf{R}\right)
\nonumber\\
&&+\nabla\left(\psi+{\frac{\epsilon}{2}}\mathbf{v}
\cdot(1-{\mathcal F} \Delta^{-1} - \alpha^2\Delta)\mathbf{v}\right)=0\,.
\end{eqnarray}
The curl of this equation yields
\begin{equation}
\epsilon{\frac{\partial{q}}{\partial t}}
+ \mathbf{v}\cdot\nabla(\epsilon{q}+f) = 0\,,
\label{qgch-vort-eqn}
\end{equation}
where the potential vorticity $q$ is now given by
\begin{equation} \label{pv-QGCH}
{q}=\hat{\mathbf{z}}\cdot\hbox{curl}\,
(1-{\mathcal F} \Delta^{-1}- \alpha^2 \Delta)\mathbf{v}\,,
\end{equation}
and we choose
\begin{equation} \label{f-def2}
f\equiv\hat{\mathbf{z}}\cdot{\rm curl}\mathbf{R}=1+\epsilon\beta{x_2}\,.
\end{equation}
The constraint $D=1$
implies $\nabla\cdot\mathbf{v}=0$ as usual and
when $\mathbf{v}=\hat{\mathbf{z}}\times\nabla\psi$ is substituted, the
equation for $q = \Delta (1-\alpha^2\Delta)\psi-{\mathcal F} \psi$ yields
\begin{equation}\label{qgb-ch}
\frac{\partial}{\partial{t}}(\Delta (1-\alpha^2\Delta)\psi - {\mathcal F}
\psi) +[\psi,\Delta (1-\alpha^2\Delta)\psi] +\beta
{\frac{\partial\psi}{\partial x_1}} = 0\,.
\end{equation}
Here $[a,b]\equiv{\partial (a,b)}/{\partial(x_1,x_2)}=J(a,b)$ is
the Jacobi bracket (Jacobian) for functions $a$ and $b$ defined on
the two dimensional domain. Steady solutions of the QG-CH equation
(\ref{qgb-ch}) satisfy
\begin{equation}
\label{steady-2dQG.CH}
J(q_e+\beta{x_2},\psi_e)=0,
\quad\hbox{for}\quad
q_e = \Delta (1-\alpha^2\Delta)\psi_e-{\mathcal F} \psi_e\,.
\end{equation}

The dispersion relation for plane-wave solutions of equation
(\ref{qgb-ch}) with frequency $\omega$ and wavenumber $\mathbf{k}$ is
\begin{equation} \label{pw-disp}
\omega(\mathbf{k})= \frac{- \beta k_1}
{{\mathcal F}+|\mathbf{k}|^2(1+\alpha ^2|\mathbf{k}|^2)}\,.
\end{equation}
Such plane-wave solutions are analogous to Rossby waves in QG. As with QG
Rossby waves, these plane-wave solutions are {\it nonlinear} solutions of
equation (\ref{qgb-ch}) in a two dimensional periodic domain. If we
choose ${\mathcal F}=O(1)$ and $\alpha ^2=o(1)$, then the effect of the
$\alpha^2$ term in this dispersion relation is to significantly reduce the
oscillation frequency and propagation speeds of those waves at wave numbers
greater than about $\alpha^{-1}$. Thus, such short waves are suppressed, and
the emerging dynamics of this modified QG theory will tend to possess
significant activity only at length scales larger than
$\alpha$. Apparently the dispersion relation for the dynamics at these
larger length scales will faithfully approximate the corresponding QG
dynamics at these scales. This scenario is, of course, consistent with our
earlier discussion of the wave number dynamics of the CH solutions in three
dimensions.

\section*{Appendix}\label{sec-app}

\paragraph{Left Representation and Right Invariant Lagrangian.}
There is a version
of this
theorem for right invariant Lagrangians, but
with the representation of $G$ on
$V$ still on the left. The proof is, of course,
identical so we shall only
state
this theorem. The set--up is the following:
\begin{itemize}
\item There is a {\it left\/} representation of Lie group $G$ on
the vector space $V$ and $G$ acts in the natural way on the {\it
right\/} on $TG \times V^\ast$: $(v_g, a)h = (v_gh, h^{-1}a)$.
\item Assume that the function $ L : T G \times V ^\ast
\rightarrow \mathbb{R}$ is right $G$--invariant.
\item In particular, if $a_0 \in V^\ast$, define the
Lagrangian $L_{a_0} : TG \rightarrow \mathbb{R}$ by
$L_{a_0}(v_g) = L( v_g, a_0)$. Then $L_{a_0}$ is right
invariant under the lift to $TG$ of the right action of
$G_{a_0}$ on $G$.
\item  Right $G$--invariance of $L$ permits us to define
$l: {\mathfrak{g}} \times V^\ast \rightarrow \mathbb{R}$ by
\[
l(v_gg^{-1}, ga_0) = L(v_g, a_0).
\]
Conversely,  this relation defines for any
$l: {\mathfrak{g}} \times V^\ast \rightarrow
\mathbb{R} $ a right $G$--invariant function
$ L : T G \times V ^\ast
\rightarrow \mathbb{R} $.
\item For a curve $g(t) \in G, $ let
$\xi (t) := \dot{g}(t) g(t) ^{ -1}$ and define the curve
$a(t)$ as the unique solution of the linear differential equation
with time dependent coefficients $\dot a(t) = \xi(t) a(t)$
with initial condition $a(0) = a_0$. The solution can be
equivalently written as $a(t) = g(t)a_0$.
\end{itemize}

\begin{thm} \label{larl}
The following are equivalent:
\begin{enumerate}
\item [{\bf i} ] Hamilton's variational principle
\begin{equation} \label{hamiltonprinciple1}
\delta \int _{t_1} ^{t_2} L_{a_0}(g(t), \dot{g} (t)) dt = 0
\end{equation}
holds, for variations $\delta g(t)$
of $ g (t) $ vanishing at the endpoints.
\item [{\bf ii}  ] $g(t)$ satisfies the Euler--Lagrange
equations for $L_{a_0}$ on $G$.
\item [{\bf iii} ]  The constrained variational principle
\begin{equation} \label{variationalprinciple1}
\delta \int _{t_1} ^{t_2}  l(\xi(t), a(t)) dt = 0
\end{equation}
holds on $\mathfrak{g} \times V ^\ast $, using variations of the form
\begin{equation} \label{variations1}
\delta \xi = \dot{\eta } - [\xi , \eta ], \quad
\delta a =  \eta a ,
\end{equation}
where $\eta(t) \in \mathfrak{g}$ vanishes at the endpoints.
\item [{\bf iv}] The Euler--Poincar\'{e} equations hold on
$\mathfrak{g} \times V^\ast$
\begin{equation} \label{eulerpoincare1}
\frac{d}{dt} \frac{\delta l}{\delta \xi} = -
 \operatorname{ad}_{\xi}^{\ast} \frac{ \delta l }{ \delta \xi}
- \left(\frac{\delta l}{\delta a}\right) \diamond a.
\end{equation}
\end{enumerate}
\end{thm}

\paragraph{Right Representation and Left Invariant Lagrangian.} The set up
is as
follows:
\begin{itemize}
\item There is a {\it right\/} representation of Lie group $G$ on
the vector space $V$ and $G$ acts in the natural way on the {\it
left\/} on $TG \times V^\ast$: $h(v_g, a) = (hv_g, ah^{-1})$.
\item Assume that the function $ L : T G \times V ^\ast
\rightarrow \mathbb{R}$ is left $G$--invariant.
\item In particular, if $a_0 \in V^\ast$, define the
Lagrangian $L_{a_0} : TG \rightarrow \mathbb{R}$ by
$L_{a_0}(v_g) = L( v_g, a_0)$. Then $L_{a_0}$ is left
invariant under the lift to $TG$ of the left action of
$G_{a_0}$ on $G$.
\item  Left $G$--invariance of $L$ permits us to define
$l: {\mathfrak{g}} \times V^\ast \rightarrow \mathbb{R}$ by
\[
l(g^{-1}v_g, a_0g) = L(v_g, a_0).
\]
Conversely,  this relation defines for any
$l: {\mathfrak{g}} \times V^\ast \rightarrow
\mathbb{R} $ a left $G$--invariant function
$ L : T G \times V ^\ast
\rightarrow \mathbb{R} $.
\item For a curve $g(t) \in G, $ let
$\xi (t) := g(t) ^{ -1}\dot{g}(t)$ and define the curve
$a(t)$ as the unique solution of the linear differential equation
with time dependent coefficients $\dot a(t) = a(t)\xi(t)$
with initial condition $a(0) = a_0$. The solution can be
equivalently written as $a(t) = a_0g(t)$.
\end{itemize}

\begin{thm} \label{rall}
 The following are equivalent:
\begin{enumerate}
\item [{\bf i} ] Hamilton's variational principle
\begin{equation} \label{hamiltonprincipleright}
\delta \int _{t_1} ^{t_2} L_{a_0}(g(t), \dot{g} (t)) dt = 0
\end{equation}
holds, for variations $\delta g(t)$
of $ g (t) $ vanishing at the endpoints.
\item [{\bf ii}  ] $g(t)$ satisfies the Euler--Lagrange
equations for $L_{a_0}$ on $G$.
\item [{\bf iii} ]  The constrained variational principle
\begin{equation} \label{variationalprincipleright}
\delta \int _{t_1} ^{t_2}  l(\xi(t), a(t)) dt = 0
\end{equation}
holds on $\mathfrak{g} \times V ^\ast $, using variations of the form
\begin{equation} \label{variationsright}
\delta \xi = \dot{\eta } + [\xi , \eta ], \quad
\delta a =  a\eta ,
\end{equation}
where $\eta(t) \in \mathfrak{g}$ vanishes at the endpoints.
\item [{\bf iv}] The Euler--Poincar\'e equations hold on
$\mathfrak{g} \times V^\ast$
\begin{equation} \label{eulerpoincareright}
\frac{d}{dt} \frac{\delta l}{\delta \xi} =
 \operatorname{ad}_{\xi}^{\ast} \frac{\delta l}{\delta \xi}
- \frac{\delta l}{\delta a} \diamond a.
\end{equation}
\end{enumerate}
\end{thm}


\section*{References} \label{sec-refs}
\begin{description}

\item Abarbanel, H.D.I., D.D. Holm, J.E. Marsden, and T.S.
Ratiu [1986]  Nonlinear stability analysis of stratified
fluid equilibria,  {\it Phil. Trans. Roy. Soc. London A\/}
{\bf 318}, 349--409; also Richardson number criterion for
the nonlinear stability of three-dimensional stratified
flow, {\it Phys. Rev. Lett.\/} {\bf 52} [1984], 2552--2555.

\item Abraham, R. and J.E. Marsden [1978]
{\it Foundations of Mechanics\/}, Second Edition,
Addison-Wesley.

\item Abraham, R., J.E. Marsden, and T.S. Ratiu [1988]
{\it Manifolds, Tensor Analysis, and Applications.\/}  Second
Edition, Applied Mathematical Sciences {\bf 75},
Springer-Verlag.

\item Alber, M.S., R. Camassa, D.D. Holm and J.E. Marsden [1994]
The geometry of peaked solitons and billiard solutions of a class of
integrable pde's, {\it Lett. Math.
Phys.\/} {\bf 32}, 137--151.

\item Alber, M.S., R. Camassa, D.D. Holm and J.E. Marsden [1995]
On the link between umbilic geodesics and soliton
solutions of nonlinear PDE's, {\it Proc. Roy. Soc} {\bf
450}, 677--692.

\item Alber, M.S., R. Camassa, D.D. Holm and J.E. Marsden [1997]
The geometry of new classes of weak billiard solutions of
nonlinear pde's, {\it in preparation}.

\item Andrews, D.G. and McIntyre, M.E. [1978a] An exact theory of
nonlinear waves on a Lagrangian-mean flow,
{\it J. Fluid Mech.} {\bf 89}, 609-646.

\item Andrews, D.G. and McIntyre, M.E. [1978b] On wave action and its
relatives,  {\it J. Fluid Mech.} {\bf 89}, 647-664.
(Corrigendum {\bf 95}, 796.)

\item Arnold, V.I. [1966a]
Sur la g\'{e}om\'{e}trie differentielle
des groupes de Lie de dimenson
infinie et ses applications \`{a}
l'hydrodynamique des fluids parfaits,
{\it Ann. Inst. Fourier, Grenoble\/} {\bf 16}, 319--361.

\item Arnold, V.I. [1966b]
On an a priori estimate in the theory of
hydrodynamical stability,
{\it Izv. Vyssh. Uchebn. Zaved. Mat. Nauk\/}
{\bf 54}, 3--5; English
Translation: {\it Amer. Math. Soc. Transl.\/}
{\bf 79} [1969], 267--269.

\item Arnold, V.I. [1966c]
Sur un principe variationnel pour les d\'ecoulements
stationnaires des liquides parfaits et ses
applications aux problemes de stabilit\'e non lin\'eaires,
{\it  J. M\'ecanique\/} {\bf 5}, 29--43.

\item Arnold, V.I. (ed.) [1988]
{\it Dynamical Systems III.\/}
Encyclopedia of Mathematics {\bf 3}, Springer-Verlag.

\item Bloch, A.M., P.S. Krishnaprasad,
J.E. Marsden, and  T.S. Ratiu [1994]
Dissipation induced instabilities,
{\it Ann. Inst. H. Poincar\'{e}, Analyse Nonlineare\/}
{\bf 11}, 37--90.

\item Bloch, A.M., P.S. Krishnaprasad,
J.E. Marsden, and T.S. Ratiu [1996]
The Euler--Poincar\'{e} equations and
double bracket dissipation,
{\it Comm. Math. Phys.}  {\bf 175}, 1--42.

\item Bretherton, F.P. [1970]
A note on Hamilton's principle for perfect fluids,
{\it J. Fluid Mech.\/} {\bf 44}, 19-31.

\item Camassa, R. and D.D. Holm [1993]
An integrable shallow water equation with peaked solitons,
{\it Phys. Rev. Lett.\/}, {\bf 71}, 1661-1664.

\item Camassa, R., D.D. Holm and J.M. Hyman [1994]
A new integrable shallow water equation,
{\it Adv. Appl. Mech.}, {\bf 31}, 1--33.

\item  Cendra, H., D.D. Holm, M.J.W. Hoyle and J.E. Marsden [1997]
Euler--Poincar\'{e} equations for Maxwell-Vlasov dynamics,
{\it preprint}.

\item Cendra, H., D.D. Holm, J. E. Marsden and T. S. Ratiu [1997]
Lagrangian Reduction, the Euler--Poincar\'{e}
Equations, and Semidirect Products,
{\it preprint}.

\item Cendra, H., A. Ibort, and J.E. Marsden [1987]
Variational principal fiber bundles: a geometric
theory of Clebsch potentials
and Lin constraints, {\it J.  Geom.  Phys.\/} {\bf 4}, 183--206.

\item Cendra, H. and J.E. Marsden [1987]
Lin constraints, Clebsch potentials and variational principles,
{\it Physica D\/} {\bf 27}, 63--89.

\item  Cendra, H.,  J. E. Marsden and T.S. Ratiu [1997]
Lagrangian Reduction, the Euler--Poincar\'{e}
Equations, and Semidirect Products. {\it To appear in the AMS Arnold
volume II.}

\item  Cendra, H.,  J. E. Marsden and T.S. Ratiu [1997]
Lagrangian reduction by stages. {\it preprint.}

\item Chetayev, N.G. [1941]
On the equations of Poincar\'{e},
{\it J. Appl. Math. Mech.\/} {\bf 5}, 253--262

\item Chandrasekhar, S. [1967]
Ellipsoidal figures of equilibrium-an historical account,
{\it Comm. on Pure and Math\/} {\bf 20}, 251--265.

\item Chandrasekhar, K. [1977]
{\it Ellipsoidal Figures of Equilibrium}.  Dover.

\item Clebsch, A. [1857]
\"{U}ber eine allgemeine Transformation der
hydrodynamischen Gleichungen,
{\it Z. Reine Angew. Math.\/} {\bf 54}, 293--312.

\item Clebsch, A. [1859]
\"{U}ber die Integration der hydrodynamischen Gleichungen,
{\it Z. Reine Angew. Math.\/} {\bf 56}, 1--10.

\item Craik, A. D. D. and Leibovich [1976]
A rational model for Langmuir circulations,
{\it J. Fluid Mech.} {\bf 73}, 401-426.

\item Ebin, D.G. and J.E. Marsden [1970]
Groups of diffeomorphisms and the motion of an incompressible
fluid, {\it Ann. Math.\/} {\bf 92}, 102--163.

\item Faraday, M. [1831]
On a peculiar class of acoustical figures; and on certain forms assumed by
groups of particles upon vibrating elastic surfaces,
{\it Phil. Trans. Roy. Soc. London}, {\bf 121}, 299-340.

\item Foias, C., D. Holm and E. Titi [1988]
 {\it manuscript in preparation\/}.

\item Gjaja, I. and D.D. Holm [1996]
Self-consistent wave-mean flow interaction
dynamics and its Hamiltonian formulation for a rotating
stratified incompressible fluid,
{\it Physica D}, {\bf 98} (1996) 343-378.

\item Goncharov, V. and V. Pavlov [1997]
Some remarks on the physical foundations of the Hamiltonian
description of fluid motions,
{\it Euro. J. Mech. B} {\bf 16}, 509-555.

\item Guillemin, V. and S. Sternberg [1980]
The moment map and collective motion,
{\it Ann. of Phys.\/} {\bf 1278}, 220--253.

\item Guillemin, V. and S. Sternberg [1984]
{\it Symplectic Techniques in Physics.\/}
Cambridge University Press.

\item Hamel, G [1904]
Die Lagrange-Eulerschen Gleichungen der Mechanik,
{\it Z. f\"{u}r Mathematik u. Physik\/} {\bf 50}, 1--57.

\item Hamel, G [1949]
{\it Theoretische Mechanik\/}, Springer-Verlag.

\item Holm, D.D. [1987]
Hamiltonian dynamics and stability analysis
of neutral electromagnetic fluids with induction,
{\it Physica D} {\bf 25} (1987) 261--287.

\item Holm, D.D. [1996a]
Hamiltonian balance equations,
{\it Physica D}, {\bf 98} (1996) 379-414.

\item Holm, D.D. [1996b]
The Ideal Craik-Leibovich Equations,
{\it Physica D}, {\bf 98} (1996) 415-441.

\item Holm, D.D, C. Foias and E. Titi [1998]
The Camassa-Holm equations: their connection to the
Navier-Stokes equations and turbulence theory,
{\it In preparation}.

\item Holm, D.D. and B.A. Kupershmidt [1983]
Poisson brackets and Clebsch representations for
magnetohydrodynamics, multifluid plasmas, and elasticity,
{\it Physica D\/} {\bf 6}, 347--363.

\item Holm, D.D., J.E. Marsden, and T.S. Ratiu [1986]
The Hamiltonian structure of continuum mechanics in
material, spatial and convective representations, {\it
S\'{e}minaire de Math\'{e}matiques sup\'{e}rie,
Les Presses de L'Univ. de Montr\'{e}al\/} {\bf 100}, 11--122.

\item Holm, D.D. and V. Zeitlin [1997]
Hamilton's principle for quasigeostrophic motion,
{\it Phys. Fluids}, to appear.

\item Holmes, P.J. and J.E. Marsden [1983]
Horseshoes and Arnold diffusion for Hamiltonian
systems on Lie groups,
{\it Indiana Univ. Math. J.\/} {\bf 32}, 273--310.

\item Kaluza, Th. [1921] Zum Unitatsproblem der Physik,
{\it Sitzungsber. Preuss. Akad. Wiss. Phys. Math. Klasse} 966--972.

\item Klein, M. [1970] {\it Paul Eherenfest}, North-Holland.

\item Klein, O. [1926] Quantentheorie und funfdimensionale
Relativitatstheorie,
{\it Z. f. Physik} {\bf 37}, 895-906.

\item Kouranbaeva, S. [1997]
Geodesic spray form of the Camassa-Holm equation, {\it preprint}.

\item Kupershmidt, B.A. and T. Ratiu [1983] Canonical maps
between semidirect products with applications to elasticity and
superfluids, {\it Comm. Math. Phys.\/} {\bf 90}, 235--250.

\item Lagrange, J.L. [1788]
{\it M\'{e}canique Analitique.\/}
Chez la Veuve Desaint

\item Leonard, N.E. and J.E. Marsden [1997]
Stability and drift of underwater vehicle dynamics:
mechanical systems with rigid motion symmetry, {\it Physica D\/}
{\bf 105}, 130--162.

\item Lie, S. [1890]
{\it Theorie der Transformationsgruppen, Zweiter Abschnitt.\/}
Teubner, Leipzig.

\item Low, F.E. [1958]
A Lagrangian formulation of the Boltzmann--Vlasov equation for
plasmas,  {\it Proc. Roy. Soc. Lond. A\/} {\bf 248}, 282--287.

\item Marsden, J.E. [1982]
A group theoretic approach to the equations of plasma physics,
{\it Can. Math. Bull.\/} {\bf 25}, 129--142.

\item Marsden, J.E. [1992]
{\it Lectures on Mechanics\/} London Mathematical
Society Lecture note series, {\bf 174}, Cambridge University Press.

\item Marsden, J.E. and T.J.R. Hughes [1983]
{\it Mathematical Foundations of Elasticity},
Prentice-Hall, Inc., Englewood Cliffs, NJ.

\item Marsden, J.E., G. Misiolek, M. Perlmutter and T.S. Ratiu [1997]
Reduction by stages and group extensions,  {\it in preparation\/}.

\item Marsden, J.E., G.W. Patrick and S. Shkoller [1997]
Variational methods in continuous and discrete
mechanics and field theory. {\it preprint}.

\item Marsden, J.E. and T.S. Ratiu [1994]  {\it Introduction to
Mechanics and Symmetry\/}, Texts in Applied Mathematics, {\bf  17},
Springer-Verlag.

\item Marsden, J.E., T.S. Ratiu, and A. Weinstein [1984a]
Semi-direct products and reduction in mechanics,
{\it Trans. Am. Math. Soc.\/} {\bf 281}, 147--177.

\item Marsden, J.E., T.S. Ratiu, and A. Weinstein [1984b]
Reduction and Hamiltonian structures on
duals of semidirect product Lie algebras,
{\it Cont. Math. AMS\/} {\bf 28}, 55--100.

\item Marsden, J.E. and J. Scheurle [1993a]
Lagrangian reduction and the double spherical pendulum,
{\it ZAMP\/} {\bf 44}, 17--43.

\item Marsden, J.E. and J. Scheurle [1993b]
The reduced Euler-Lagrange equations,
{\it Fields Institute Comm.\/} {\bf 1}, 139--164.

\item Marsden, J.E., G.W. Patrick and S. Shkoller [1997]
Variational methods in continuous and discrete
mechanics and field theory. {\it preprint}.

\item Marsden, J.E. and S. Shkoller [1997] Multisymplectic geometry,
covariant Hamiltonians and water waves {\it Math. Proc. Camb. Phil.
Soc., to appear}.

\item Marsden, J.E. and A. Weinstein [1974]  Reduction of
symplectic manifolds with symmetry,  {\it Rep. Math. Phys.\/}
{\bf 5}, 121--130.

\item Marsden, J.E. and A. Weinstein [1982]
The Hamiltonian structure of the Maxwell-Vlasov equations,
{\it Physica D\/} {\bf 4}, 394--406.

\item Marsden, J.E. and A. Weinstein [1983]
Coadjoint orbits, vortices and Clebsch variables
for incompressible fluids, {\it Physica D\/} {\bf 7}, 305--323.

\item Marsden, J.E., A. Weinstein, T.S. Ratiu,
R. Schmid, and R.G. Spencer [1983]
Hamiltonian systems with symmetry, coadjoint
orbits and plasma physics, in
Proc. IUTAM-IS1MM Symposium on
{\it Modern Developments in Analytical Mechanics\/},
Torino 1982, {\it Atti della Acad. della Sc. di Torino\/}
{\bf 117}, 289--340.

\item  Marsden, J.E. and Wendlandt, J.M. [1997] Mechanical
systems with symmetry, variational principles, and integration algorithms,
in \textit{Current and Future Directions in Applied Mathematics},
edited by M. Alber, B. Hu and J. Rosenthal, Birkh\"auser, pp 219-261.

\item Misiolek, G. [1993]
Stability of Flows of Ideal Fluids and the Geometry
of the Group of Diffeomorphisms.
{\it Indiana Univ. Math. Journal\/} {\bf 42}, 215-235.

\item Misiolek, G. [1996] Conjugate points in ${\cal D}(T^2)$.
Proc. Amer. Math. Soc. {\bf 124} 977--982.

\item Misiolek, G. [1997a] Conjugate points in the
Bott-Virasoro group and the KdV equation. {\it Proc. Amer.
Math. Soc.} {\bf 125}, 935--940.

\item Misiolek, G. [1997b] A shallow water equation as a geodesic
flow on the Bott-Virasoro group. {\it J. Geom. Phys.}, to
appear.

\item Newcomb, W.A. [1958]
Appendix in Bernstein, B. [1958]
Waves in a plasma in a magnetic field.
{\it Phys. Rev.\/} {\bf 109}, 10--21.

\item Newcomb, W.A. [1962]
Lagrangian and Hamiltonian methods in magnetohydrodynamics,
{\it Nuc. Fusion. Suppl., part 2}, 451--463.

\item Ono, T. [1995a]
Riemannian geometry of the motion of an ideal
incompressible magnetohydrodynamical fluid,
{\it Physica D} {\bf 81}, 207--220.

\item Ono, T. [1995b]
A Riemannian geometrical description for Lie-Poisson systems and its
application to idealized magnetohydrodynamics
{\it J. Phys. A} {\bf 28}, 1737-1751.

\item Ovsienko, V.Y. and B.A. Khesin [1987]
Korteweg-de Vries superequations as an Euler equation.
{\it Funct. Anal.
and Appl.\/} {\bf 21}, 329--331.

\item Pedlosky, J. [1987] {\it Geophysical Fluid Dynamics}, 2nd
Edition, Springer, New York.

\item Poincar\'{e}, H. [1885]
Sur l'\'{e}quilibre d'une masse fluide anim\'{e}e d'un mouvement de
rotation,  {\it Acta. Math.\/} {\bf 7}, 259.

\item Poincar\'{e}, H. [1890] {\it Th\'eorie des tourbillons},
Reprinted by \'Editions Jacques Gabay, Paris.

\item Poincar\'{e}, H. [1890]
Sur le probl\`{e}me des trois corps et les \'{e}quations de la
dynamique, {\it Acta Math.\/} {\bf 13}, 1--271.

\item Poincar\'e, H. [1892--1899],
{\it Les M\'ethodes Nouvelles de la M\'ecanique Celeste.\/}
3 volumes. English translation {\it New Methods of
Celestial Mechanics.\/} History of Modern Physics and
Astronomy {\bf 13}, Amer. Inst. Phys., 1993.

\item Poincar\'{e}, H. [1892] Les formes d'\'{e}quilibre d'une
masse fluide en rotation,  {\it Revue G\'{e}n\'{e}rale des
Sciences\/} {\bf 3}, 809--815.

\item Poincar\'{e}, H. [1901a] Sur la stabilit\'{e} de
l'\'{e}quilibre  des figures piriformes affect\'{e}es par une
masse fluide en rotation, {\it Philosophical Transactions A\/}
{\bf 198}, 333--373.

\item Poincar\'{e}, H. [1901b] Sur une forme nouvelle des
\'{e}quations de la m\'{e}chanique,  {\it C.R. Acad. Sci.\/} {\bf
132}, 369--371.

\item Poincar\'{e}, H. [1910] Sur la precession des corps
deformables, {\it Bull Astron\/} {\bf 27}, 321--356.

\item Ratiu. T.S.  [1980], {\it Thesis\/}, University of
California at Berkeley.

\item Ratiu. T.S. [1981]  Euler-Poisson equations on Lie
algebras and the $N$-dimensional heavy rigid body,
{\it Proc. Natl. Acad. Sci. USA \/} {\bf 78}, 1327--1328.

\item Ratiu, T.S. [1982]  Euler-Poisson equations on Lie
algebras and  the $N$-dimensional heavy rigid body,
{\it Am. J. Math.\/} {\bf 104}, 409--448, 1337.

\item Simo, J.C., D.R. Lewis, and J.E. Marsden [1991]
Stability of relative equilibria I: The reduced
energy momentum method,
{\it Arch. Rat. Mech. Anal.\/} {\bf 115}, 15-59.

\item Sudarshan, E.C.G. and N. Mukunda [1974]
{\it Classical Mechanics: A Modern Perspective.\/}
Wiley, New York, 1974; Second Edition, Krieber,
Melbourne--Florida, 1983.

\item Thirry, Y. [1948]
Les equations de la theorie unitaire de Kuluza,
{\it Comptes Rendus} (Paris) {\bf 226} 216--218.

\item Vinogradov, A.M. and B.A. Kupershmidt [1977]
The structures of Hamiltonian mechanics,
{\it Russ. Math. Surv.\/} {\bf 32}, 177--243.

\item Wendlandt, J.M. and J.E. Marsden [1997]  Mechanical
integrators derived from a discrete variational principle,  {\it
Physica D\/} {\bf  106}, 223--246.

\item Whitham, G.B. [1974]
{\it Linear and Nonlinear Waves},
Wiley-Interscience, pp. 461-462.

\item Zeitlin, V. and Kambe, T. [1993]
2-dimensional ideal magnetohydrodynamics and differential geometry,
{\it J. Phys. A} {\bf 26}, 5025-5031

\item  Zeitlin, V. and Pasmanter, R.A. [1994] On the differential
geometry approach to the geophysical flows,
{\it Phys. Lett. A} {\bf 189}, 59--63.

\end{description}

\end{document}